%% file: Draft_V14.tex
\documentclass{article}

\usepackage{jheppub}

\usepackage{amsmath, amssymb, amsthm} 
\usepackage{mathtools}
\usepackage{thmtools}
\usepackage{bm}
\usepackage{dsfont}
\usepackage{braket}
\usepackage{graphicx}
\usepackage{parskip}
\usepackage{xcolor}

\numberwithin{equation}{section}
\usepackage{soul}
\usepackage{enumitem}
\setlist[itemize]{noitemsep}
\setlist[description]{noitemsep}

\usepackage[dvipsnames]{xcolor}
\usepackage{tikz}
\usetikzlibrary{positioning, tikzmark}

\usepackage[numbers, sort&compress]{natbib}
\usepackage{hyperref}
\hypersetup{
    linkcolor=[rgb]{0.91, 0.24, 0.43}, 
    citecolor=[rgb]{0.28, 0.24, 0.75}
    }

\newcommand{\dd}{\text{d}}

\newcommand{\pp}{\partial}

\newcommand{\ol}{\overline}

\newcommand{\zb}{\overline{z}}
\renewcommand{\[}{\left[}
\renewcommand{\]}{\right]}
\renewcommand{\(}{\left(}
\renewcommand{\)}{\right)}

\renewcommand{\epsilon}{\varepsilon}
\newcommand{\dzz}{\dd z \, \dd\zb \,}

\newtheorem{identity}{Identity}

\usepackage{multirow}

\usepackage{tabularray} 
\UseTblrLibrary{booktabs}

\begin{document}

\date{\today}

\title{Holomorphic structure of massive scalar fields in $\text{(A)dS}_2$}

\author[a,b,c]{Calvin Y.-R. Chen,}
\author[c]{Lukas W. Lindwasser,}
\author[d]{and Massimo Porrati}

\affiliation[a]{Max Planck-IAS-NTU Center for Particle Physics, Cosmology and Geometry, Taipei 10617, Taiwan}
\affiliation[b]{Leung Center for Cosmology and Particle Astrophysics, Taipei 10617, Taiwan}
\affiliation[c]{Department of Physics and Center for Theoretical Physics,
National Taiwan University, Taipei 10617, Taiwan}
\affiliation[d]{Center for Cosmology and Particle Physics
Department of Physics, New York University
726 Broadway, New York, NY 10003, USA}
    
\emailAdd{cyrchen@ntu.edu.tw}
\emailAdd{llindwasser@ntu.edu.tw}
\emailAdd{massimo.porrati@nyu.edu}

\abstract{
Scalar field theories in $\text{(A)dS}_{2}$ with integer scaling dimensions $\Delta = k+1$ are characterised by the existence of a pair of (anti-)holomorphic higher-spin currents.
We explore the consequences of this to describe their quantisation and subsets of their linear and non-linear symmetries, taking care to treat $\text{AdS}_{2}$ and $\text{dS}_{2}$ separately.
In particular, we point out that the theories admit mode expansions reminiscent of standard two-dimensional conformal field theories in complex coordinates, with which we are able to construct a large set of symmetry transformations which includes Virasoro symmetry.
We further leverage holomorphicity of the currents to show that a significant subset of the full symmetries of theories with $k>0$ is captured by a chiral algebra, which is a subalgebra of the one in the $k=0$ (massless) theory.
This allows us to identify integrable deformations for $k \in \{0,1,2\}$.
We finally observe that a lack of integrable deformations for $k>2$ is a consequence of a known conjecture. 
}

\newpage

\maketitle

\section{Introduction}

It is well-known that holomorphicity and conformal transformations in two dimensions are closely connected \cite{Belavin:1984vu}. 
On any given Riemann surface, we may choose complex coordinates $(z,\zb)$, for which two-dimensional conformal transformations $z \mapsto f(z)$ are generated by holomorphic functions $f(z)$. 
As a result of this, in a conformally invariant theory, complex coordinates are a natural choice which allow us to fully leverage the power of complex analysis. 
Particularly noteworthy are local functionals $\mathcal{O}(z)$ of the field which are holomorphic on-shell, i.e. for which $\ol{\pp} \mathcal{O} = 0$ after imposing the equations of motion\footnote{We use the term `holomorphic' as is commonly done in the physics literature, i.e. to refer to a lack of dependence on the complex conjugate coordinate $\zb$. 
In particular, these functions need not actually be \textit{holomorphic} in the mathematical sense.}. 
For us, a particularly interesting consequence of this is holomorphic splitting. 
Holomorphicity of the functional $\mathcal{O}$ implies that there exists a local operation on the fundamental field which isolates the holomorphic sector of the classical phase space---this is frequently used to deduce a decomposition of the classical solution space into holomorphic and anti-holomorphic parts.
For instance, every local conformal field theory in two dimensions has a traceless, symmetric, and conserved stress tensor $T_{\mu\nu}$ that, in complex coordinates, splits into the components $T(z) \equiv T_{zz}$ and $\ol{T}(\zb) \equiv T_{\zb\zb}$ which are respectively holomorphic and anti-holomorphic.
All such conformal field theories then in principle admit this type of decomposition; famous examples include massless scalar field theory, Liouville theory \cite{Liouville1855}, and the Wess--Zumino--Witten model \cite{Witten:1984}.

Let us zero in on a simple two-dimensional conformal field theory: the massless scalar field $\phi$ in two-dimensional flat space. 
The reader is strongly encouraged to keep the following discussion in the back of their mind as we move on.
This is described by the action
\begin{equation}
    S_{\text{massless}} = \int \dzz \pp \phi \,\ol{\pp} \phi,
    \label{eq: flat action} 
\end{equation}
from which one derives the equation of motion
\begin{equation}
    \pp \ol{\pp}\phi = 0.
    \label{eq: flat eom}
\end{equation}
The (not canonically normalised) (anti-)holomorphic stress tensor components of the massless free scalar \eqref{eq: flat action} are
\begin{equation}
    T(z) = \(\pp\phi\)^{2},\quad \ol{T}(\zb) = \(\ol{\pp}\phi\)^{2}.
    \label{eq: mless stress}
\end{equation}
Their (anti-)holomorphicity is sufficient to find holomorphic splitting of the scalar field:
\begin{equation}
    \phi(z,\zb)=\varphi(z)+\tilde{\varphi}(\zb),
    \label{eq: mless decomp}
\end{equation}
with arbitrary functions $\varphi$ and $\tilde{\varphi}$.
In the case of the massless free scalar, there exists a different set of currents 
\begin{equation}
    j(z,\zb) \equiv \pp \phi,\quad \ol{j}(z,\zb) \equiv \ol{\pp} \phi,
\end{equation}
which generate the symmetry of the action \eqref{eq: flat action} associated with shifts by a constant $c$, i.e. $\phi(z,\zb) \mapsto \phi(z,\zb) + c$, and are the lowest-order operators linear in the field $\phi$ which are invariant under such shifts. 
The currents $j$ and $\ol{j}$ are respectively holomorphic and anti-holomorphic.
This follows from (and is, in fact, equivalent to) the equation of motion for $\phi$:
\begin{equation}
    \ol{\pp} j(z,\zb) = 0,\quad \pp \ol{j}(z,\zb) = 0.
\end{equation}
This is also sufficient to show the decomposition \eqref{eq: mless decomp}. 
Furthermore, because the (anti-)holomorphic stress tensor components \eqref{eq: mless stress} can be built respectively from $j(z)$ and $\ol{j}(\zb)$\footnote{This is similar to the Sugawara construction of the stress tensor in the Wess--Zumino--Witten model \cite{Witten:1984}.}, (anti-)holomorphicity of $j(z)$ and $\ol{j}(\zb)$ implies conformal invariance.
In Lorentzian signature, the decomposition of the phase space is the familiar statement that the field in question is made up of non-interacting left- and right-moving waves propagating on the light cone.
One might hence expect that, as soon as the theory is deformed by a mass term, the nice properties of the massless theory described above disappear. 

\hspace{15pt} The purpose of this manuscript is to show that, to the contrary, certain massive fields in two-dimensional Anti-de Sitter and de Sitter space admit a holomorphic splitting and retain many of the features which distinguish massless scalar fields in flat spacetime, and are yet not invariant under local conformal transformations.
To be more precise, these massive fields of interest are the ones with integer scaling dimensions in $\text{(A)dS}_{2}$ studied in \cite{Farnsworth:2024yeh, Anninos:2023lin}, which are closely connected to the discrete series representation in $\text{dS}_{2}$ as for instance reviewed in \cite{Joung:2007je, Basile:2016aen, Sun:2021thf, Schaub:2024rnl} and are associated with extended shift symmetries discussed in \cite{Bonifacio:2018zex, Hinterbichler:2022vcc, Blauvelt:2022wwa, Bonifacio:2023prb, Hinterbichler:2024vyv}. 
In fact, at least in de Sitter, one might expect this holomorphic splitting from the representation theory of the isometry group.
We will refer to these as the shift-symmetric or discrete series scalars in $\text{AdS}_{2}$ and $\text{dS}_{2}$, even though we will shortly see that neither term is quite accurate.

Let us be more explicit.
For illustrative purposes, we will be deliberately vague at certain points, as some of the expressions for $\text{AdS}_{2}$ and $\text{dS}_{2}$ will differ.
Let us take $g$ to be the metric on $\text{(A)dS}_{2}$, and consider the action 
\begin{equation}
    S_{\text{massive}} = -\frac{1}{2}\int \dd^{2} x\sqrt{-g}\(g^{\mu\nu}\partial_\mu\phi\partial_\nu\phi+m^2\phi^2\),
    \label{eq: massive scalar action}
\end{equation}
of a scalar field $\phi$ with mass $m$.
In \cite{Bonifacio:2018zex}, it was shown that this action is invariant under shifts of the field $\phi$ by certain order-$k$ polynomials in embedding coordinates for $\text{(A)dS}_{2}$ when the mass is tuned to $m=m_{k}$, with 
\begin{equation}
    m_{k}^{2} =-\frac{k(k+1)}{2}R,\quad k \in \mathbb{Z}_{\geq 0},
    \label{eq: shift pt}
\end{equation}
and where $R$ is the Ricci scalar of $\text{(A)dS}_{2}$.
Distinguishing these discrete series scalar fields from scalar fields on $\text{(A)dS}_2$ with generic mass is the existence of the field strength 
\begin{equation}
    \mathcal{F}_{\mu_1\cdots\mu_{k+1}} = \nabla_{(\mu_1}\cdots\nabla_{\mu_{k+1})_T}\phi,
    \label{eq: field strength}
\end{equation}
where $(\dots )_{T}$ denotes projection onto the totally symmetric and traceless part.
In complex coordinates $(z,\zb)$, which are to be defined explicitly later, this splits into a pair of fields $F$ and $\ol{F}$ as
\begin{equation}
    \mathcal{F} = F(z,\zb) \,\dd z \otimes \dots \otimes \dd z + \ol{F}(z,\zb)\, \dd\zb \otimes \dots \otimes \dd \zb.
\end{equation}
Note that the massless scalar corresponds to $k=0$, in which case $F$ and $\ol{F}$ reduce to $j$ and $\ol j$.
The field strength $\mathcal{F}$ is associated with the extended shift symmetries, and is the lowest-order in derivatives operator which is linear in the field $\phi$ and invariant under the shift symmetries.
It is also conserved in the sense that 
\begin{equation}
    \nabla^\lambda \mathcal{F}_{\lambda \mu_1\cdots\mu_k}=0
\end{equation}
on-shell.
This implies the components of it are (anti-)holomorphic as a consequence of the equations of motion:
\begin{equation}
    \ol{\pp} F(z,\zb) =0,\quad \pp \ol{F}(z,\zb) =0.
    \label{eq: F (anti-)holomorphicity}
\end{equation}
Unlike for the massless theory in flat space, the expressions on the left-hand sides are now not equivalent to the equations of motion, but rather to certain order-$k$ differential operators acting thereon.
The field strengths $F$ and $\ol{F}$ also cannot be used to construct (anti-)holomorphic stress tensor components, so conformal invariance is not given.
Nonetheless, we can use \eqref{eq: F (anti-)holomorphicity} to argue that the classical space of solutions for the discrete series scalars decomposes as
\begin{equation}
    \phi(z,\zb) = \mathcal{D}_{k}\varphi(z) + \ol{\mathcal{D}}_{k}\tilde{\varphi}(\zb),
    \label{eq: holomorphic splitting}
\end{equation}
where $\varphi$ and $\tilde{\varphi}$ are arbitrary functions, and $\mathcal{D}_{k}$ and $\ol{\mathcal{D}}_{k}$ are order-$k$ differential operators to be defined shortly.
The entire remainder of this manuscript is devoted to exploring the consequences of this simple fact.

The first part of our results is based on the observation that \eqref{eq: F (anti-)holomorphicity} implies the following Laurent series
\begin{equation}
    F(z) = \sum_{n} \frac{\beta_{n}}{z^{n+k+1}},
\end{equation}
for some operators $\beta_{n}$.
From explicit computation of the mode expansion in (Lorentzian) conformal coordinates, we find that the finite-norm modes are reproduced by
\begin{equation}
    \beta_{n} = 
    \begin{cases}
        0, & |n| \leq k \\
        \lambda_{n} \alpha_{n},& |n| \geq k+1,
    \end{cases}
\end{equation}
where the $\lambda_{n}$ are fixed, numerical constants and $\alpha_{n}$ are operators obeying \textit{canonical} commutation relations $[\alpha_{m},\alpha_{n}] = m\delta_{m+n,0}$.
The mode expansion for $\ol{F}(\zb)$ follows \textit{mutatis mutandis}\footnote{This involves a second set of oscillator modes $\tilde{\alpha}_{n}$, which is independent from $\alpha_{n}$ in $\text{dS}_{2}$ and is related to $\alpha_{n}$ in $\text{AdS}_{2}$ due to the boundary conditions on the conformal boundary (the real axis).
This is reminiscent of the oscillator mode expansions of the massless free scalar conformal field theory without and with boundary or the open and closed string respectively.}.
Using these mode expansions, we are able to construct operators which implement the background isometries $\text{PSL}(2,\mathbb{R})$ and $\text{SO}(3)$ for $\text{AdS}_{2}$ and $\text{dS}_{2}$ respectively.
The observation from \cite{Farnsworth:2024yeh} that $F$ acts as a conformal primary under (global) transformations then follows from assigning it the holomorphic weights $(h,\ol{h}) = (k+1,0)$.
Furthermore, the fact that the current $F$ transforms under global conformal symmetry and is holomorphic raises the question of whether $F$ also transforms under \textit{local} conformal transformations in two dimensions. 
We argue that the previously identified operators implementing the isometries cannot be embedded into a larger set of operators implementing local conformal transformations.
Nonetheless, linear automorphisms on the space of oscillator modes are linear symmetries of the theory\footnote{We thank the anonymous referee for pointing this out.}, which includes non-locally acting Virasoro generators unrelated to the isometries.
We do not rule out the possibility of, but were unable to find generators which extend the algebra of global conformal transformations to the Virasoro algebra.

The consequences of \eqref{eq: F (anti-)holomorphicity} however go far beyond this.
The second half of this manuscript is concerned with the following observation: Given that \textit{partial} derivatives commute, the fact that $F$ is holomorphic \eqref{eq: F (anti-)holomorphicity} implies that any arbitrary functional $\mathcal{L}$ of $F$ itself, its derivatives $\pp^{n}F$, powers thereof $(\pp^{n}F)^{m}$, combined with arbitrary $z$-dependence will also be holomorphic:
\begin{equation}
    \ol{\pp}\mathcal{L}\(\{F,\pp F, \pp^{2}F,\dots\},z\) = 0.
\end{equation}
It goes without saying that similar holds for $\ol{F}$.
This describes an enormously large symmetry algebra!
We show that the structure of this algebra is determined by its flat-space limit, in which $F \rightarrow \pp^{k+1} \phi$.
This implies that the chiral symmetry of the discrete scalar theory at level $k$ is a subalgebra of the one of the massless theory at level $k=0$, which was studied in \cite{Lindwasser:2024qyh}.

In that case, it is known that the existence of infinite-dimensional mutually commuting subalgebras directly implies the existence of an infinite hierarchy of deformations which preserve an infinite number of the symmetries of the original theory, and are integrable \cite{Zamolodchikov:1989hfa,Lindwasser:2024qyh}. We are able to use the fact that the algebra for theories with level $k > 0$ is a subalgebra of the massless theory on flat space (for which infinite-dimensional commuting subalgebras are known) to determine integrable deformations of the discrete series scalars with $k \in \{0,1,2\}$. 
Since terms such as $\pp F$ are not components of a covariant derivative acting on the field strength, these integrable deformations break the background $\text{(A)dS}_{2}$ by construction.
This is therefore not in contradiction with the no-go result in \cite{Antunes:2025iaw} for $\text{AdS}_{2}$.
Finally, we point out that the apparent obstruction to finding integrable deformations for theories with $k>2$ is a consequence of a conjecture \cite{Sanders:1998,Heredero:2019arc}.

\subsection*{Organisation}

This manuscript is organised as follows.
We start in sections \ref{sec: ads2} and \ref{sec: ds2} by studying the discrete series scalars in $\text{AdS}_{2}$ and then $\text{dS}_{2}$.
Despite various shared features, there are sufficiently many non-trivial differences that we decided to split the discussions into two separate sections running essentially in parallel. 
In either case, we first discuss the basic properties of $\text{(A)dS}_{2}$ in conformal coordinates, then the solutions to the classical equations of motion, and canonical quantisation thereof.
We then analytically continue to $\text{E(A)dS}_{2}$ in complex coordinates and present the mode expansions hinted at in the introduction.
In section \ref{sec: lin sym}, we use the operators appearing in the mode expansions to construct operators realising the $\text{(A)dS}_{2}$ isometries, the Ward identities of which constrain the form of correlation functions.
We argue that these cannot be embedded within a Virasoro algebra, but nonetheless construct a different set of operators which do realise that. 
We finish by commenting on the relation between the zero-modes and the shift symmetries of the fields.
In section \ref{sec: chiral}, we describe the chiral algebra of symmetries underlying the discrete series scalar fields.
For $k \leq 2$, we identify infinite-dimensional commuting subalgebras, which each define an infinite hierarchy of integrable deformations breaking the background isometries.
We offer some concluding remarks and comment on several future directions and works in progress in section \ref{sec: conclusion}.

Euclidean $\text{(A)dS}_{2}$ are examples of Riemann surfaces.
In appendix \ref{app: riemann surfaces}, we introduce differential operators in complex isothermal coordinates on generic Riemann surfaces, and prove several identities which they satisfy.
Relegated to appendix \ref{app: holomorphicity} is an explicit derivation for the holomorphicity of the field strength (in $\text{AdS}_{2}$).
Explicit expressions for the transformation properties of the mode functions in $\text{dS}_{2}$ are listed in appendix \ref{app: ds2 isometry action}.
Finally, we explicitly show that the discrete series scalar action is invariant of under chiral symmetry transformations  in appendix \ref{app: explicit invariance}.

\subsection*{Conventions}

We consider two-dimensional manifolds with mostly-plus convention, i.e. $(\text{time},\text{space})=(-,+)$ in Lorentzian signature and $(+,+)$ in Euclidean signature.
We denote the radius of AdS by $L$ and the Hubble scale of dS by $H$.
Some (but not all!) expressions can be analytically continued from AdS to dS and \textit{vice versa} by taking $-L^{2} \leftrightarrow 1/H^{2}$.
We will mostly use complex coordinates on Euclidean $\text{(A)dS}_{2}$.
Instead of using explicit indices, it is useful to denote (anti-)holomorphic tensors $T_{z \dots z}$ ($T_{\zb \dots \zb}$) by $T$ ($\ol{T}$).
Similarly, partial derivatives $\pp_{z}$ ($\pp_{\zb}$) and covariant derivatives $\nabla_{z}$ ($\nabla_{\zb}$) are denoted by $\pp$ ($\ol{\pp}$) and $\nabla$ ($\ol{\nabla}$) respectively.
Tensor indices are (anti-)symmetrised with unit weight and are denoted by round (square) brackets, and the subscript $T$ further denotes the traceless projection. 
The theories of interest are scalar fields with masses given by \eqref{eq: shift pt}.
We will refer to these as shift-symmetric or discrete series scalar fields, even though neither terms are completely accurate.
When discussing the chiral symmetry algebra of these theories in section \ref{sec: chiral}, it is convenient to denote the number of partial derivatives of the field strength $F$ by a subscript, i.e. $F_{n} \equiv \pp^{n}F$. 


\section{Discrete series scalars in $\text{AdS}_{2}$ \label{sec: ads2}}

In $\text{AdS}$, there exists a series of scalar fields with discrete values of masses which have no flat-space analogue, possessing a current $\mathcal{F}$ associated with enhanced shift symmetries.
In actuality, these are not physical symmetries of the theory, even though their actions are indeed left invariant by shifts of the field by order-$k$ polynomials in embedding space \cite{Bonifacio:2018zex}.
To follow standard terminology in $\text{dS}$, we shall refer to this series as the discrete series scalars\footnote{Incidentally, there is another notion of \textit{discrete series representation} in $\text{AdS}$, which in two dimensions corresponds to all scaling dimensions $\Delta>1/2$, see e.g. \cite{Higuchi:2021fxg}. 
Analytically continuing to $\text{dS}$, these theories with $\Delta=k+1\in\mathbb{N}$ actually realise the discrete series representation in $\text{dS}$ \cite{Joung:2007je, Sun:2021thf}. 
We shall discuss these in section \ref{sec: ds2}.
Note the isometries of $\text{AdS}_2$ and $\text{dS}_2$ are both isomorphic to $\text{SO}(2,1)$.}.

We begin in section \ref{ssec: AdS lorentz} by a discussion of these theories in conformal coordinates, before moving to complex coordinates in Euclidean signature in section \ref{ssec: UHP}.


\subsection{Strip \label{ssec: AdS lorentz}}

\begin{figure}
    \centering
    \resizebox{0.95\columnwidth}{!}{
        \input{figures/ads.tikz}
    }
    \caption{$\text{(E)AdS}_{2}$ in different coordinate systems. 
    Depicted on the left is the Penrose diagram of describing the conformal compactification of Lorentzian $\text{AdS}_{2}$ with metric \eqref{eq: conformal ads metric}. 
    After analytic continuation and under the transformation \eqref{eq: complex coordinates ads}, this is mapped to the UHP, depicted on the right. 
    The two disjoint conformal boundaries of $\text{AdS}_{2}$ at $\rho = \pm \pi/2$ in conformal coordinates are mapped to the negative and positive parts of the real axis, respectively in red and blue.}
    \label{fig: ads}
\end{figure}
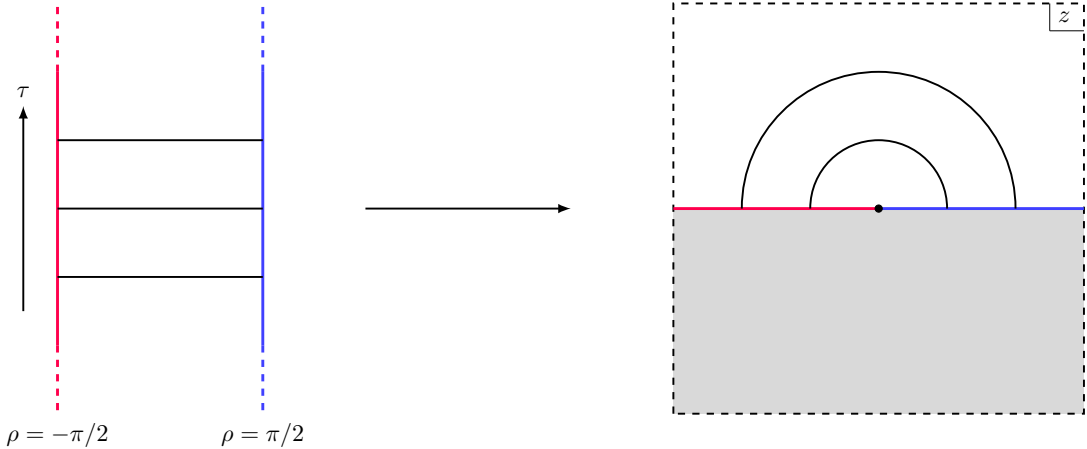

Consider Lorentzian Anti-de Sitter space in two dimensions ($\text{AdS}_{2}$) with radius $L$ and scalar curvature $R = -2/L^{2}$. 
Conformal coordinates $(\tau,\rho)$ cover the conformal compactification of the global patch, which is described by the metric 
\begin{equation}
    \dd s^{2} = \frac{L^{2}}{\cos(\rho)^{2}} (- \dd \tau^{2} + \dd \rho^{2}),
    \label{eq: conformal ads metric}
\end{equation}
with $\tau \in \mathbb{R}$ and $\rho \in (-\pi/2,\pi/2)$.
Topologically this is $\mathbb{R} \times I$ with the conformal boundary being the disjoint union of the time-like surfaces at $\rho = \pm \pi/2$, see figure \ref{fig: ads}.

For the massive scalar field $\phi$ in $\text{AdS}_{2}$ with action \eqref{eq: massive scalar action}, the shift-symmetric points are now defined by
\begin{equation}
    m_{k}^{2} L^{2} =k(k+1),\quad k \in \mathbb{Z}_{\geq 0}.
\end{equation}
In this case, we impose Dirichlet boundary conditions, so that $\phi(\tau,\pm\pi/2)$ is fixed. 
Instead of working with the mass $m$ directly, it is convenient to define the $\text{AdS}_{2}$ scaling dimension
\begin{equation}
    \Delta = \frac{1}{2} \(1 + \sqrt{1 + 4m^{2}L^{2}}\),
\end{equation}
which takes integer values $\Delta=k+1$ when $m=m_{k}$.

As its classical equation of motion, the scalar field $\phi$ obeys the Klein--Gordon (KG) equation
\begin{equation}
    \begin{aligned}
        0 &= (\Box_{g} - m^{2})\phi \\
        &= \frac{\cos(\rho)^{2}}{L^{2}}\(-\pp_{\tau}^{2} + \pp_{\rho}^{2} - \frac{\Delta(\Delta-1)}{\cos(\rho)^{2}} \) \phi.
    \end{aligned}
    \label{eq: ads kg eq}
\end{equation}
Let us briefly review properties of the solutions to \eqref{eq: ads kg eq}, following \cite{Higuchi:2021fxg} specified to integer $\Delta$.
First note that the standard KG inner product associated with this theory is
\begin{equation}
\label{eq: AdS KG inner}
    (\phi_{1},\phi_{2}) = i \int_{-\pi/2}^{\pi/2} \dd\rho\(\ol{\phi_{1}} \pp_{\tau}\phi_{2} - \pp_{\tau}\ol{\phi_{1}}\phi_{2} \) .
\end{equation}
Due to the presence of a globally defined timelike Killing vector $\pp/\pp\tau$, the space of solutions to the KG equation \eqref{eq: ads kg eq} admits a unique orthogonal decomposition into positive- and negative-frequency solutions.
The former are spanned by the mode functions
\begin{equation}
    f_{\omega}(\tau,\rho) = \psi_{\omega}(\rho) e^{-i\omega\tau},\quad \omega >0,
\end{equation}
which are positive-definite with respect to \eqref{eq: AdS KG inner}. 
The spatial parts $\psi_{\omega}$ satisfy
\begin{equation}
    \(- \pp_{\rho}^{2}  + \frac{k(k+1)}{\cos(\rho)^{2}}\) \psi_{\omega}(\rho) = \omega^2\psi_\omega(\rho).
\end{equation}
Solutions to this with positive energy and which are square-integrable and $C^1$ on $\rho \in (-\pi/2,\pi/2)$ are subject to the quantisation condition $\omega \in \mathbb{Z}$ and $\omega \geq k+1$.
Explicitly,  
\begin{equation}
    \psi_{\omega}(\rho) = \frac{2^{k}k!}{\sqrt{\pi\omega\prod_{j=1}^k(\omega^2-j^2)}} \cos(\rho)^{k+1} C_{\omega-k-1}^{(k+1)}(\sin( \rho)),
    \label{eq: mode functions}
\end{equation}
where $C_n^{(\alpha)}(x)$ are Gegenbauer polynomials, and the normalisation was chosen such that $f_{\omega}$ are orthonormal with respect to the KG inner product \eqref{eq: AdS KG inner}: $(f_{\omega},f_{\omega'}) = \delta_{\omega\,\omega'}$.
Since the $\psi_{\omega}$'s are real, it is clear that the negative-frequency solutions are spanned by
\begin{equation}
    f_{-\omega}(\tau,\rho)=\ol{f_{\omega}}(\tau,\rho)  = \psi_{\omega}(\rho)e^{i\omega \tau},\quad \omega >0.   
\end{equation}
The mode expansion for solutions to the massive $\text{AdS}_{2}$ wave equation in conformal coordinates hence takes the form
\begin{equation}
    \begin{aligned}
        \phi(\tau,\rho) =& \sum_{\omega = k+1}^{\infty} a_{\omega} f_{\omega}(\tau,\rho) + a_{\omega}^{\dag} \ol{f_{\omega}}(\tau,\rho) \\
        =& \sum_{\omega=k+1}^{\infty} \psi_{\omega}(\rho) \(a_{\omega} e^{-i\omega \tau} + a_{\omega}^{\dag} e^{i\omega \tau}\),
    \end{aligned}
    \label{eq: mode expansion}
\end{equation}
where we have defined the operators 
\begin{equation}
    a_{\omega} = (f_{\omega},\phi),\quad a^{\dag}_{\omega} = -(\ol{f_{\omega}},\phi).
\end{equation}
The unique orthogonal decomposition of the solution space to the KG equation defined by $\pp/\pp\tau$ fixes $a_{\omega}^{\dag} = a_{-\omega}$.
By virtue of $f_{\omega}$ being orthonormal with respect to the KG inner product, these satisfy
\begin{equation}
    [a_{\omega},a_{\omega}^\dagger]=1.
    \label{eq: ads ccr}
\end{equation}
There is a unique vacuum state $|\Omega\rangle$ defined by the condition
\begin{equation}
    a_\omega|\Omega\rangle=0,
\end{equation}
and we can build our Fock space by repeatedly applying the creation operators $a_{\omega}^{\dag}$.


\subsection{Upper half-plane \label{ssec: UHP}}

We will now switch to describing the discrete series scalars in two-dimensional Euclidean Anti-de Sitter space ($\text{EAdS}_2$) using complex coordinates on the upper half-plane (UHP).

We first analytically continue to $\text{EAdS}_2$ coordinates via $\tau = -i\tau_{E}$.
Then,
\begin{equation}
    \dd s^{2} = \frac{L^{2}}{\cos(\rho)^{2}} (\dd \tau_{E}^{2} + \dd \rho^{2}).
\end{equation}
We will now introduce the complex coordinate\footnote{Note that these coordinates are distinct from the analytic continuation of the complex coordinates used in \cite{Farnsworth:2024yeh}, which maps $\text{EAdS}_{2}$ to the open disk.
Our choice will turn out to be more suitable for the mode expansions discussed shortly.}
\begin{equation}
    z = e^{\tau_{E} -i(\rho-\pi/2)},
    \label{eq: complex coordinates ads}
\end{equation}
such that
\begin{equation}
    \rho = \frac{i}{2} \log(z/\zb) + \frac{\pi}{2},\quad \tau_{E} = \frac{1}{2}\log(z\zb).
\end{equation}
Under this transformation, the entirety of $\text{EAdS}_2$ is mapped to the complex UHP, with the conformal boundary now being at the real axis.
In these coordinates, the metric takes the form 
\begin{equation}
    \dd s^{2} = -4L^{2}\frac{\dd z\,\dd \zb}{(z-\zb)^{2}}.
    \label{eq: UHP metric}
\end{equation}
These are isothermal coordinates \eqref{eq: general 2d metric} with $\Omega^{2} = -4L^{2}/(z-\zb)^{2}$.
Various geometric properties of this space, alongside useful differential operators and identities they satisfy are discussed in appendix \ref{app: riemann surfaces}.

$\text{EAdS}_{2}$ has three Killing vectors, which form the generators of $\text{PSL}(2,\mathbb{R})$.
In these coordinates, they take the form
\begin{subequations}
    \begin{align}
        \ell_{+1}&=-z^2\pp-\zb^2\ol\pp,\\ 
        \ell_0 &= -z\pp-\zb\ol\pp,\\
        \ell_{-1}&=-\pp-\ol\pp,
    \end{align}
    \label{eq: AdS ell}%
\end{subequations}
which satisfy the algebra
\begin{equation}
    [\ell_m,\ell_n]=(m-n)\ell_{m+n},\qquad m,n \in \{-1,0,1\}.
    \label{eq: psl2r algebra} 
\end{equation}
In these coordinates, the action and KG equation for the discrete series scalars take the form
\begin{equation}
    S = \int \dzz \(\pp\phi \ol{\pp}\phi - \frac{k(k+1)}{(z-\zb)^{2}} \phi^{2} \) 
\end{equation}
and
\begin{equation}
    \(\pp\ol{\pp} + \frac{k(k+1)}{(z-\zb)^{2}}\)  \phi = 0.
    \label{eq: uhp KG}
\end{equation}
The field strength $\mathcal{F}$ in \eqref{eq: field strength} is traceless by construction, so the only independent components of it are
\begin{equation}
    F \equiv \mathcal{F}_{z \dots z},\quad \ol{F} \equiv \mathcal{F}_{\zb \dots \zb}.
    \label{eq: field strength components}
\end{equation}
Covariant conservation of the field strength translates to \eqref{eq: F (anti-)holomorphicity}.
We show this explicitly in appendix \ref{app: holomorphicity}.

\subsubsection{Mode expansions}

The mode expansion of $\phi$ takes a particularly simple form on the UHP, in a way that manifests the holomorphic splitting these theories exhibit. 
As we will see shortly, the mode functions described in section \ref{ssec: AdS lorentz}, when converted to the complex UHP, strongly resemble a mode expansion typically encountered in $2d$ CFTs.

To describe these, we will make use of the order-$k$ differential operators $\mathcal{D}_k$ and $\ol{\mathcal{D}}_k$ constructed in appendix \ref{app: riemann surfaces}.
These are defined in \eqref{eq: mathcalD expressions}, the explicit forms of which for $\text{EAdS}_{2}$ are given in \eqref{eq: nabla mathcalD ads}.
More importantly, they can be shown to satisfy the identities \ref{idty: nabla mathcalD ads} and \ref{idty: ads cov d anti composition identity} in appendix \ref{sapp: eads2 identities}, which can be summarised as
\begin{equation}
    \begin{aligned}
        &\nabla^{k+1}\mathcal{D}_kf(z)=\partial^{2k+1}f(z), && \nabla_{\bar z}^{k+1}\mathcal{D}_kf(z)=0\\
        &\nabla^{k+1}\ol{\mathcal{D}}_k\tilde f(\bar z)=0, && \nabla_{\bar z}^{k+1}\ol{\mathcal{D}}_k\tilde f(\bar z)=\ol{\partial}^{2k+1}\tilde f(\bar z),
    \end{aligned}
    \label{eq: nabla mathcalD composition identities}
\end{equation}
with $f(z)$ and $\tilde{f}(\bar z)$ holomorphic and anti-holomorphic, respectively. 
At the heart of holomorphic splitting of the discrete series scalars is the following Ansatz
\begin{equation}
    \phi(z,\bar z)=\mathcal{D}_k\varphi(z) + \ol{\mathcal{D}}_k\tilde{\varphi}(\bar z),
    \label{eq: phi general solution varphi}
\end{equation}
for arbitrary holomorphic $\varphi(z)$ and anti-holomorphic $\tilde{\varphi}(\bar z)$.
Identity \ref{idty: commutator mathcalD ol pp} guarantees that this is a solution to the KG equation \eqref{eq: uhp KG} before imposing reality and square-integrability\footnote{We note \textit{en passant} that this solution may be written covariantly in terms of a real rank-$k$ symmetric and traceless tensor $W^{\mu_1\cdots\mu_k}$ as follows
\begin{equation}
    \phi=\nabla_{\mu_1}\cdots\nabla_{\mu_k}W^{\mu_1\cdots\mu_k},
\end{equation}
whose independent components satisfy $\nabla_{\zb}W^{z\cdots z}=0$ and $\nabla_zW^{\zb\cdots \zb}=0$.}. 
To demonstrate holomorphic splitting explicitly, we construct the functions $\varphi(z)$ and $\tilde{\varphi}(\zb)$ which reproduce the mode expansion \eqref{eq: mode expansion}.
This is achieved by
\begin{subequations}
    \begin{align}
        \varphi(z) =&-\frac{(-i)^{k+1}}{\sqrt{4\pi}}\sum_{|n|>k}\frac{1}{n}\frac{1}{\sqrt{\prod_{j=1}^k(n^2-j^2)}}\frac{\alpha_n}{z^{n-k}}, \\
        \tilde{\varphi}(\bar z) =&-\frac{i^{k+1}}{\sqrt{4\pi}}\sum_{|n|>k}\frac{1}{n}\frac{1}{\sqrt{\prod_{j=1}^k(n^2-j^2)}}\frac{\alpha_n}{\bar z^{n-k}},
    \end{align}
    \label{eq: UHP phi mode expansion}%
\end{subequations}
where we have introduced the notation 
\begin{equation}
    \alpha_n=\sqrt{n}a_n,\quad \alpha_{-n}=\sqrt{n}a_n^\dagger
\end{equation}
for $n>k$. 
These newly defined oscillator operators $\alpha_n$ satisfy the commutation relations
\begin{equation}
    [\alpha_m,\alpha_n]=m\delta_{m+n,0}.
\end{equation}
The field strengths $F(z)$ and $\ol F(\zb)$ can be readily computed using \eqref{eq: nabla mathcalD composition identities}
\begin{subequations}
    \begin{align}
        F(z)&=\frac{(-i)^{k+1}}{\sqrt{4\pi}}\sum_{|n|>k} \sqrt{\textstyle\prod_{j=1}^k(n^2-j^2)} \frac{\alpha_{n}}{z^{n+k+1}}\\
        \ol F(\bar z)&=\frac{i^{k+1}}{\sqrt{4\pi}}\sum_{|n|>k} \sqrt{\textstyle\prod_{j=1}^k(n^2-j^2)} \frac{\alpha_{n}}{\bar z^{n+k+1}}
    \end{align}
    \label{eq: UHP F mode expansion}%
\end{subequations}
Note the mode expansions \eqref{eq: UHP F mode expansion} make explicit that $F(z)$ and $\ol{F}(\zb)$ satisfy the following condition on the conformal boundary $z=\bar z$ of the UHP
\begin{equation}
    F(z)\sim (-1)^{k+1}\ol{F}(\bar z)\quad (\text{as } z\to\bar z).
\end{equation}
The (anti-)holomorphic sectors are therefore related, as is typical for fields with Dirichlet boundary conditions.
\subsubsection{Correlation functions}

In $\text{AdS}_2$ there is a unique choice of vacuum for scalar fields with $\Delta >3/2$, and therefore a unique two-point correlation function for $\phi(z,\zb)$. 
This two-point correlation function can be found in several standard texts, see for instance \cite{Penedones:2016voo}. 
In UHP coordinates, the scalar two-point correlation function for this field is
\begin{equation}
    \langle\phi(z,\zb)\phi(z',\zb')\rangle = \frac{2}{\pi}\frac{4^kk!(k+1)!}{(2k+2)!}\frac{1}{\zeta^{k+1}}{}_2F_1\left(k+1,k+1;2k+2;-\frac{4}{\zeta}\right),
    \label{eq: AdS phi 2pt}
\end{equation}
where
\begin{equation}
    \zeta=\frac{1}{L^2}|z-z'|^2\Omega(z,\zb)\Omega(z',\zb'),\quad \Omega(z,\zb)=\frac{2iL}{z-\zb}
\end{equation}
is the chordal distance\footnote{The chordal distance between two embedding space points $X$ and $Y$, this is given by $\zeta=(X-Y)^2/L^2$.}.
Correlation functions involving the field strengths $F(z)$ and $\ol{F}(\zb)$ are of particular interest due to their on-shell (anti-)holomorphicity. 
These can be calculated from \eqref{eq: AdS phi 2pt} by applying the appropriate number of covariant derivatives. 
Explicitly,
\begin{subequations}
    \begin{align}
        \langle F(z)F(z')\rangle&=(-1)^{k+1}\frac{1}{4\pi}(2k+1)!\frac{1}{(z-z')^{2k+2}}\\
        \langle \ol{F}(\zb)\ol{F}(\zb')\rangle&=(-1)^{k+1}\frac{1}{4\pi}(2k+1)!\frac{1}{(\zb-\zb')^{2k+2}}\\
        \langle F(z)\ol{F}(\zb')\rangle&=\frac{1}{4\pi}(2k+1)!\frac{1}{(z-\zb')^{2k+2}}.
        \label{eq: AdS FFbar cor}
    \end{align}
    \label{eq: AdS F cor}%
\end{subequations}
The form of the correlation functions \eqref{eq: AdS F cor} involving only $F(z)$ and $\ol{F}(\zb)$ are suggestive of correlation functions found in $2d$ CFTs with Dirichlet boundary conditions, see e.g. \cite{Blumenhagen:2009zz}.
Note in particular \eqref{eq: AdS FFbar cor}: In the coincident limit $z'=z$ this is neither vanishing nor singular (except on the conformal boundary $z=\zb$).
The former is typical of one-point functions in boundary CFTs, while the latter is a reflection of the conformal boundary conditions.
As we will see in section \ref{sssec: SL2R AdS}, the form of these correlation function is fixed by (anti-)holomorphicity as well as the $\text{EAdS}_2$ isometry Ward identities, which apply the same constraints global conformal symmetry would on correlation functions between primary operators with scaling dimension $\Delta=k+1$ and spin $s=\pm(k+1)$ (cf. \eqref{eq: conformal weights}). 


\section{Discrete series scalars in $\text{dS}_2$ \label{sec: ds2}}

In this section, we study the analogous theory of discrete series scalars in $\text{dS}_2$, which have been discussed in earlier works \cite{Anninos:2023lin, Hinterbichler:2024vyv}---see also \cite{Bros:2010wa, Epstein:2014jaa, Letsios:2024snc}.
As we will explain, the discrete series scalars in $\text{dS}_2$ share many of the features of the theory in $\text{AdS}_2$, but with essential differences.

Once again, we begin in section \ref{ssec: dS lorentz} by discussing the discrete series scalar theories in conformal coordinates.
In section \ref{ssec: S2}, we will also analytically continue to Euclidean signature and switch to complex coordinates $(z,\zb)$.
We will never refer to the theories defined on $\text{dS}_{2}$ and $\text{AdS}_{2}$ simultaneously, so we trust the distinction will be clear from context.


\subsection{Cylinder \label{ssec: dS lorentz}}

\begin{figure}
    \centering
    \resizebox{0.95\columnwidth}{!}{
        \input{figures/ds.tikz}
    }
    \caption{$\text{(E)dS}_{2}$ in different coordinate systems. 
    Depicted on the left is the Penrose diagram of (the conformal compactification of) Lorentzian $\text{dS}_{2}$ in conformal coordinates with metric \eqref{eq: conformal ds metric}. 
    Under the transformation \eqref{eq: complex coordinates ds}, this is mapped to the extended complex plane, depicted on the right.}
    \label{fig: ds}
\end{figure}
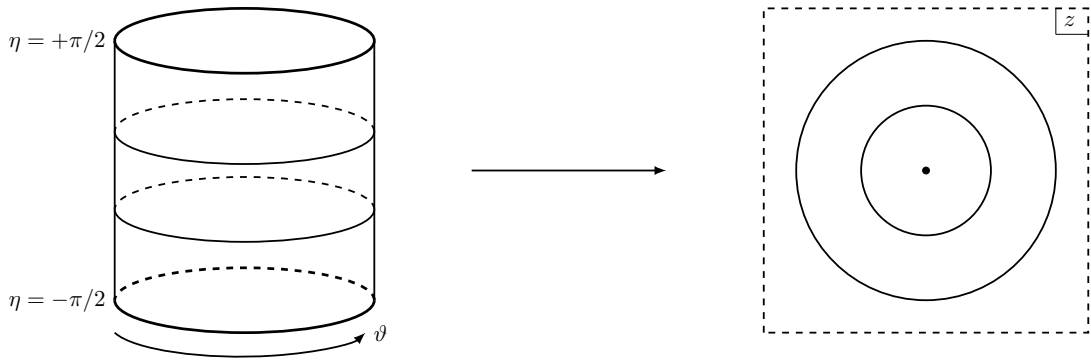

Consider Lorentzian de Sitter space in two dimensions ($\text{dS}_{2}$) with Hubble scale $H$ and scalar curvature $R = 2H^{2}$. 
Conformal coordinates $(\eta,\vartheta)$ cover the conformal compactification of the global patch, which is described by the metric 
\begin{equation}
    \dd s^2=\frac{1}{H^{2}\cos(\eta)^2}(-\dd\eta^2+\dd\vartheta^2)
\end{equation}
with $\eta\in(-\pi/2,\pi/2)$ and $\vartheta\in(-\pi,\pi]$.
Topologically this is $ \mathbb{R} \times S^{1}$ with the conformal boundary being the disjoint union of the space-like surfaces at $\eta=\pm\pi/2$, see figure \ref{fig: ds}.

For the massive scalar field $\phi$ in $\text{dS}_{2}$ with action \eqref{eq: massive scalar action}, the shift-symmetric points are now defined by
\begin{equation}
    m_{k}^{2} = -k(k+1)H^{2}.
\end{equation}
Instead of working with the mass $m$ directly, it is once again convenient to define the $\text{dS}_{2}$ scaling dimension
\begin{equation}
    \Delta = \frac{1}{2}+\sqrt{\frac{1}{4} -\frac{m^{2}}{H^{2}}},
\end{equation}
which takes integer values $\Delta=k+1$ when $m=m_{k}$.

In terms of this, the scalar field $\phi$ obeys the KG equation
\begin{equation}
    \begin{aligned}
        0 &= (\Box_{g} - m^{2})\phi \\
        &= H^{2}\cos(\eta)^{2}\(-\pp_{\eta}^{2} + \pp_{\vartheta}^{2} + \frac{\Delta(\Delta-1)}{\cos(\eta)^{2}} \) \phi.
    \end{aligned}
    \label{eq: ds kg eq}
\end{equation}
Once again, let us briefly review properties of the solutions, following \cite{Higuchi:2022nfy}, but specified to integer $\Delta=k+1$.
See also \cite{Letsios:2025pqo} for a related discussion of the quantisation of the discrete series fermions.
This is where the discussion starts to differ from $\text{AdS}_{2}$.
First note that the standard KG inner product associated with this theory is
\begin{equation}
    (\phi_{1},\phi_{2}) = i \int_{-\pi}^{\pi} \dd\vartheta \(\ol{\phi_{1}} \pp_{\eta}\phi_{2} - \pp_{\eta}\ol{\phi_{1}}\phi_{2} \),
    \label{eq: dS KG inner}
\end{equation}
which is invariant under dS isometries and independent of $\eta$.
To describe the space of solutions to the KG equation \eqref{eq: ds kg eq}, a good starting point is to decompose the field $\phi$ into its Fourier modes with respect to the angular coordinate $\vartheta$
\begin{equation}
    f_{n}(\eta,\vartheta) = \psi_{n}(\eta) e^{in\vartheta},\quad n \in \mathbb{Z}.
\end{equation}
These are modes with definite angular momentum $n$. 
However, now $\pp/\pp\vartheta$ is not a global timelike Killing vector, so contrary to $\text{AdS}_{2}$, this does not define a unique decomposition of the solution space into positive- and negative-frequency/norm modes.
Plugging the Ansatz above back into \eqref{eq: ds kg eq}, we see that the temporal parts now satisfy
\begin{equation}
    \(-\pp_\eta^2+\frac{k(k+1)}{\cos(\eta)^2}\)\psi_{n}(\eta)=n^2\psi_{n}(\eta).
    \label{eq: psim eq}
\end{equation}
This equation is invariant under $n \mapsto -n$, so the temporal parts $\psi_{n}$ can be made to depend only on $|n|$.
We will now describe the solutions for $\psi_{|n|}$.

\subsubsection{Finite-norm modes}

Let us first focus on solutions to this with non-zero and finite KG norm.
On top of $|n| \in \mathbb{Z}$, these are also subject to the condition $|n|\geq k+1$.
They specifically take the form 
\begin{equation}
    \psi_{|n|}(\eta)=  e^{-i|n|\eta} \chi_{|n|}(\eta)
\end{equation}
with
\begin{equation}
    \chi_{|n|}(\eta) = \frac{k!i^{k+1}}{\sqrt{4\pi|n|\textstyle\prod_{j=1}^k(n^2-j^2)}}P_k^{(-|n|,|n|)}(-i\tan(\eta)),
    \label{eq: ds mode functions eta}
\end{equation}
where $P_m^{(\alpha,\beta)}(x)$ are Jacobi polynomials, and the normalisation was chosen such that $f_{n}$ are orthonormal with respect to the KG inner product \eqref{eq: dS KG inner}: $(f_{n},f_{n'}) = \delta_{n,n'}$.
The positive-norm modes are therefore spanned by
\begin{equation}
    f_{n}(\eta,\vartheta) = \psi_{|n|}(\eta) e^{in\vartheta},\quad |n| \geq k+1.
    \label{eq: ds mode functions}
\end{equation}
The crucial difference with $\text{AdS}_2$ is that the $\psi_{|n|}$'s are not real, so conjugation does not just flip the sign of $n$.
The mode expansion for solutions to the massive $\text{dS}_2$ wave equation in conformal coordinates \eqref{eq: ds kg eq} with integer $k$ hence takes the form
\begin{equation}
    \begin{aligned}
        \phi_{\text{phys.}}(\eta,\vartheta)=&\sum_{|n| \geq k+1} a_{n}f_{n}(\eta,\vartheta) + a_{n}^{\dag} \ol{f_{n}}(\eta,\vartheta) \\
        =& \sum_{n \geq k+1} \chi_{n}(\eta)\(a_{n}e^{-in(\eta-\vartheta)} + a_{-n}e^{-in(\eta+\vartheta)}\) + \sum_{n \geq k+1} \ol{\chi_{n}}(\eta)\(a_{n}^{\dag}e^{in(\eta-\vartheta)} + a_{-n}^{\dag}e^{in(\eta+\vartheta)}\).
    \end{aligned}
    \label{eq: dS mode expansion}
\end{equation}
Here, we have defined the operators 
\begin{equation}
    a_n=(f_{n},\phi),\quad a_{n}^{\dag} = -(\ol{f_{n}},\phi)
\end{equation}
for $|n| \geq k+1$.
By virtue of $f_{n}$ being orthonormal with respect to the KG inner product, these satisfy
\begin{equation}
    [a_n,a_n^\dagger]=1,\quad [a_{-n},a_{-n}^\dagger]=1.
    \label{eq: ds ccr}
\end{equation}
As alluded to in the introduction, it is suggestive to identify the operators with positive and negative $n$ respectively as separate sets of creation and annihilation operators.
In particular, comparing to \eqref{eq: ds mode functions}, it is clear these are reminiscent of left- and right-moving modes (albeit with more complicated $\eta$-dependence).
It is clear that going from AdS to dS is not just a matter of analytically continuing $L^2\mapsto -1/H^{2}$ as a parameter in the theory.

\subsubsection{Zero-norm modes \label{sssec: zero norm modes}}

The solutions $\psi_{|n|}(\eta)$ of \eqref{eq: psim eq} for $|n|\leq k$ can be made purely real with a particular choice of normalisation, which implies they have zero norm with respect to \eqref{eq: dS KG inner}. 
These modes generalise the modes which are constant and linear in $\eta$ present in the theory when $k=0$ (the massless scalar).
    
In particular, the modes which generalise the constant shift-mode present when $k=0$ are obtained by taking the mode functions \eqref{eq: ds mode functions eta} up to an overall constant and continuing $n$ to $ |n|\leq k$.
Explicitly, we take these to be
\begin{equation}
    f^{(x)}_{n}(\eta,\vartheta) = \frac{k!i^{k+ |n|}}{\sqrt{4\pi}} P_{k}^{(-|n|,|n|)}\(-i\tan (\eta)\) e^{-i|n|\eta}e^{in\vartheta}.
\end{equation}
There is a second set of independent solutions, which take the form
\begin{equation}
     g_{n}(\eta,\vartheta)= -\frac{1}{\Gamma\(k+3/2\)} \(\frac{1}{2}\cos (\eta)\)^{k+1} {}_{2}F_{1}\(k+1+n,k+1-n;k+\frac{3}{2};\frac{1-\sin (\eta)}{2}\)e^{in\vartheta}.
\end{equation}
For reasons which will become evident in sections \ref{sssec: ds hilbert space} and \ref{ssec: S2}, it is useful to define the following linear combination
\begin{equation}
     f^{(p)}_{n}(\eta,\vartheta)= g_{n}(\eta,\vartheta) + \frac{(-1)^{k+n}\pi}{(k-n)!(k+n)!} f^{(x)}_{n}(\eta,\vartheta).
\end{equation}
The split between these two classes of modes is ambiguous, but any (real) linear combination of them is zero-norm, and we have chosen a particular linear combination such that their inner product is normalised in the following way: 
\begin{equation}
    \(f^{(x/p)}_{n},f^{(x/p)}_{n'}\) = 0,\quad \(f^{(x)}_{n},f^{(p)}_{n'}\) = i\delta_{n,n'}.
    \label{eq: zero mode inner products}
\end{equation}
The $\eta$-dependence of these modes is fully real, so these satisfy
\begin{equation}
    \ol{f^{(x/p)}_{n}}(\eta,\vartheta) = f^{(x/p)}_{-n}(\eta,\vartheta).
    \label{eq: xp reality}
\end{equation}
The corresponding mode expansions are therefore
\begin{equation}
    \phi_{\text{pos.}}(\eta,\vartheta) = \sum_{|n|\leq k}x_nf_n^{(x)}(\eta,\vartheta), \quad \phi_{\text{mom.}}(\eta,\vartheta) = \sum_{|n|\leq k}p_nf_n^{(p)}(\eta,\vartheta),
    \label{eq: xp mode expansion}
\end{equation}
where we have $x_n^\dagger=x_{-n}$ and $p_n^\dagger=p_{-n}$ because of \eqref{eq: xp reality}. 
The zero-norm sector of the field is a combination of these two sectors: 
\begin{equation}
    \phi_{0}(\eta,\vartheta) = \phi_{\text{pos.}}(\eta,\vartheta) + \phi_{\text{mom.}}(\eta,\vartheta).
    \label{eq: zero mode expansion}
\end{equation}
For illustrative purposes, we listed the mode functions in the zero-mode sector for the lowest three $k$ in table \ref{tab: xp modes}.
From this, it is clear that we ought to define $x_{m}$ and $p_{m}$ as
\begin{equation}
    x_{n} = -i\(\phi_{0},f_{n}^{(p)}\) ,\quad p_{n} = i\(\phi_{0},f_{n}^{(x)}\)
\end{equation}
By \eqref{eq: zero mode inner products}, these will satisfy
\begin{equation}
    \quad [x_n,p_{n'}]=i\delta_{n+n',0}.
    \label{eq: ds ccr xp}
\end{equation}

The full mode expansion for $\phi(\eta,\vartheta)$ is the sum of \eqref{eq: dS mode expansion} and \eqref{eq: zero mode expansion},
\begin{equation}
    \phi(\eta,\vartheta)=\phi_0(\eta,\vartheta) + \phi_{\text{phys.}}(\eta,\vartheta).
    \label{eq: dS full mode expansion}
\end{equation}
As we shall explain in more detail in section \ref{sssec: dS shift}, the modes in $\phi_{\text{pos.}}$ are interpreted as shift-modes for the extended shift symmetries present in this theory. 
Importantly, these shift-modes are annihilated by the $(k+1)$-th symmetric and traceless covariant derivative
\begin{equation}
    \nabla_{(\mu_1}\cdots\nabla_{\mu_{k+1})_T}\phi_{\text{pos.}}(\eta,\vartheta)=0,
\end{equation}
which means that the field strength tensor $\mathcal{F}_{\mu_1\cdots\mu_{k+1}}$ is invariant under shifts of the form
\begin{equation}
    \label{eq: dSshift symmetry}
    \phi(\eta,\vartheta) \mapsto \phi(\eta,\vartheta) + \sum_{|n|\leq k}c_n f_n^{(x)}(\eta,\vartheta).
\end{equation}
Furthermore, the commutation relations \eqref{eq: ds ccr xp} imply that the operators $p_n$ act as the generators of shift transformations. 
In particular,
\begin{equation} 
    \delta_{n}\phi(\eta,\vartheta)=i[p_n,\phi(\eta,\vartheta)]=f_{-n}^{(x)}(\eta,\vartheta)
\end{equation}
for $|n|\leq k$. 
\begin{table}[t]
    \centering
    \begin{tblr}{|l|c|c|}
    \toprule 
    \SetRow{gray!25} & \SetCell{c} $x$-modes & $p$-modes  \\ 
    \midrule 
    \SetCell[r=1]{c,2cm}{$k=0$} &  $\displaystyle\frac{1}{2\sqrt{\pi}}$ &$\displaystyle\frac{1}{\sqrt{\pi}}\eta$  \\ 
    \midrule
    \SetCell[r=2]{c,2cm}{$k=1$} & $\displaystyle\frac{1}{2\sqrt{\pi}}\tan(\eta)$ & $-\displaystyle\frac{1}{\sqrt{\pi}}\big(1+\eta\tan(\eta)\big)$ \\ 
    & $\displaystyle\frac{1}{2\sqrt{\pi}}e^{\pm i\vartheta}\sec(\eta)$ & $\displaystyle\frac{1}{2\sqrt{\pi}}e^{\pm i\vartheta}\big(\sin(\eta)+\eta\sec(\eta)\big)$ \\
    \midrule 
    \SetCell[r=3]{c,2cm}{$k=2$} & $\displaystyle\frac{1}{2\sqrt{\pi}}\big(1+3\tan(\eta)^2\big)$ & $\displaystyle\frac{1}{4\sqrt{\pi}}\big(\eta+\tan(\eta)+3\eta\tan(\eta)^2\big)$ \\
    & $\displaystyle\frac{3}{2\sqrt{\pi}}e^{\pm i\vartheta}\sec(\eta)\tan(\eta)$ & $\displaystyle -\frac{1}{6\sqrt{\pi}}e^{\pm i\vartheta}\sec(\eta)\big(3-\cos(\eta)^2+3\eta\tan(\eta)\big)$\\ 
    & $\displaystyle\frac{3}{2\sqrt{\pi}}e^{\pm 2i\vartheta}\sec(\eta)^2$ & $\displaystyle\frac{1}{96\sqrt{\pi}}e^{\pm 2 i\vartheta}\sec(\eta)^2\big(\sin(4\eta)+8\sin(2\eta)+12\eta\big)$  \\ 
    \bottomrule
    \end{tblr}
    \caption{The explicit forms of the mode functions in $\phi_{\text{pos.}}(\eta,\vartheta)$ and $\phi_{\text{mom.}}(\eta,\vartheta)$ when $k \in \{0,1,2\}$.
    Note that for $k=0$, these are precisely the zero-modes discussed e.g. in \cite{DeBievre:1998yx}.}
    \label{tab: xp modes}
\end{table}

\subsubsection{Quantisation \label{sssec: ds hilbert space}}

With the mode expansion of the field in hand, we are now able to canonically quantise the theory.
The quantisation of the discrete series in $\text{dS}_{2}$ was already discussed extensively in \cite{Anninos:2023lin}.
Our aim here is to give a complementary perspective which is closer to the discussion of the massless scalar in $\text{dS}_{2}$ e.g. in \cite{Kirsten:1993ug, Tolley:2001gg}.

Canonical quantisation on curved space typically proceeds as follows, see e.g. \cite{Birrell:1982ix, Wald:1995yp} for reviews.
Given a space $\mathcal{S}$ of complex solutions to the wave equation, we may pick a complex structure on $\mathcal{S}$, which defines a split into positive- and negative-norm subspaces:
\begin{equation}
    \mathcal{S} = \mathcal{S}_{+} \oplus \mathcal{S}_{-}.
    \label{eq: canonical splitting}
\end{equation}
This splitting defines a vacuum state $\ket{\Omega}$ as the state annihilated by all annihilation operators for the modes in $\mathcal{S}_+$.
The single-particle Hilbert space $\mathcal{H}_{1}$ is obtained by acting on this with the creation operators for the modes in $\mathcal{S}_{+}$.
The full Hilbert space is then the associated Fock space. 
The splitting \eqref{eq: canonical splitting} should furthermore be invariant under $\text{dS}_2$ isometries, so that the vacuum as defined above is dS-invariant.
This ensures any two inertial observers will agree on the particle content of a given state in the Fock space.
For a scalar field in $\text{dS}_{2}$ with generic mass, this is straightforward.
There is a one-parameter family of dS-invariant vacua which is in one-to-one correspondence with particular choices of the space of positive-norm solutions to the $\text{dS}_{2}$ KG equation.
Among them, there is a unique choice which has Hadamard form and the same short-distance/high-energy behaviour as the scalar field in flat space \cite{Allen:1985ux, Allen:1987tz}---this is the Euclidean or Bunch--Davies vacuum.

\begin{figure}
    \centering
    \resizebox{0.5\columnwidth}{!}{
        \input{figures/modemapping.tikz}
    }
    \caption{The action of the elements in the $\text{dS}_{2}$ isometry group $\text{SO}(3)$ on the subspaces of the solutions to the $\text{dS}_{2}$ KG equation \eqref{eq: ds kg eq}.
    Note that elements in $\mathcal{Z}$ are mixed into $\mathcal{V}_{+}$ and $\mathcal{V}_{-}$ under the isometries. }
    \label{fig: modes mapping}
\end{figure}
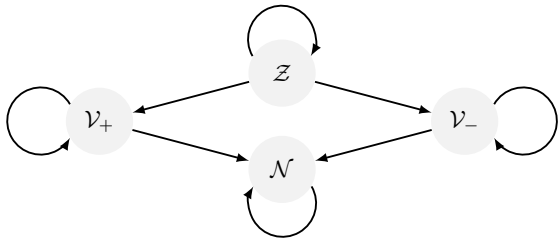

For the discrete series scalars however, the zero-norm modes described in section \ref{sssec: zero norm modes} introduce difficulties with this construction. 
In particular, there is no longer any dS-invariant splitting of the solution space as in \eqref{eq: canonical splitting}. 
This issue has been discussed for the $k=0$ massless scalar in \cite{Allen:1985ux,Folacci:1992xc,Kirsten:1993ug,Tolley:2001gg, DeBievre:1998yx, Gazeau:1999mi} as well as when $k>0$ in \cite{Bros:2010wa, Anninos:2023lin}. 
In particular, we borrow notation from \cite{Tolley:2001gg} and denote subspaces of the full solution space $\mathcal{S}$ by $\mathcal{V}_{+} \ni f_{n}$ and $\mathcal{V}_{-} \ni \ol{f_{n}}$ for $|n| \geq k+1$, and $\mathcal{N} \ni f_{n}^{(x)}$ and $\mathcal{Z} \ni f_{n}^{(p)}$ for $|n| \leq k$.
The explicit expressions for the action of the $\text{dS}_{2}$ isometries on the operators $a_n$, $a_n^\dagger$, $x_n$, and $p_n$ is given in appendix \ref{app: ds2 isometry action}, and we may use this to deduce the action of the isometries on the mode functions $f_{n}$, $\ol{f_{n}}$, $f_{n}^{(x)}$, and $f_{n}^{(p)}$.
This is summarised in figure \ref{fig: modes mapping}.  
Clearly, the na\"ive split $\mathcal{S} = \mathcal{V}_{+} \oplus \mathcal{V}_{-}$ excluding the zero-modes is \textit{not} invariant under the $\text{dS}_2$ isometries\footnote{Alternatively, we could have included $\mathcal{N}$ and $\mathcal{Z}$ into $\mathcal{V}_{+}$ and $\mathcal{V}_{-}$, now writing $\mathcal{S}=\tilde{\mathcal{V}}_+\oplus\tilde{\mathcal{V}}_-$, by noting that the combinations 
\begin{equation}
    \tilde{f}_{n} =\tfrac{1}{\sqrt{2}} f_{n}^{(x)} - \tfrac{i}{\sqrt{2}}f_{n}^{(p)}\in\tilde{\mathcal{V}}_+,\quad \ol{\tilde{f}_{n}} =\tfrac{1}{\sqrt{2}} \ol{f_{n}^{(x)}} + \tfrac{i}{\sqrt{2}}\ol{f_{n}^{(p)}}\in\tilde{\mathcal{V}}_-
\end{equation}
have positive and negative norm, respectively, due to \eqref{eq: zero mode inner products}.
However, the action of the $\text{dS}_{2}$ isometries would mix $\tilde{\mathcal{V}}_+$ and $\tilde{\mathcal{V}}_-$.}.
To restore dS invariance, one might define the Hilbert space of the discrete series scalar theory at level $k$ from the generic mass theory in the limit $m^2\to m_{k}^{2}$. 
Notably however, Wightman functions evaluated in this vacuum diverge in this limit \cite{Anninos:2023lin,Farnsworth:2024yeh}. 
In Euclidean signature, it is understood that the infrared divergence in the Euclidean two-point function is tied to the zero-mode in the Euclidean action \cite{Folacci:1992xc, Bros:2010wa, Anninos:2023lin}. 
These obstructions are typically taken to mean that dS invariance is broken spontaneously \cite{Allen:1985ux}.

Despite this, it is possible to define a dS-invariant vacuum, provided we remove the modes in $\mathcal{N}$ and $\mathcal{Z}$ in a way that respects the $\text{dS}_2$ isometries. 
One way typically chosen in the literature \cite{Tolley:2001gg,Anninos:2023lin,Farnsworth:2024yeh} is to start by imposing an equivalence relation 
\begin{equation}
    \phi(\eta,\vartheta)\sim \phi(\eta,\vartheta) + \sum_{|n|\leq k} c_nf_n^{(x)}(\eta,\vartheta).
\end{equation}
This renders the subspaces $\mathcal{V}_{+}/\mathcal{N}$ and $\mathcal{V}_{-}/\mathcal{N}$ dS-invariant, with the former being equipped with a positive-definite inner product \eqref{eq: dS KG inner}. 
Practically, this amounts to only considering operators which are invariant under shifts \eqref{eq: dSshift symmetry}, such as $\mathcal{F}_{\mu_1\cdots\mu_{k+1}}$. 
We emphasise that imposing this equivalence relation is, at this point, not yet equivalent to gauging the shift symmetry via introducing a gauge field. 
We are reducing the number of operators we consider, rather than increasing.
Next, we also restrict the set of states to physical states $\ket{\text{phys}}$ satisfying
\begin{equation}
    p_{n} \ket{\text{phys}} = 0,\quad |n|\leq k.
    \label{eq: shift physical states}
\end{equation}
The reason for this is two-fold.
The modes in $\mathcal{Z}$ still mix $\mathcal{V}_{+}/\mathcal{N}$ and $\mathcal{V}_{-}/\mathcal{N}$ under $\text{dS}_2$ isometries, which is in tension with the complex structure on $\mathcal{S}$. 
Furthermore, now that we have effectively removed the operators $x_n$ from our algebra, each $p_n$ is in the centre of the operator algebra, and it is now natural to diagonalise them on the space of states. 
Each state $\ket{\mathbf{q},\psi}$ therefore is labelled by generally complex eigenvalues $\mathbf{q}=(q_{0},\dots,q_k)$ of $\mathbf{p}=(p_{0},\dots,p_k)$, where because $p^\dagger_n=p_{-n}$ and all $p_n$'s commute, the corresponding eigenvalue of $p_{-n}$ for $n>0$ is $\ol q_n$. It is however not possible to define normalisable eigenstates of such operators. 
Instead, they are only normalisable in the generalised sense,
\begin{equation}
    \braket{\mathbf{q},\psi|\mathbf{q'},\psi} \propto \delta(q_0-q_0')\prod_{i=1}^{k}\delta^2(q_i-q_i',\ol{q}_i-\ol{q}_i').
\end{equation}
Restricting to the states with $\mathbf{q} = 0$ by imposing \eqref{eq: shift physical states} resolves both of these problems.
In this way, all expectation values of shift-invariant operators with respect to physical states $\ket{\text{phys}}$ do not receive contributions from modes in $\mathcal{N}$ and $\mathcal{Z}$, only from modes in the solution space $\mathcal{S}'=(\mathcal{V}_+/\mathcal{N})\oplus(\mathcal{V}_-/\mathcal{N})$, which admits a dS-invariant splitting.

With this in mind, we can now describe the full Hilbert space.
We define the dS-invariant vacuum $\ket{\Omega}$ to satisfy
\begin{subequations}
    \begin{align}
        a_{n} \ket{\Omega} &= 0,\quad |n| \geq k+1 \\
        p_{n} \ket{\Omega} &= 0,\quad |n| \leq k.
    \end{align}
    \label{eq: dS vacuum}%
\end{subequations}
The full Hilbert space is then the Fock space generated by successive applications of $a_n^\dagger$ on the vacuum $\ket{\Omega}$. 
As in the generic massive scalar in $\text{dS}_{2}$, this is not a unique choice. 
As we will see in section \ref{sssec: dS crltn}, this is the choice which reproduces physical correlation functions that are consistent with those obtained from taking the limit $m^2\to m_{k}^{2}$ of the generic mass theory with the Bunch--Davies vacuum. 
Note that the aforementioned problems with infrared divergences of correlation functions are absent when one considers shift-invariant operators---this was confirmed in \cite{Farnsworth:2024yeh}. 


\subsection{Riemann sphere \label{ssec: S2}}

We now analytically continue to Euclidean signature via $\eta = -i\eta_{E}$ and switch to complex coordinates on Euclidean de Sitter space in two dimensions ($\text{EdS}_2$), resulting in the invariant interval
\begin{equation}
    \dd s^{2} = \frac{1}{H^{2}\cosh(\eta_E)^{2}} (\dd \eta_{E}^{2} + \dd \vartheta^{2}).
    \label{eq: conformal ds metric}
\end{equation}
Here, we have taken $\eta_E\in\mathbb{R}$. 
Next, we introduce the complex coordinate
\begin{equation}
    z = e^{\eta_{E} -i\vartheta},
\end{equation}
such that
\begin{equation}
    \vartheta = \frac{i}{2} \log(z/\zb),\quad \eta_{E} = \frac{1}{2}\log(z\zb).
    \label{eq: complex coordinates ds}
\end{equation}
Note that these are now related to the coordinates used in \cite{Farnsworth:2024yeh} up to a dimensionful rescaling.
Under this transformation, the entirety of $\text{EdS}_2$ is mapped to the Riemann sphere\footnote{Also known as the extended complex plane, the complex plane with the point at infinity, or the complex projective line.}, with invariant interval
\begin{equation}
    \dd s^2=\frac{4}{H^2}\frac{\dd z\dd\bar z}{(1+z\bar z)^2}.
    \label{eq: S2 metric}
\end{equation}
These are isothermal coordinates \eqref{eq: general 2d metric} with $\Omega^{2} = 4H^{-2}/(1+z\zb)^{2}$.
We remind the reader that various geometric properties of this space, alongside useful differential operators and identities they satisfy are discussed in appendix \ref{app: riemann surfaces}.

$\text{EdS}_{2}$ has three Killing vectors, which form the generators of $\text{PSU}(2)=\text{SO}(3)$.
In these coordinates, they take the form
\begin{subequations}
    \begin{align}
        \xi_0&=iz\pp-i\zb\ol\pp \\
        \xi_1&=-\frac{1}{2}\(1+z^2\)\pp-\frac{1}{2}\(1+\zb^2\)\ol\pp \\
        \xi_2&=-\frac{i}{2}\(1-z^2\)\pp+\frac{i}{2}\(1-\zb^2\)\ol\pp,
    \end{align}
    \label{eq: psu2 killing vs}%
\end{subequations}
which satisfy the algebra
\begin{equation}
    [\xi_0,\xi_1]=-\xi_2,\quad [\xi_1,\xi_2]=-\xi_0,\quad [\xi_2,\xi_0]=-\xi_1.
\end{equation}
It is more convenient however to consider the following linear combinations of the Killing vectors above:
\begin{subequations}
    \begin{alignat}{2}
        \ell_{+1}&\equiv \xi_1+i\xi_2 &&=-z^2\pp-\ol\pp \\
        \ell_0& \equiv \enspace i\xi_0 &&= - z\pp+\zb\ol\pp\\
        \ell_{-1}&\equiv \xi_1-i\xi_2 && = -\pp-\zb^2\ol\pp.
    \end{alignat}
    \label{eq: dS ell}%
\end{subequations}
These satisfy the algebra
\begin{equation}
    [\ell_m,\ell_n]=(m-n)\ell_{m+n},\quad m,n \in \{-1,0,1\},
    \label{eq: isometries complex}
\end{equation}
which is isomorphic to the Lie algebra of $\text{SL}(2,\mathbb{R})$. 
In these coordinates, the action and the KG equation take the form
\begin{equation}
    S = \int \dzz \(\pp\phi \ol{\pp}\phi - \frac{k(k+1)}{(1+z\zb)^{2}} \phi^{2} \)
\end{equation}
and
\begin{equation}
    \(\pp\ol{\pp} + \frac{k(k+1)}{(1+z\zb)^{2}}\)  \phi = 0.
    \label{eq: s2 KG}
\end{equation}
As in $\text{EAdS}_{2}$, the only independent components of the field strength $\mathcal{F}$ in \eqref{eq: field strength} are $F \equiv \mathcal{F}_{z \dots z}$ and $\ol{F} \equiv \mathcal{F}_{\zb \dots \zb}$, which are conserved in the sense of \eqref{eq: F (anti-)holomorphicity}.

\subsubsection{Mode expansions}

The mode expansion of $\phi$ takes a particularly simple form on the Riemann sphere.
The mode functions described in section \ref{ssec: dS lorentz}, when converted to the Riemann sphere, also strongly resemble a mode expansion typically encountered in $2d$ CFT.
As a preview, this includes integer powers of $z$ and $\zb$ alongside a finite number of modes with logarithmic dependence on $z$ and $\zb$
\begin{equation}
    \frac{1}{z^{n-k}},\quad \frac{1}{\zb^{n-k}},\quad \frac{1}{z^{n-k}}\log z,\quad \frac{1}{\zb^{n-k}}\log\zb,
\end{equation}
which are subsequently acted on by certain differential operators. 

Let us make this explicit.
As we argued for $\text{AdS}_{2}$, the general solution to the KG equation \eqref{eq: s2 KG}, before imposing reality and periodicity, may be written as 
\begin{equation}
    \phi(z,\ol z)=\mathcal{D}_k\varphi(z) + \ol{\mathcal{D}}_k\tilde{\varphi}(\ol z),
\end{equation}
for any holomorphic $\varphi(z)$ and anti-holomorphic $\tilde{\varphi}(\ol z)$ functions.
In particular, the choices of $\varphi(z)$ and $\tilde{\varphi}(\ol z)$ which reproduce the mode expansion \eqref{eq: dS full mode expansion} are
\begin{subequations}
    \begin{align}
        \varphi(z) =&\frac{1}{\sqrt{4\pi}}\sum_{|n|\leq k}\(\frac{1}{2}i^{k+n}\frac{x_n}{z^{n-k}}+\frac{(-i)^{k+1+n}}
        {(k-n)!(k+n)!}\frac{p_n}{z^{n-k}}\log z
        \)+\frac{i^{k+1}}{\sqrt{4\pi}}\sum_{|n|>k}\frac{1}{n}\frac{1}{\sqrt{\textstyle\prod_{j=1}^k(n^2-j^2)}}\frac{\alpha_n}{z^{n-k}} \\
        \tilde{\varphi}(\ol z) =&\frac{1}{\sqrt{4\pi}}\sum_{|n|\leq k}\(\frac{1}{2}i^{k+n}\frac{x_{-n}}{\ol z^{n-k}}+\frac{(-i)^{k+1+n}}
        {(k-n)!(k+n)!}\frac{p_{-n}}{\ol z^{n-k}}\log \ol z\)+\frac{i^{k+1}}{\sqrt{4\pi}}\sum_{|n|>k}\frac{1}{n}\frac{1}{\sqrt{\textstyle\prod_{j=1}^k(n^2-j^2)}}\frac{\tilde{\alpha}_n}{\ol z^{n-k}}.
    \end{align}
    \label{eq: ds varphi mode expansion}%
\end{subequations}
We remind the reader that $\mathcal{D}_k$ and $\ol{\mathcal{D}}_k$ are the order-$k$ differential operators defined in appendix \ref{app: riemann surfaces}, the explicit forms of which for $\text{EdS}_{2}$ are given in \eqref{eq: nabla mathcalD ds}.
In appendix \ref{sapp: eds2 identities}, they are shown to satisfy the identities \ref{idty: nabla mathcalD composition ds}
and \ref{idty: cov d anti composition identity ds}, which are summarised by \eqref{eq: nabla mathcalD composition identities}. 
We also introduced the notation
\begin{equation}
    \begin{aligned}
        \alpha_n&=\sqrt{n} a_n,\quad && \alpha_{-n}=\sqrt{n}a_n^\dagger, \\
        \tilde{\alpha}_n&=\sqrt{n}a_{-n},\quad && \tilde{\alpha}_{-n}=\sqrt{n}a_{-n}^\dagger
    \end{aligned}
\end{equation}
for $n > k$.
The non-trivial commutation relations are then summarised as
\begin{equation}
    \label{eq: ds ccrs}
    [\alpha_m,\alpha_n]=m\delta_{m+n,0} =  [\tilde\alpha_m,\tilde\alpha_n],\quad [x_m,p_n]=i\delta_{m+n,0},
\end{equation}
for any value of $m,n$. 

The logarithmic dependence of $\varphi(z)$ and $\tilde{\varphi}(\zb)$ as written introduces an ambiguity in the choice of branch. The logarithmic dependence in $z$ and $\zb$ within $\phi(z,\zb)$ however appears always via the combination $\log(z)+\log(\zb)$, which we define as $\log(|z|^2)$ on the principal branch. 

The field strengths $F(z)$ and $\ol F(\bar z)$ can be readily computed using \eqref{eq: nabla mathcalD composition identities}:
\begin{subequations}
    \begin{align}
        F(z)&=-\frac{1}{\sqrt{4\pi}}\sum_{|n|\leq k}i^{k+1+n}\frac{p_n}{z^{n+k+1}}-\frac{i^{k+1}}{\sqrt{4\pi}}\sum_{|n|>k}\sqrt{\textstyle\prod_{j=1}^k(n^2-j^2)} \frac{\alpha_{n}}{z^{n+k+1}}\\
        \ol F(\ol z)&=-\frac{1}{\sqrt{4\pi}}\sum_{|n|\leq k}i^{k+1+n}\frac{ p_{-n}}{\ol z^{n+k+1}}-\frac{i^{k+1}}{\sqrt{4\pi}}\sum_{|n|>k} \sqrt{\textstyle\prod_{j=1}^k(n^2-j^2)} \frac{\tilde{\alpha}_{n}}{\ol z^{n+k+1}}.
    \end{align}
    \label{eq: mode expansion eds2}%
\end{subequations}
Here we see that the shift-modes $\phi_{\text{pos.}}$, associated with the operators $x_n$, are annihilated by $\nabla_z^{k+1}$ and $\nabla_{\bar z}^{k+1}$. 
The momentum modes $\phi_{\text{mom.}}$ on the other hand still contribute to $F(z)$ and $\ol F(\bar z)$. 
Clearly, the choice of vacuum in section \ref{sssec: ds hilbert space} ensures regularity of expectation values at $z=0$ and $z=\infty$ as for a primary field of weight $k$, see e.g. \cite{Blumenhagen:2013fgp}.
In contrast to the discrete series scalars in $\text{EAdS}_2$, where $F$ and $\ol{F}$ were related to each other on the conformal boundary, the two field strengths are independent operators in $\text{EdS}_2$.

\subsubsection{Correlation functions \label{sssec: dS crltn}}

As discussed in section \ref{sssec: ds hilbert space}, the two-point correlation function for $\phi(z,\zb)$ in $\text{dS}_2$ suffers from an infrared divergence due to the presence of zero-modes. 
Indeed, for generic $m$ or $\Delta$ in \eqref{eq: ds kg eq}, the two-point correlation function in the Bunch--Davies vacuum takes the following form in complex coordinates on $\text{EdS}_{2}$ 
\begin{equation}
    \langle \phi(z,\zb)\phi(z',\zb')\rangle=\frac{1}{4\pi}\Gamma(\Delta)\Gamma(1-\Delta){}_2F_1\(\Delta,1-\Delta;1,\frac{(1+z\zb')(1+z'\zb)}{(1+z\zb)(1+z'\zb')}\).
    \label{eq: BD 2pt}
\end{equation}
This is divergent as $\Delta$ approaches an integer due to the factor $\Gamma(1-\Delta)$. 
Nevertheless, this correlation function can be used to construct well-defined two-point correlation functions of $F$ and $\ol{F}$ by applying the appropriate number of covariant derivatives to \eqref{eq: BD 2pt}, as was shown in \cite{Farnsworth:2024yeh}.
Explicitly,
\begin{subequations}
    \begin{align}
        \langle F(z)F(z')\rangle&=(-1)^{k+1}\frac{1}{4\pi}(2k+1)!\frac{1}{(z-z')^{2k+2}}\\
        \langle \ol{F}(\zb)\ol{F}(\zb')\rangle&=(-1)^{k+1}\frac{1}{4\pi}(2k+1)!\frac{1}{(\zb-\zb')^{2k+2}}\\
        \langle F(z)\ol{F}(\zb')\rangle&=0.
        \label{eq: dS FFbar cor}
    \end{align}
    \label{eq: dS F cor}%
\end{subequations}
These may also be derived directly from the mode expansions \eqref{eq: mode expansion eds2} using the canonical commutation relations \eqref{eq: ds ccrs} and the definition of the vacuum \eqref{eq: dS vacuum}. 
The form of the correlation functions \eqref{eq: dS F cor} is, once again, reminiscent of correlation functions of 2d CFTs on the full complex plane.
Note that \eqref{eq: dS FFbar cor} now \textit{does} vanish everywhere due to the absence of boundary conditions.
The relation to the $\text{EdS}_{2}$ isometries will be discussed in section \ref{sssec: PSU2 dS}.


\section{Linear symmetries \label{sec: lin sym}}

In this section, we discuss (some of) the linear symmetries the discrete scalar theories on $\text{(A)dS}_{2}$ exhibit when $k > 0$ in Euclidean signature.
The oscillator mode expansions we found in sections \ref{sec: ads2} and \ref{sec: ds2} present a convenient framework for us to describe the symmetries of the discrete series scalar theories---in particular, any automorphism of the space of oscillator mode will be a symmetry of the free theory.
It is clear that a subset of these generators is responsible for the hidden conformal symmetry of the current $F$ observed in \cite{Farnsworth:2024yeh}. 
In light of the current $F$ also being holomorphic, we are particularly interested in the question whether this global conformal symmetry also extends to Virasoro symmetry.

Even though the discussions for $\text{EAdS}_{2}$ and $\text{EdS}_{2}$ mirror each other, there are sufficiently many differences that we thought it best to separate them into distinct subsections \ref{ssec: lin sym ads} and \ref{ssec: lin sym ds}.
In either case, we start by discussing how the isometries are represented as operators acting on the Hilbert space of states, and their consequences for the correlation functions \eqref{eq: AdS F cor} and \eqref{eq: dS F cor} discussed in \cite{Farnsworth:2024yeh}.
We then show that both theories exhibit Virasoro symmetry by explicitly constructing the Virasoro generators out of the available oscillators.
Finally, we discuss a (would-be) symmetry under shift transformations, which is related to the existence of the conserved current $\mathcal{F}_{\mu_1\cdots\mu_{k+1}}$.


\subsection{Anti-de Sitter \label{ssec: lin sym ads}}

Let us begin by discussing the linear symmetries in $\text{AdS}_{2}$.

\subsubsection{$\text{PSL}(2,\mathbb{R})$ transformations \label{sssec: SL2R AdS}}

The group of continuous isometries of $\text{EAdS}_2$ which preserve the UHP is $\text{PSL}(2,\mathbb{R})$, which are parameterised by (a subset of) M\"{o}bius transformations
\begin{equation}
    w(z)=\frac{az+b}{cz+d},\quad ad-bc=1,
\end{equation}
with $a,b,c,d\in\mathbb{R}$.
As mentioned earlier in section \ref{ssec: UHP}, the Killing vectors $\ell_{-1}$, $\ell_0$, $\ell_1$ of $\text{EAdS}_2$ in \eqref{eq: AdS ell} form the algebra \eqref{eq: psl2r algebra}.
These are associated with the infinitesimal coordinate transformations 
\begin{equation}
    w(z)=z+\varepsilon(z),\quad \ol{w}(\zb)=\zb+\varepsilon(\zb),
\end{equation}
for an infinitesimal $\varepsilon(z)\in\{1,z,z^2\}$. 
These are, of course, the same symmetry transformations as the global conformal transformations on the flat infinite strip.

The scalar field $\phi(z,\zb)$ transforms as a scalar with respect to these coordinate transformations, i.e.
\begin{equation}
    \phi'(w,\ol{w})=\phi(z,\zb)
\end{equation}
under $z \mapsto w(z)$.
Infinitesimally, this is
\begin{equation}
    \delta_\epsilon\phi(z,\zb)=-\epsilon(z)\partial\phi(z,\zb)-\epsilon(\zb)\ol{\pp}\phi(z,\zb).
\end{equation}
Because $F(z)$ and $\ol{F}(\zb)$ are components of a rank $k+1$ traceless and symmetric tensor $\mathcal{F}_{\mu_1\cdots\mu_{k+1}}$, they transform as $\text{SL}(2,\mathbb{R})$ primaries with holomorphic weights 
\begin{equation}
    (h_F,\ol{h}_F)=(k+1,0),\quad (h_{\ol{F}},\ol{h}_{\ol{F}})=(0,k+1),
    \label{eq: conformal weights}
\end{equation}
or
\begin{equation}
    F'(w) = \(\frac{\pp w(z)}{\pp z}\)^{-(k+1)} F(z),\quad  \ol{F}'(\ol{w}) = \(\frac{\pp \ol{w}(\ol{z})}{\pp \ol{z}}\)^{-(k+1)} \ol{F}(\zb),
    \label{eq: conf primary transformation finite}
\end{equation}
respectively.
Infinitesimally, this is
\begin{equation}
   \delta_\epsilon F(z) = - \epsilon(z) \,\pp F(z) - (k+1) \pp \epsilon(z) \, F(z),
    \label{eq: conf primary transformation}
\end{equation}
and similarly for $\ol{F}(\zb)$.

As operators acting on the Lorentzian Hilbert space, $\ell_{-1},\ell_0,\ell_1$ can be obtained directly from the stress tensor or deduced from the requirement
\begin{equation}
    \delta_{z^{n+1}}\phi=-[\ell_n,\phi].
\end{equation}
In terms of the oscillators $\alpha_n$, we find that
\begin{subequations}
    \begin{align}
        \ell_{-1}&=\sum_{n=k+1}^\infty\sqrt{\frac{(n-k)(n+k+1)}{n(n+1)}}\alpha_{-1-n}\alpha_n,\\
        \ell_0&=\sum_{n=k+1}^\infty\alpha_{-n}\alpha_n,\\
        \ell_{+1}&=\sum_{n=k+1}^\infty\sqrt{\frac{(n+k)(n-k-1)}{n(n-1)}}\alpha_{1-n}\alpha_{n}.
    \end{align}
    \label{eq: AdS ell op}%
\end{subequations}
The correlation functions \eqref{eq: AdS F cor} are uniquely determined up to an overall constant by the Ward identities implied by \eqref{eq: conf primary transformation} as well as the invariance of the vacuum $\ell_n|\Omega\rangle=0$.

\subsubsection{Virasoro algebra \label{sssec: virasoro ads}}

The fact that the correlation functions involving $F$ and $\ol{F}$ satisfy the same Ward identities as those of a $2d$ CFT was pointed out in \cite{Farnsworth:2024yeh} for the equivalent theory in $\text{EdS}_{2}$---we will return to this in section \ref{ssec: lin sym ds}. 
The fact that $F$ and $\ol{F}$ are also (anti-)holomorphic furthermore suggests that this theory may have an extension of the aforementioned symmetry isomorphic to $2d$ local conformal transformations.

It is straightforward however to show that this theory cannot be invariant under local conformal transformations of the form \eqref{eq: conf primary transformation} with $\epsilon(z)=z^{n+1}$ and $n\notin \{-1,0,1\}$. 
Indeed, any such transformation (say $n=-2$) will excite a mode $z^{-(n'+k+1)}$ inside the gap of non-normalisable modes $-k\leq n'\leq k$, meaning that this transformation takes $F$ out of the phase space---see figure \ref{fig: virasoro obstruction}. 
This implies in particular that there is no way to represent this transformation as an operator acting on the Hilbert space. 
More generally, it is clear that in order to avoid exciting any of the $2k+1$ non-normalisable modes, a local infinitesimal transformation $\delta\phi$ must be at least an order $2k+1$ differential operator acting on $\phi$---we will see symmetry transformations of this form in section \ref{sec: chiral}. If this theory has a symmetry isomorphic to local conformal symmetry, it cannot act on the fields in the standard way, i.e. as a conformal coordinate transformation.
\begin{figure}
    \centering
    \input{figures/virasoro.tikz}
    \caption{Action of various generators on the modes in mode expansion \eqref{eq: UHP F mode expansion} of the holomorphic current $F(z)$. 
    Included in grey are the modes which are non-normalisable (in $\text{AdS}_{2}$) and are excluded from the mode expansion.
    Shown in blue, below the expression, is the action of the $\text{PSL}(2,\mathbb{R})$ generator $\ell_{-1}$---this does not involve the modes in the gap.
    An extension of the global to local conformal transformations would require the existence of an operator such as $\ell_{-2}$, whose action on the mode expansion is indicated by orange dotted arrows above the expression. 
    This excites non-normalisable modes in the gap, e.g. $\delta_{z^{-1}}z=kz^{-1}$.}
    \label{fig: virasoro obstruction}
\end{figure}

Despite this, there are various ways to construct Virasoro operators out of the oscillators $\alpha_n$ with $|n|\geq k+1$. 
One such construction is to take 
\begin{equation}
    L_{m}=\sum_{n=k+m+1}^\infty\sqrt{\frac{(n-m-k)(n-k)}{(n-m)n}}\alpha_{m-n}\alpha_n + \frac{1}{2}\sum_{n=1}^{m-1}\sqrt{\frac{(m-n)n}{(k+m-n)(k+n)}}\alpha_{k+m-n}\alpha_{k+n}
    \label{eq: AdS virasoro op}
\end{equation}
for $m\geq 0$ and we define $L_{-m}=L_m^\dagger$. 
These operators satisfy the Virasoro algebra
\begin{equation}
    [L_m,L_n]=(m-n)L_{m+n} + \frac{c}{12}m(m^2-1)\delta_{m+n,0}.
    \label{eq: virasoro alg}
\end{equation}
The central charge $c$ can be computed in a standard way, by considering the expectation value of $[L_2,L_{-2}]$ in the vacuum
\begin{equation}
    \langle\Omega|[L_2,L_{-2}]|\Omega\rangle=\frac{1}{4(k+1)^2}\langle\Omega|\alpha_{k+1}^2\alpha_{-k-1}^2|\Omega\rangle=\frac{1}{2}.
\end{equation}
Together with the fact that $L_0|\Omega\rangle=0$, this implies that $c=1$. 
A simple way to see that the operators \eqref{eq: AdS virasoro op} satisfy the Virasoro algebra \eqref{eq: virasoro alg} is to define
\begin{equation}
    \hat{\alpha}_n = \sqrt{\frac{n}{n+k}}\alpha_{k+n},\quad\hat{\alpha}_{-n} = \sqrt{\frac{n}{n+k}}\alpha_{-k-n}
\end{equation}
for $n>0$ and taking $\alpha_{0} =0$.
These new operators satisfy the commutation relations
\begin{equation}
    [\hat{\alpha}_m,\hat{\alpha}_n]=m\delta_{m+n,0}
\end{equation}
now for $m,n\neq0$. 
In terms of $\hat{\alpha}_n$, the Virasoro operators take the canonical form
\begin{subequations}
    \begin{align}
       L_m&=\frac{1}{2}\sum_{n=-\infty}^{\infty}\hat{\alpha}_{m-n}\hat{\alpha}_n,\quad m\neq 0\\
       L_0&=\sum_{n=1}^\infty\hat{\alpha}_{-n}\hat{\alpha}_n
    \end{align}
\end{subequations}
from which \eqref{eq: virasoro alg} easily follows.

A few comments are in order.
First note that when $k=0$, these Virasoro operators coincide with the isometries \eqref{eq: AdS ell op}, i.e. $L_m=\ell_m$ for $m\in\{-1,0,1\}$.
In this case, they therefore provide the expected extension of global conformal transformations to local conformal transformations.
However, when $k>0$, these Virasoro operators are no longer related to the isometries, and instead act on the local fields $\phi(z,\zb)$, $F(z)$, $\ol{F}(\zb)$, etc. as non-local transformations.
For $k=0$, the Virasoro generators $L_{m}$ are the operators in the mode expansion of the (anti-)holomorphic stress tensor---the analogous object which generates the operators $L_{m}$ for $k>0$ is non-local and presumably not a useful object to consider.
Further, by construction, these operators define symmetry transformations, because if $\phi(z,\zb)$ is a solution to the equations of motion and vanishes at the conformal boundary, $[L_m,\phi(z,\zb)]$ does also. 
In the current description of the theory in terms of the local field $\phi(z,\zb)$ however, this symmetry is not useful in the conventional sense, because it does not imply local Ward identities that help constrain correlation functions between local operators built from $\phi(z,\zb)$. 
They nonetheless provide an alternative way to organise the full Hilbert space in terms of Verma modules rather than multi-particle Hilbert spaces. 


\subsubsection{Shift transformations \label{sssec: ads shift}}

The existence of the current $\mathcal{F}_{\mu_1\cdots\mu_{k+1}}$ is related to a would-be symmetry under field-independent shifts 
\begin{equation}
    \phi(z,\zb)\mapsto\phi(z,\zb)+\phi_{\text{shift}}(z,\zb),
    \label{eq: shift trans}
\end{equation}
where $\phi_{\text{shift}}$ is a local solution to the equations of motion which is annihilated by the $(k+1)$-th symmetric and traceless covariant derivative:
\begin{equation}
    \(\pp\ol{\pp} + \frac{k(k+1)}{(z-\zb)^{2}}\)  \phi_{\text{shift}} = 0,\quad \nabla^{k+1}\phi_{\text{shift}}=0,\quad \ol{\nabla}^{k+1}\phi_{\text{shift}}=0.
    \label{eq: shift eqns ads}
\end{equation}
The field strength $\mathcal{F}_{\mu_1\cdots\mu_{k+1}}$ is therefore invariant under such shift transformations. 
As a solution to the equation of motion \eqref{eq: uhp KG}, we may once again write the holomorphic splitting
\begin{equation}
    \phi_{\text{shift}}(z,\zb)=\mathcal{D}_k\varphi_{\text{shift}}(z)+\ol{\mathcal{D}}_k\tilde{\varphi}_{\text{shift}}(\zb),
    \label{eq: phi varphi shift}
\end{equation}
which solves the other conditions in \eqref{eq: shift eqns ads} with
\begin{equation}
    \varphi_{\text{shift}}(z) =i^k\sum_{|n|\leq k}\frac{c_n}{z^{n-k}},\quad \tilde{\varphi}_{\text{shift}}(\zb) =(-i)^k\sum_{|n|\leq k}\frac{c_n}{\zb^{n-k}}.
    \label{eq: varphi shift}
\end{equation}
Interestingly, these shift-modes \textit{fill in the gap} in the mode expansion \eqref{eq: UHP phi mode expansion} of $\phi$ in powers of $z,\zb$. 
One may have expected $2(2k+1)$ independent modes from \eqref{eq: varphi shift} for each independent power of $z$ and $\zb$---the other half of solutions to \eqref{eq: uhp KG} with $|\omega|\leq k$ have logarithmic dependence in $z$ and $\zb$. 
These were discussed in section \ref{sec: ds2}.
However, identity \ref{idty: ads shift mode identity} relates the powers of $z$ and $\zb$ considered above, and halves the number of independent modes.
This leaves us with a $(2k+1)$-dimensional parameter space of solutions to \eqref{eq: shift eqns ads}, which constitute solutions with energies $|\omega|\leq k$.
Importantly, these shift-modes are not square-integrable when $k>0$---they scale as 
\begin{equation}
    \phi_{\text{shift}}\sim\cos(\rho)^{-k}f(\tau),\quad \text{as } \rho\to\pm\pi/2
\end{equation}
in conformal coordinates. 
This is the reason why we omitted a discussion of these modes of solutions to \eqref{eq: uhp KG} in section \ref{ssec: AdS lorentz} (which is to be contrasted with the analogous discussion of $\text{EdS}_{2}$ in \ref{ssec: dS lorentz}).
In $\text{(E)AdS}_2$ then, the symmetry under shift transformations is explicitly broken by the condition at the conformal boundary. 
This is reminiscent of the massless scalar on a strip with Dirichlet boundary conditions, in which the shift symmetry $\phi\to\phi + c$ is broken by the boundary conditions. 
Within the family of discrete series scalars, only the massless $k=0$ theory sits within the Breitenlohner--Freedman window \cite{Breitenlohner:1982jf}, so in that case and in that case only it is possible to choose Neumann boundary conditions which restore this shift symmetry. 
For $k>0$ however, there is a unique boundary condition \cite{Higuchi:2021fxg}, and so the shift symmetry cannot be restored.


\subsection{de Sitter \label{ssec: lin sym ds}}

We now turn to discussing the linear symmetries in $\text{dS}_{2}$.

\subsubsection{$\text{PSU}(2)$ transformations \label{sssec: PSU2 dS}}

The group of continuous isometries of $\text{EdS}_2$ which preserve the Riemann sphere is $\text{PSU}(2)$, parameterised by the (subset of) M\"{o}bius transformations 
\begin{equation}
    w(z)=\frac{a z-b}{\ol{b} z+\ol{a}}, \quad |a|^2+|b|^2=1,
    \label{eq: dS mobius}
\end{equation}
with $a,b\in\mathbb{C}$.
The infinitesimal versions of the transformations \eqref{eq: dS mobius} are generated by the Killing vectors \eqref{eq: dS ell}, which obey the algebra \eqref{eq: isometries complex}.
As was discussed for $\text{EAdS}_{2}$ in section \ref{sssec: SL2R AdS}, $F(z)$ and $\ol{F}(\zb)$ are components of a rank $k+1$ traceless and symmetric tensor $\mathcal{F}_{\mu_1\cdots\mu_{k+1}}$, and hence have finite and infinitesimal transformations \eqref{eq: conf primary transformation finite} and \eqref{eq: conf primary transformation} respectively, now with $w(z)=z+\epsilon(z)\in\text{SO}(3)$ for infinitesimal $\epsilon$.
With respect to the linear combinations \eqref{eq: isometries complex}, $F(z)$ and $\ol F(\zb)$ obey Ward identities of $\text{SL}(2,\mathbb{R})$ primaries with holomorphic weights \eqref{eq: conformal weights}.

Starting with the mode expansion \eqref{eq: mode expansion eds2}, the construction of the $\text{SL}(2,\mathbb{R})$ generators closely mirrors the discrete series scalars in $\text{EAdS}_{2}$ discussed in section \ref{sssec: SL2R AdS}.
Instead of a single set of oscillators in $\text{EAdS}_{2}$, we now have two sets of independent oscillators $\alpha_n$ and $\tilde{\alpha}_n$, along with position and momentum operators $x_n$ and $p_n$ associated with modes which transform differently with respect to $\text{EdS}_2$ Killing vectors \eqref{eq: dS ell}. 
The operator forms of $\ell_1$, $\ell_0$, and $\ell_{-1}$ may be deduced from the transformation properties of each mode function.
In particular,
\begin{subequations}
    \begin{align}
        \ell_{1}=&\sum_{n=k+1}^\infty\sqrt{\frac{(n-k-1)(n+k)}{n(n-1)}}\alpha_{1-n} \alpha_{n}                      +\sum_{n=k+1}^\infty\sqrt{\frac{(n-k)(n+k+1)}{n(n+1)}}\tilde{\alpha}_{-1-n} \tilde{\alpha}_{n} \nonumber\\
        &+\sum_{n=-(k-1)}^{k}(k+1-n)x_{1-n}p_{n}+\frac{2k+1}{\sqrt{(k+1)(2k+1)!}}\left(i^{k}\tilde{\alpha}_{-k-1}+i^{-k}\alpha_{k+1}\right)p_{-k}\\
        \ell_0=&\sum_{n=k+1}^\infty \alpha_{-n} \alpha_n-\sum_{n=k+1}^\infty \tilde{\alpha}_{-n} \tilde{\alpha}_{n} +i\sum_{n=-k}^knx_{-n}p_{n}\\
        \ell_{-1}=&\sum_{n=k+1}^\infty\sqrt{\frac{(n-k)(n+k+1)}{n(n+1)}}\alpha_{-1-n} \alpha_{n} +\sum_{n=k+1}^\infty\sqrt{\frac{(n-k-1)(n+k)}{n(n-1)}}\tilde{\alpha}_{1-n} \tilde{\alpha}_{n}\nonumber\\
        &+\sum_{n=-k}^{k-1}(k+1+n)x_{-1-n}p_{n}+\frac{2k+1}{\sqrt{(k+1)(2k+1)!}}\left(i^{-k}\tilde{\alpha}_{k+1}+i^k\alpha_{-k-1}\right)p_k.
    \end{align}
    \label{eq: ell dS op}%
\end{subequations}
We collect explicit expressions of the commutators of these generators with the mode operators in appendix \ref{app: ds2 isometry action}.
A quick inspection of these operators reveals that the choice of vacuum \eqref{eq: dS vacuum} is dS-invariant in the sense that $\ell_n|\Omega\rangle$=0. 

\subsubsection{Virasoro algebra}

Similar to $\text{EAdS}_{2}$, we will find that, even though the two-point correlation functions \eqref{eq: dS F cor} in $\text{EdS}_{2}$ are reminiscent of those found in a $2d$ CFT, the isometries \eqref{eq: ell dS op} cannot be embedded within a larger set of coordinate transformations generated by $\ell_n$ to form a Virasoro algebra. 

In this case, the infinitesimal transformation \eqref{eq: conf primary transformation} with $\epsilon(z)=z^{n+1}$ and $n\notin \{-1,0,1\}$ mixes $\alpha_{m}$ (for $|m-n|\leq k$) with $p_{m-n}$ modes in the expansion \eqref{eq: mode expansion eds2}. 
To represent this transformation as an operator acting on the Hilbert space for $k>0$, it is therefore necessary to have a term like $\sim ix_{-m'}\alpha_m \subset \ell_n$.
Crucially, because $F(z)$ does not depend on any $x_m$, the action of this term on $F(z)$, i.e. 
\begin{equation}
    \delta_{z^{n+1}}F(z)\equiv -[\ell_n,F(z)]\sim-[ix_{-m'}\alpha_m,F(z)]+\dots
\end{equation}
mixes $F(z)$ with $\phi(z,\zb)$. 
This implies that $\delta_{z^{n+1}}F(z)$ is a non-local transformation, contradicting the assumed representation of $\delta_{z^{n+1}}$ in \eqref{eq: conf primary transformation}. 
Note that when $k=0$, there is no contradiction because the transformation \eqref{eq: conf primary transformation} with $\epsilon(z)=z^{n+1}$ annihilates the mode $\alpha_{n}z^{-(n+1)}$, instead of mixing with $p_0$. 
It is therefore impossible to represent this transformation as an operator acting on the Hilbert space if and only if $k>0$.

Nevertheless, it is possible to construct operators which satisfy the Virasoro algebra, in an identical manner to section \ref{sssec: virasoro ads}. 
In $\text{EdS}_2$, we now have two sets of oscillators $\alpha_n$ and $\tilde{\alpha}_n$, so there will be two sets of Virasoro generators $L_n$ and $\tilde{L}_n$.  

\subsubsection{Shift transformations \label{sssec: dS shift}}

As we saw in section \ref{ssec: dS lorentz}, in $\text{EdS}_{2}$, there is no restriction on the set of allowed modes as in $\text{EAdS}_2$ owing to the lack of spatial boundary conditions. 
In particular, this means that the functions $\phi_{\text{shift}}(z,\zb)$ locally satisfying 
\begin{equation}
    \(\pp\ol{\pp} + \frac{k(k+1)}{(1+z\zb)^{2}}\)  \phi_{\text{shift}} = 0,\quad \nabla^{k+1}\phi_{\text{shift}}=0,\quad \ol{\nabla}^{k+1}\phi_{\text{shift}}=0.
    \label{eq: shift eqns ds}
\end{equation}
and associated with shift transformations of the form \eqref{eq: shift trans} are allowed in the mode expansion. In $\text{EdS}_{2}$, the general solution to \eqref{eq: shift eqns ds} is spanned by the modes in $\phi_{\text{pos.}}$ given in \eqref{eq: xp mode expansion}, constituting solutions with definite angular momentum $|n|\leq k$, which on the Riemann sphere admits the holomorphic splitting \eqref{eq: phi varphi shift} now with 
\begin{equation}
    \varphi_{\text{shift}}(z) =\sum_{|n|\leq k}i^{k+n}\frac{c_n}{z^{n-k}},\quad \tilde{\varphi}_{\text{shift}}(\zb) =\sum_{|n|\leq k}i^{k+n}\frac{c_{-n}}{\zb^{n-k}}.
    \label{eq: shift soln ds}
\end{equation}
As discussed in section \ref{sssec: ads shift} for $\text{EAdS}_{2}$, there are $2k+1$ independent modes which contribute to the shift transformation \eqref{eq: shift trans} rather than the expected $2(2k+1)$ due to identity \ref{idty: ds shift mode identity}.

We may alternatively also characterise these shift-modes as the space of solutions to the KG equation \eqref{eq: s2 KG} which are globally defined on the Riemann sphere. 
All local solutions to \eqref{eq: s2 KG} with definite angular momentum $|n|\geq k+1$ diverge either at the origin or infinity. 
The momentum modes are similarly not globally defined on the Riemann sphere due to their logarithmic dependence on $|z|$. 
Only the shift-modes defined by \eqref{eq: shift soln ds} are globally defined. 
In particular, this means that the dimension of the kernel of the KG-operator $\(\pp\ol\pp+ k(k+1)/(1+z\zb)^2\)$ on the Riemann sphere is $2k+1$. 
These are precisely the modes with zero Euclidean action discussed in \cite{Anninos:2023lin}, which form a spin-$k$ representation of $\mathrm{SO}(3)\cong \mathrm{PSU}(2)$.

The canonical commutation relations \eqref{eq: ds ccrs} imply that the shift transformations are generated by the momentum operators $p_n$. 
Indeed, the action of a single $p_n$ on $\phi(z,\zb)$ is
\begin{equation}
    \begin{aligned}
        \delta_n\phi(z,\zb)&=i[p_n,\phi(z,\zb)] \\
        &=\frac{1}{2}\frac{1}{\sqrt{4\pi}}i^{k-n}\mathcal{D}_k\frac{1}{z^{-n-k}}+\frac{1}{2}\frac{1}{\sqrt{4\pi}}i^{k+n}\ol{\mathcal{D}}_k\frac{1}{\zb^{n-k}}\\
        &= \frac{i^{k-n}k!}{\sqrt{4\pi}}z^nP_k^{(n,-n)}\(\frac{1-z\zb}{1+z\zb}\)
    \end{aligned}
\end{equation}
for each $|n|\leq k$, where $P_m^{(\alpha,\beta)}(x)$ is a Jacobi polynomial. 
Real shift transformations are generated by the Hermitian combinations $p_n+p_{-n}$ and $i(p_n-p_{-n})$.

As discussed in section \ref{sssec: ds hilbert space}, this theory is made consistent after modding out by the shift symmetries. 
This amounts to considering as observables only shift-invariant operators, i.e. $F(z)$ and $\ol{F}(\zb)$ and derivatives thereof. 
When $k>0$, all of the isometries $\ell_1$, $\ell_0$, and $\ell_{-1}$ in \eqref{eq: ell dS op} clearly have explicit dependence on the position operators $x_n$, which reflects the fact that the stress tensor $T_{\mu\nu}$, which generates these operators, is not invariant under shift transformations and hence not an observable.


\section{Chiral symmetries \label{sec: chiral}}

In this section, we will describe the general structure of non-linear, chiral symmetries present in this theory at the classical level due to the holomorphicity of $F(z)$, ignoring global considerations coming from boundary conditions. 
Such symmetries are important for understanding the integrable structure of the theory, as well as determining deformations which preserve some of the symmetry. 

We begin by describing the symmetries in \ref{ssec: sym trans}.
In subsection \ref{ssec: chiral algebra}, we derive the Lie algebra of symmetries, and show for $k>0$ that it is a subalgebra of the algebra present in the $2d$ massless scalar theory, i.e. when $k=0$. 
In subsection \ref{ssec: int deforms}, we show that at least for $k\in \{0,1,2\}$, there exist infinite-dimensional mutually commuting subalgebras. 
As we explain, these subalgebras define infinitely many integrable deformations of the theory. 
An identical description can be given due to the anti-holomorphicity of $\ol{F}(\zb)$, which we omit. 
The discussion will use isothermal coordinates \eqref{eq: general 2d metric}, remaining agnostic of the choice of defining function $\Omega^{2}$ and hence the distinction between $\text{EAdS}_{2}$ and $\text{EdS}_{2}$.

\subsection{Symmetry transformations \label{ssec: sym trans}}

The existence of a much larger algebra of symmetries than the linear symmetries described in section \ref{sec: lin sym} can be deduced from a rather simple observation. 
Because $F$ is holomorphic, any power of $z$ times any power of any partial $z$-derivative $z^i(\partial^{m}F)^n$ is also holomorphic. 
In the same way, any \textit{functional} $\mathcal{L}(\{F,\partial F,\partial^2F,\dots\},z)$ of $F(z)$ and its derivatives, which may be formally of arbitrary order, is holomorphic. 
This implies an enormous set of symmetries---one for each functional $\mathcal{L}$---which we now describe. 
To avoid technical subtleties, we will assume that the functionals considered here depend on $F$ with at most $n$ (finite) derivatives. 
We will use the shorthand $\mathcal{L}(\{F_{n}(z)\},z)$ to indicate such a functional, where $F_{n} \equiv \pp^{n}F$.

As a warm-up, let us consider the massless scalar on flat space as discussed in \cite{Lindwasser:2024qyh}.
Since this is conformally coupled, this is locally equivalent to the $k=0$ theory.
The holomorphic current is $j=\partial\phi$ associated with constant $\mathbb{R}$-shifts. 
For that theory, any functional $\mathcal{L}(\{j_{n}(z)\},z)$ of $j(z)$ and its derivatives $j_{n} \equiv \pp^{n}j$ implies an infinitesimal symmetry transformation
\begin{equation}
    \begin{aligned}
        &\phi\mapsto\phi+\epsilon\delta_\mathcal{L}\phi\\
        &\delta_\mathcal{L}\phi = \frac{1}{2}\mathcal{E}(\mathcal{L})=\frac{1}{2}\sum_{i=0}^\infty(-1)^i\partial^i\frac{\partial}{\partial j_{i}(z)}\mathcal{L}(\{j_{n}(z)\},z),
    \end{aligned}
    \label{eq: phi chiral trans k=0}
\end{equation}
where $\epsilon$ is some infinitesimal parameter.
Here, $\mathcal{E}(\mathcal{L})$ is a linear operator called the Euler operator which gives the Euler--Lagrange equation for $j(z)$ as derived from the functional $\mathcal{L}(\{j_{n}(z)\},z)$ interpreted as a $1d$ Lagrangian (see below).
In appendix \ref{app: explicit invariance}, we show that for all $k\geq 0$, the action 
\begin{equation}
    S = \int \dzz \(\pp\phi \ol{\pp}\phi + \frac{\Omega^{2}m_{k}^{2}}{4}\,\phi^{2} \)
    \label{eq: isothermal action}
\end{equation}
for the $\text{E(A)dS}_{2}$ discrete series scalar in complex coordinates is invariant under the infinitesimal transformation
\begin{equation}
    \begin{aligned}
        &\phi\mapsto \phi +\epsilon\delta_{\mathcal{L}}\phi\\
        &\delta_\mathcal{L}\phi=\frac{1}{2}(-1)^k\mathcal{D}_k\mathcal{E}(\mathcal{L})=\frac{1}{2}(-1)^k\mathcal{D}_k\sum_{i=0}^\infty(-1)^i\partial^i\frac{\partial}{\partial F_{i}(z)}\mathcal{L}(\{F_{n}(z)\},z),
    \end{aligned}
    \label{eq: phi chiral trans}
\end{equation}
where $\epsilon$ is some infinitesimal parameter and $\mathcal{D}_k$ is the, by now familiar, order-$k$ differential operator defined in appendix \ref{app: riemann surfaces}.
Note that there is an infinitesimal symmetry transformation associated to \textit{any} functional $\mathcal{L}(\{F_{n}(z)\},z)$.

Let us make a few comments on the Euler operator $\mathcal{E}$.
An important property of it is that $\mathcal{E}(\mathcal{L})=0$ for any value of $F(z)$ and its derivatives if and only if $\mathcal{L}$ is a total derivative \cite{Olver:1993}. 
Because of the linearity of $\mathcal{E}$, any two functionals $\mathcal{L}$ and $\mathcal{L}'=\mathcal{L}+\partial f$ will generate the same infinitesimal transformation. 
In particular this means that the set of functionals which generate the algebra of symmetries is the set of all functionals \textit{modulo} total derivatives, or in other words, the space of all $1d$ Lagrangians. 
A different perspective on this is to note that the generators of the symmetry transformation are the integrated currents $Q_\mathcal{L}\propto\oint dz \,\mathcal{L}$---we have however found it useful to describe the Lie algebra in terms of the currents $\mathcal{L}$ themselves.
With this in mind, we will say that two functionals $\mathcal{L}$ and $\mathcal{L}'$ are equivalent if and only if they generate the same Euler--Lagrange equation, i.e. $\mathcal{E}(\mathcal{L})=\mathcal{E}(\mathcal{L}')$, or $\mathcal{L}=\mathcal{L}'+\partial f$ for some functional $f$.
Let us denote this equivalence relation by $\mathcal{L}\sim\mathcal{L}'$.
We shall henceforth refer to $\mathcal{L}(\{F_{n}(z)\},z)$ as a \textit{generator} of the Lie algebra, and call each element a \textit{Lagrangian} (as opposed to a \textit{functional}, which will refer to a $f(\{F_n(z)\},z)$ not equipped with the equivalence relation).
Given this, we see that a Lagrangian $\mathcal{L}(\{F_{n}(z)\},z)$ which depends on $F(z)$ and its derivatives up to order $n$ may be equivalent to a Lagrangian $\mathcal{L}'(\{F_{n'}(z)\},z)$ with $n'<n$. 
We therefore define the \textit{order of a Lagrangian} as the minimum value of $n$ which can be obtained by adding total derivatives.
For example, we say that $\mathcal{L}(\{F_4(z)\},z)=F_2(z)F_4(z)$ is of order 3, because $F_2F_4\sim -F_3F_3$. 
We assume without loss of generality that $\mathcal{L}(\{F_n(z)\},z)$ has been reduced such that $n$ is the order of $\mathcal{L}$.


\subsection{Chiral symmetry algebra \label{ssec: chiral algebra}}

The symmetry transformations \eqref{eq: phi chiral trans} associated with two Lagrangians $\mathcal{L}$ and $\mathcal{L}'$ do not in general commute. 
In this section, we will show that the commutator $[\delta_\mathcal{L},\delta_{\mathcal{L}'}]$ generates a transformation $\delta_{\mathcal{L}''}$ associated with a new Lagrangian $\mathcal{L}''$, defined by
\begin{equation}
    [\delta_\mathcal{L},\delta_{\mathcal{L}'}]=\delta_{\mathcal{L}''},\quad \mathcal{L}''=\frac{1}{4}\(\partial^{k+1}\mathcal{E}(\mathcal{L})\partial^k\mathcal{E}(\mathcal{L}')-\partial^{k+1}\mathcal{E}(\mathcal{L}')\partial^k\mathcal{E}(\mathcal{L})\).
    \label{eq: F comm}
\end{equation}
Here, we have written $\mathcal{L}''$ in a way which is manifestly anti-symmetric in $(\mathcal{L}\leftrightarrow\mathcal{L}')$, but we note that
\begin{equation}
    \mathcal{L}'' \sim \frac{1}{2}\partial^{k+1}\mathcal{E}(\mathcal{L})\partial^k\mathcal{E}(\mathcal{L}')
\end{equation}
up to total derivatives.

To show this, one may proceed via direct computation. 
Instead, it is more instructive to proceed first by considering the action of the symmetry transformation on $F(z)$.
Because of identity \ref{idty: nabla mathcalD ads}, this takes the form
\begin{equation}
    \delta_\mathcal{L} F(z)=\frac{1}{2}(-1)^k\partial^{2k+1}\mathcal{E}(\mathcal{L}).
    \label{eq: F chiral trans}
\end{equation}
There is an interesting relation between the transformation \eqref{eq: F chiral trans} and the infinitesimal transformation associated with the current $j(z)$ in the theory of the $2d$ massless scalar discussed above. 
Associated with a general Lagrangian $\mathcal{L}(\{j_{n}(z)\},z)$, the infinitesimal transformation of $j_{k}(z)$ is
\begin{equation}
    \delta_\mathcal{L} j_{k}(z)=\partial^{k+1}\delta_\mathcal{L}\phi=\frac{1}{2}\partial^{k+1}\mathcal{E}(\mathcal{L}).
    \label{eq: jk chiral trans}
\end{equation}
Next, let us consider the subset of Lagrangians $\mathcal{L}(\{j_{k+n}(z)\},z)$ which only depend on $j_{k},\dots,j_{k+n}$. 
In other words, we are now considering functionals of $j_{k}(z_{})$ rather than $j(z)$. 
The Euler operator for functionals of $j(z)$ appearing in \eqref{eq: jk chiral trans} acting on functionals of $j_{k}(z)$ simplifies to
\begin{equation}
    \begin{aligned}
        \mathcal{E}(\mathcal{L})&=\sum_{i=k}^{\infty}(-1)^i\partial^i\frac{\partial}{\partial j_i(z)}\mathcal{L}(\{j_{k+n}(z)\},z) \nonumber\\
        &=(-1)^k\partial^k\sum_{i=0}^{\infty}(-1)^i\partial^i\frac{\partial}{\partial j_{k+i}(z)}\mathcal{L}(\{j_{k+n}(z)\},z)\nonumber\\
        &=(-1)^k\partial^k\mathcal{E}^k(\mathcal{L}),
    \end{aligned}
    \label{eq: j to jk}
\end{equation}
where $\mathcal{E}^k$ denotes the Euler operator acting on functionals of $j_{k}(z)$ (treating $j_{k}(z)$ as the function one varies to get the Euler--Lagrange equation from the $1d$ Lagrangian $\mathcal{L}(\{j_{k+n}(z)\},z)$). 
With respect to functionals of $j_{k}(z)$, this transforms as
\begin{equation}
    \delta_\mathcal{L} j_k(z)=\frac{1}{2}(-1)^k\partial^{2k+1}\mathcal{E}^k(\mathcal{L}),
    \label{eq: jk trans jk}
\end{equation}
which, up to the relabelling $j_{k}(z)\mapsto F(z)$ and $\mathcal{E}^k\mapsto\mathcal{E}$, is structurally identical to \eqref{eq: F chiral trans}. 
Because of the equivalence between \eqref{eq: F chiral trans} and \eqref{eq: jk trans jk}, the commutator $[\delta_\mathcal{L},\delta_{\mathcal{L}'}]$ for the discrete series scalar with $k>0$ in $\text{EAdS}_2$ is isomorphic to the commutator which would be computed in the massless scalar theory for the subset of functionals $\mathcal{L}(\{j_{k+n}(z)\},z)$ of $j_{k}(z)$.
In this sense, the Lie algebra of symmetries of the discrete series scalar for $k>0$ is a subalgebra of the Lie algebra of symmetries present in the massless scalar theory.

We may now use this isomorphism to our advantage and calculate the commutator $[\delta_\mathcal{L},\delta_{\mathcal{L}'}]$ for the transformations \eqref{eq: F chiral trans} explicitly. 
In \cite{Lindwasser:2024qyh}, it was shown that for the massless scalar, the commutator $[\delta_\mathcal{L},\delta_{\mathcal{L}'}]$ for the transformations \eqref{eq: phi chiral trans k=0} generates a new transformation $\delta_{\mathcal{L}''}$ associated with a new Lagrangian $\mathcal{L}''$ defined by
\begin{equation}
    [\delta_\mathcal{L},\delta_{\mathcal{L}'}]=\delta_{\mathcal{L}''},\quad \mathcal{L}''=\frac{1}{4}\(\partial\mathcal{E}(\mathcal{L})\mathcal{E}(\mathcal{L}')-\partial\mathcal{E}(\mathcal{L}')\mathcal{E}(\mathcal{L})\).
    \label{eq: phi comm k=0}
\end{equation}
To get the commutator $[\delta_\mathcal{L},\delta_{\mathcal{L}'}]$ for the transformations \eqref{eq: F chiral trans}, we simply take \eqref{eq: phi comm k=0} and restrict $\mathcal{L}$ and $\mathcal{L}'$ to be functionals of $j_{k}(z)$ using \eqref{eq: j to jk}, and make the replacements $j_{k}(z)\mapsto F(z)$ and $\mathcal{E}^k\mapsto\mathcal{E}$.
This precisely leads to \eqref{eq: F comm}.


\subsection{Integrable deformations \label{ssec: int deforms}}

Being infinite-dimensional, the chiral symmetry algebra \eqref{eq: F comm} provides a lot of structure for this theory, enabling us to explore the types of interactions that preserve subalgebras of \eqref{eq: F comm}. 
Of particular interest are interactions which preserve an \textit{infinite number} of mutually commuting symmetries, associated with some sequence of Lagrangians $\mathcal{L}_{i}$ for $i=1,2,\dots,\infty$, which satisfy
\begin{equation}
    \partial^{k+1}\mathcal{E}(\mathcal{L}_{i})\partial^k\mathcal{E}(\mathcal{L}_{j})-\partial^{k+1}\mathcal{E}(\mathcal{L}_{j})\partial^k\mathcal{E}(\mathcal{L}_{i})\sim 0,
    \label{eq: commute integrable}
\end{equation}
for any $i,j$. By construction, these infinite-dimensional mutually commuting subalgebras would be symmetry algebras of a would-be integrable deformation of \eqref{eq: isothermal action}. The task is to find an interaction $S_{\text{int.}}$ which is invariant under this subalgebra, i.e. $\delta_{\mathcal{L}_{i}}S_{\text{int.}}=0$ for all $i$. 

Luckily, the existence of an infinite-dimensional mutually commuting subalgebra directly implies the existence of infinitely many integrable deformations of the action \eqref{eq: isothermal action}. 
Indeed, consider the action
\begin{equation}
    S_i=\int\dzz\(\pp\phi \ol{\pp}\phi + \frac{\Omega^{2}m_{k}^{2}}{4} \phi^{2} + \lambda \,\mathcal{L}_{i}(\{F_{n}\},z)\).
    \label{eq: int deform}
\end{equation}
We claim this is invariant under all transformations $\delta_{\mathcal{L}_{j}}\phi$ given by \eqref{eq: phi chiral trans} for arbitrary $\lambda$.
To show this, it is sufficient to show that $\delta_{\mathcal{L}_{j}}\mathcal{L}_{i}\sim 0$ is a total derivative for any $i,j$.
Indeed,
\begin{equation}
    \begin{aligned}
        \delta_{\mathcal{L}_{j}}\mathcal{L}_{i}&=\sum_{m=0}^{\infty}\(\delta_{\mathcal{L}_{j}}F_m(z)\)\frac{\pp}{\pp F_m(z)}\mathcal{L}_{i}(\{F_{n}(z)\},z)\\
        &=\frac{1}{2}(-1)^k\sum_{m=0}^{\infty}\pp^m\pp^{2k+1}\mathcal{E}(\mathcal{L}_{j})\frac{\pp}{\pp F_m(z)}\mathcal{L}_{i}(\{F_{n}(z)\},z)\\
        &\sim\frac{1}{2}(-1)^k\sum_{m=0}^{\infty}(-1)^m\pp^{2k+1}\mathcal{E}(\mathcal{L}_{j})\pp^m\frac{\pp}{\pp F_m(z)}\mathcal{L}_{i}(\{F_{n}(z)\},z)\\
        &= \frac{1}{2}(-1)^k\pp^{2k+1}\mathcal{E}(\mathcal{L}_{j})\mathcal{E}(\mathcal{L}_{i})\\
        &\sim \frac{1}{2}\partial^{k+1}\mathcal{E}(\mathcal{L}_{j})\partial^k\mathcal{E}(\mathcal{L}_{i}),
    \end{aligned}
\end{equation}
which vanishes because $\mathcal{L}_{i}$ and $\mathcal{L}_{j}$ commute in the sense of \eqref{eq: commute integrable}. 

After an Abelian deformation of the theory as in \eqref{eq: int deform}, the $\text{E(A)dS}_2$ isometry is generically broken. 
Such deformations are nonetheless interesting---a famous example of this is the Korteweg--de Vries (KdV) model \cite{Korteweg:1895lrm, Gardner:1967wc,10.1063/1.1665772}.
There may be other deformations not of the form \eqref{eq: int deform} which preserve an infinite-dimensional mutually commuting subalgebra \textit{and} the spacetime isometries. 
It was argued in \cite{Antunes:2025iaw} that this is not possible in $\text{AdS}_{2}$, but it is unclear whether this applies to $\text{dS}_{2}$.
When $k=0$ in the flat space limit, there are a few such well-known examples of integrable deformations which preserve Poincar\'{e} symmetry, such as the sine-Gordon model \cite{Coleman:1974bu}, Liouville model \cite{Liouville1855}, and the Bullough--Dodd (or Tzitz\'{e}ica) model \cite{Tzitzica:1907, Dodd:1977bi}. 

In the remaining part of this section, we shall focus on finding such infinite-dimensional mutually commuting subalgebras.
For simplicity, we will consider mutually commuting subalgebras of Lagrangians $\mathcal{L}_n(\{F_i(z)\})$ with no explicit $z$-dependence.
In $\text{EAdS}_{2}$, this guarantees that the deformation \eqref{eq: int deform} preserves the real translation isometry of the complex UHP. 
Furthermore, a useful organising tool is to assume that the generators $\mathcal{L}_n$ have definite weight: If we assign the weights $[\partial]=1$ and $[F]$, then we demand that $\mathcal{L}_n(\{F_{i}(z)\})$ depends on $F(z)$ and its derivatives such that $\mathcal{L}_n$ has definite weight $[\mathcal{L}_n]$, i.e. 
\begin{equation}
    \[\sum_{j=0}^\infty([F]+j)F_{j}(z)\frac{\pp}{\pp F_{j}(z)}\] \mathcal{L}_n(\{F_{i}(z)\})=[\mathcal{L}_n]\mathcal{L}_n(\{F_{i}(z)\}).
\end{equation}
Finally, we will also restrict ourselves to finding subalgebras consisting of Lagrangians $\mathcal{L}_n$ which are non-quadratic functionals of $F(z)$, so that the integrable deformation \eqref{eq: int deform} defines a non-linear equation of motion. 
As we explain in the following sections, we are only able to find infinite-dimensional mutually commuting subalgebras when $k=0,1,2$. 
When $k>2$, there is an apparent obstruction to finding such subalgebras which we relate to a longstanding conjecture in the integrable equations literature. 

\subsubsection{$k=0$ \label{sssec: k=0}}

Let us first consider $k=0$.
The $k=0$ theory is, due to its conformal symmetry, equivalent to the $2d$ massless scalar in flat space, for which a lot is known about its infinite-dimensional mutually commuting subalgebras \cite{Lindwasser:2024qyh,Lindwasser:2025slu}. 
We briefly review the infinite-dimensional mutually commuting subalgebras found in \cite{Lindwasser:2024qyh}. 
Because we have $F(z)=j(z)=\partial\phi$ in this case, we may simply restate the results therein by relabelling $j\to F$.

There are six known infinite-dimensional mutually commuting subalgebras of $z$-independent functionals $\mathcal{L}_n$ with definite weight.
These are associated with the following choices for weights $[F] \in \{0,\pm 1,\pm 2\}$.
Below, we list the first three Lagrangians of lowest order in the subalgebra\footnote{The subscript $n$ in $\mathcal{L}_n$ now denoting the order. In general, there is a basis of Lagrangians in any such subalgebra for which each generator has a distinct order \cite{Lindwasser:2025slu}.}:
\begin{itemize}
    \item $[F]=2$: There is essentially one infinite-dimensional subalgebra, which is unique up to an affine transformation $F\mapsto a F+b$ with $a,b\in\mathbb{C}$.
    This subalgebra is associated with the KdV equation. 
    We list here the first three Lagrangians of lowest order in the subalgebra, which are unique up to an affine transformation $F\mapsto a F+b$ with $a,b\in\mathbb{C}$:
    \begin{subequations}
        \begin{align}
            &\mathcal{L}_1= F_1^2+2 F^3,\\
            &\mathcal{L}_2= F_2^2 + 10 FF_1^2 + 5F^4,\\
            &\mathcal{L}_3= F_3^2 + 14 FF_2^2 + 70 F^2F_1^2 + 14 F^5.
        \end{align}
    \end{subequations}
    \item $[F]=1$: There are two distinct subalgebras, which are unique up to an affine transformation $F\mapsto a F+b$ with $a,b\in\mathbb{C}$. 
    The first is associated with the symmetries of the sine-Gordon model, the Liouville model, and the modified KdV equation \cite{Miura:1968}. 
    The first three Lagrangians of lowest order are 
    \begin{subequations}
        \begin{align}
            &\mathcal{L}_1= F_1^2+ F^4,\\
            &\mathcal{L}_2= F_2^2 + 10 F^2F_1^2 + 2F^6,\\
            &\mathcal{L}_3= F_3^2 + 14 F^2F_2^2 -7F_1^4+ 70 F^4F_1^2 + 5 F^8.
        \end{align}
    \end{subequations}
    The second subalgebra with $[F]=1$ is associated with the symmetries of the Bullough--Dodd (or Tzitz\'{e}ica) model, of which the first three Lagrangians of lowest order are 
    \begin{subequations}
        \begin{align}
            &\mathcal{L}_2= F_2^2-5F_1^3+45F^2F_1^2+27F^6,\\
            &\mathcal{L}_3= F_3^2-21F_1F_2^2+63F^2F_2^2-21F_1^4-126F^2F_1^3+1134F^4F_1^2+243F^8, \\
            &\begin{aligned}
                \mathcal{L}_5= &\,F_5^2-33F_1F_4^2+44F_3^3+99F^2F_4^2-990FF_2F_3^2-594F_1^2F_3^2+561F_2^4 +8118F_1^3F_2^2 \\
                &\, +3168FF_1F_2^3-1782F^2F_1F_3^2+\frac{24156}{5}F_1^6 - 58806F^2F_1^2F_2^2-15444F^3F_2^3 \\
                &\, +3564F^4F_3^2 + 44550F^2F_1^5-26730F^4F_1F_2^2 + 58806F^6F_2^2-267300F^4F_1^4 \\
                &\, - 53460F^6F_1^3 + 481140F^8F_1^2 +26244F^{12}.
            \end{aligned}
        \end{align}
    \end{subequations}
    \item $[F]=0$: There is one subalgebra, which is unique up to an affine transformation $F\mapsto a F+b$ with $a,b\in\mathbb{C}$.
    Its first three lowest-order Lagrangians being
    \begin{subequations}
        \begin{align}
            &\mathcal{L}_1= \frac{F_1^2}{(1-F^2)^3}, \\
            &\mathcal{L}_2= \frac{F_2^2}{(1-F^2)^5}-\frac{5}{3}\frac{1+8F^2}{(1-F^2)^7}F_1^4,\\
            &\mathcal{L}_3= \frac{F_3^2}{(1-F^2)^7}-14\frac{F}{(1-F^2)^8}F_2^3-14\frac{2+13F^2}{(1-F^2)^9}F_1^2F_2^2 +\frac{14}{5}\frac{11+248F^2+416F^4}{(1-F^2)^{11}}F_1^6.
        \end{align}
    \end{subequations}
    \item $[F]=-1$: There is one subalgebra, which is unique up to an affine transformation $F_1\mapsto a F_1+b$ with $a,b\in\mathbb{C}$, with the first three lowest-order Lagrangians being
    \begin{subequations}
        \begin{align}
            &\mathcal{L}_1= (1+F_1^2)^{1/2}, \\
            &\mathcal{L}_2= \frac{F_2^2}{(1+F_1^2)^{5/2}},\\
            &\mathcal{L}_3= \frac{F_3^2}{(1+F_1^2)^{7/2}}+\frac{7}{4}\frac{1-4F_1^2}{(1+F_1^2)^{11/2}}F_2^4.
        \end{align}
        \label{eq: weight -1 algebra}%
    \end{subequations}
    \item $[F]=-2$: There is one subalgebra, which is unique up to an affine transformation $F_2\mapsto a F_2+b$ with $a,b\in\mathbb{C}$, with the first three lowest-order Lagrangians being
    \begin{subequations}
        \begin{align}
            &\mathcal{L}_2= (1+F_2)^{1/3}, \\
            &\mathcal{L}_3= \frac{F_3^2}{(1+F_2)^{7/3}},\\
            &\mathcal{L}_5= \frac{F_5^2}{(1+F_2)^{11/3}} + \frac{44}{9}\frac{F_4^3}{(1+F_2)^{14/3}}-\frac{220}{9}\frac{F_3^2F_4^2}{(1+F_2)^{17/3}} + \frac{6545}{243}\frac{F_3^6}{(1+F_2)^{23/3}}.
        \end{align}
        \label{eq: weight -2 algebra}%
    \end{subequations}
\end{itemize}
This concludes the classification. 

\subsubsection{$k>0$ \label{sssec: k>0}}

Let us now turn to $k > 0$.
When $k>0$, recall that the Lie algebra \eqref{eq: F comm} is isomorphic to the subalgebra of the massless scalar symmetries, generated by Lagrangians $\mathcal{L}(\{j_k(z)\},z)$ which are functionals of $j_k(z)$. 
Because of this, as long as there are infinite-dimensional mutually commuting subalgebras in the massless theory made up of Lagrangians $\mathcal{L}(\{j_k(z)\},z)$ which are functionals of $j_k(z)$, there is automatically a similar subalgebra of \eqref{eq: F comm}.

When $k=1$ then, there are automatically two infinite-dimensional subalgebras, related to the subalgebras \eqref{eq: weight -1 algebra} and \eqref{eq: weight -2 algebra} for $k=0$ in section \ref{sssec: k=0}. 
They correspond to definite-weight Lagrangians if we assign $[F]=0$ and $[F]=-1$ respectively, and their first three lowest-order Lagrangians are obtained by relabelling $F_{n} \mapsto F_{n-1}$ (identifying $F_{0}=F$) in \eqref{eq: weight -1 algebra} and \eqref{eq: weight -2 algebra} respectively.

The same logic can be applied when $k=2$. 
In this case, only the subalgebra \eqref{eq: weight -2 algebra} from section \ref{sssec: k=0} survives. 
This subalgebra has definite weight Lagrangians after assigning $[F]=0$, and is obtained from \eqref{eq: weight -2 algebra} and simultaneously relabelling $F_{n} \mapsto F_{n-2}$ (identifying $F_{0} = F$).

When $k>2$, there are no longer any mutually commuting subalgebras that are known to be infinite-dimensional. 
Instead, what is known for the massless scalar theory are the necessary conditions on the functional form of the lowest-order Lagrangian in such a subalgebra \cite{Lindwasser:2025slu}. 
For our $k>0$ theory, this means for instance that if an infinite-dimensional mutually commuting subalgebra with Lagrangians $\mathcal{L}_{n_i}$ with $i=1,2,\dots,\infty$ has a lowest-order Lagrangian $\mathcal{L}_{n_1}(\{F_{n_1}(z)\})$ with order $n_1$, it must solve the transcendental equation
\begin{equation}
    \left(\mathcal{L}_{n_1}{}_2F_1\left(\frac{1}{2},-\frac{1}{2(k+n_1)};1-\frac{1}{2(k+n_1)};a_{n_1-1}(\mathcal{L}_{n_1})^{-2(k+n_1)}\right)\right)^2=\left(\frac{cF_{n_1}}{\sqrt{a_{n_1-1}}}+b_{n_1-1}\right)^2,
    \label{eq: lowest order ell condition}
\end{equation}
where $a_{n_1-1}$ and $b_{n_1-1}$ are undetermined functionals of $F$ of order $n_1-1$, and $c$ is a free constant parameter. 
Other necessary conditions on infinite-dimensional mutually commuting subalgebras of \eqref{eq: F comm} can be read off from \cite{Lindwasser:2025slu}.
Because these are only necessary conditions, it may be that a subalgebra containing $\mathcal{L}_{n_1}$ that solves \eqref{eq: lowest order ell condition} is either not infinite-dimensional or not mutually commuting. 
Indeed, a Lagrangian which solves \eqref{eq: lowest order ell condition} when $n_1=0$ is
\begin{equation}
    \mathcal{L}_0 = (1 +F)^{\frac{1}{k+1}},
    \label{eq: test ell}
\end{equation}
for some choice of free parameters in \eqref{eq: lowest order ell condition}. 
A non-exhaustive search for Lagrangians which commute with \eqref{eq: test ell} finds solutions when $k \in \{0,1,2\}$, but none when $k>2$.

This apparent obstruction for finding infinite-dimensional mutually commuting subalgebras when $k>2$ is actually related to a conjecture in the integrable equations literature \cite{Heredero:2019arc,Sanders:1998}. 
In particular, consider an integrable hierarchy of evolution equations for the function $u(x,t_i)$ of the form 
\begin{equation}
    u_{t_i} =f_i(u,u_x,u_{xx},\dots,u_{n_i},x)
    \label{eq: hierarchy}
\end{equation}
where subscripts denote partial derivatives in $t_i$ and $x$, and the $f_i$ form an infinite sequence of functionals $i=1,2,\dots,\infty$. 
This sequence of equations forms an integrable hierarchy in the sense that the different notions of time $t_i$, each associated with a Hamiltonian $H_i$, are mutually commuting so that $\{H_i,H_j\}=0$ for all $i,j$, for each equation \eqref{eq: hierarchy}.
There are powerful classification theorems for integrable equations of the form \eqref{eq: hierarchy} when $f_i$ is of order 2, 3 or 5 \cite{SISvinolupov_1992,Svinolupov1982,Sokolov1985,Mikhailov1991,Mikhailov2009,meshkov2013integrable}. 
Going beyond order-5 is practically challenging. 
Nevertheless, it is believed that 2, 3 and 5 are the only relevant orders to study to classify all integrable hierarchies of the form \eqref{eq: hierarchy}. 
For instance, it is known that if $f_i(u,u_x,u_{xx},\dots,u_{n_i},x)$ is a polynomial functional with homogeneous weight, then the integrable hierarchy it is a member of must have a member of orders 2, 3, or 5 \cite{Sanders:1998}. The aforementioned conjecture states that \textit{every} integrable hierarchy \eqref{eq: hierarchy} has a member of order 2, 3, or 5 \cite{Heredero:2019arc}. 
As discussed in \cite{Lindwasser:2024qyh,Lindwasser:2025slu}, the problem of classifying odd order integrable hierarchies of the form \eqref{eq: hierarchy} is related to the problem of finding all infinite-dimensional mutually commuting subalgebras for the massless scalar theory, where an order-$n$ Lagrangian $\mathcal{L}_n$ in such a subalgebra defines an integrable equation \eqref{eq: hierarchy} of order $2n+1$. 

A corollary of this conjecture is that the discrete series scalar with $k>2$ has no infinite-dimensional mutually commuting subalgebras, because such a subalgebra would imply an integrable hierarchy whose lowest-order member has an order greater than 5. 
Incidentally, barring exceptions due to dimension-dependent identities, there exist similar obstructions to finding interactions which preserve $\text{(A)dS}$ isometries, shift symmetries with level $k>2$, and parity invariance \cite{Hinterbichler:2014cwa, Griffin:2014bta, Bonifacio:2018zex}. 
In flat space, there are also analogous obstructions to constructing $S$-matrices with the enhanced soft limits implied by such extended shift symmetries \cite{Cheung:2016drk}.


\section{Concluding remarks \label{sec: conclusion}}

In this work, we used complex coordinates on $\text{E(A)dS}_{2}$ to describe certain linear and non-linear symmetries of the discrete series scalar in $\text{(A)dS}_{2}$.
Due to the holomorphicity of the field strength components $F(z)$ and $\ol{F}(\zb)$, this field admits a Laurent series as its mode expansion, reminiscent of mode expansions typically seen in two-dimensional conformal field theories in the complex UHP and the extended complex plane. 
In terms of those, we constructed the operators which implement the $\text{(A)dS}_{2}$ isometries associated with the Ward identities constraining the correlation functions observed in \cite{Farnsworth:2024yeh}.
The fact that the current transforms as a primary under (transformations identical to the) \textit{global} conformal transformations (in complex coordinates) on flat space and is also holomorphic raises the question whether these can be extended to \textit{local} conformal transformations. 
We answer this in the negative, but show that there nonetheless exists a set of operators which obeys the Virasoro algebra.
In the other half of this manuscript, we describe the full chiral algebra of symmetries of the discrete series scalar implied by the (anti-)holomorphicity of the field strength.
After showing that they are described by subalgebras of the massless theory in flat space, we were able to lift previous results to find integrable deformations of the theories with shift symmetries at levels $k=0$, $k=1$, and $k=2$.
Finally, we state the apparent absence of integrable deformations for $k>2$ as a consequence of a known conjecture in the integrability literature.

An important feature of the oscillator mode expansions for the discrete series scalar at level $k$ was that it excluded the $2k+1$ modes around $n=0$ (whether these were excised from the mode expansion because they are non-normalisable in $\text{AdS}_{2}$ or manually modded out because they are zero modes in $\text{dS}_{2}$).
As we discussed in section \ref{sec: lin sym}, these missing modes precisely characterise the shift symmetries of $F(z)$ and $\ol{F}(\zb)$.
In the flat space limit, these remain excluded, which is in stark contrast with the massless scalar field on flat space.
Therefore, even though from \eqref{eq: shift pt} one finds $m_{k} \to 0$ in the flat space limit, the resulting theory is therefore \textit{not} equivalent to the flat space massless scalar. 
Instead, the flat space limit of the level-$k$ discrete series scalar is equivalent to the flat space massless scalar after modding out by polynomial shift transformations of degree $k$.

\hspace{15pt} Several future directions present themselves.
For instance, in this work, we studied shift symmetries of scalar fields in $\text{(A)dS}_{2}$.
The analogous shift symmetries for Dirac fermions in higher dimensions were studied in \cite{Bonifacio:2023prb, Letsios:2025pqo}.
In upcoming work, we will carry out the analysis above for fermions in two dimensions. 
Supposing that these fermions also admit a holomorphic splitting, it is natural to ask whether this is related to the holomorphic splitting in the bosonic sector via supersymmetry and if one might be able to find a bosonisation map between the theories. 
More generally, one could ask what other theories admit a holomorphic splitting.

Recall that all constant non-zero curvature Riemann surfaces can be mapped locally to the complex UHP or the extended complex plane with metrics \eqref{eq: UHP metric} and \eqref{eq: S2 metric}, respectively. 
It is clear then that similarly holomorphic yet massive scalar fields can be found on all constant, non-zero curvature Riemann surfaces. 
It will be interesting to explore the properties of the discrete series scalars on remaining surfaces of higher genus.
Without invariance under conformal coordinate transformations, one expects the holomorphicity to be lost when you allow the metric to fluctuate for $k>0$. 
A nice setting which is expected to retain holomorphicity with a fluctuating metric is when minimally coupling these fields to the metric of Jackiw--Teitelboim gravity \cite{Jackiw:1984je, TEITELBOIM198341}.

Also note that one may view the discrete series scalars as a kind of deformation of a worldsheet coordinate on constant curvature worldsheet configurations. 
Paired with the analogous fermions described above and the option to define these theories on higher-genus surfaces, there is an interesting prospect of realising these theories within (super)string theory, or perhaps some deformation thereof. 

The two-dimensional shift-symmetric theories described here actually secretly realise chiral $p$-forms in $D=2p+2$ with $p=0$ introduced in \cite{Henneaux:1988gg}.
A common theme throughout this work has been that even though the theories are defined in terms of the massive scalar field $\phi$, all of the signs seem to point towards the idea that the fundamental field is $F$.
It is difficult to find good action formulations for $F$ directly---na\"ive guesses such as $F_{\mu_{1}\dots \mu_{k+1}}F^{\mu_{1}\dots \mu_{k+1}}$ do not propagate the right degrees of freedom.
This is an avatar of the typical obstruction of finding action formulations for chiral forms, formulations of which are e.g.  discussed in \cite{Marcus:1982yu,Floreanini:1987as,Tseytlin:1990va,Pasti:1996vs,Sen:2015nph,Mkrtchyan:2019opf,Evnin:2022kqn,Arvanitakis:2022bnr,Avetisyan:2022zza,Evnin:2023ypu}.
The relevant bulk theory for the $\mathfrak{sl}(2,\mathbb{R})$ discrete series is higher-spin gravity in $\text{AdS}_{3}$ \cite{Blencowe:1988gj,Gonzalez:2014tba}.
Various formulations derived from performing different boundary reductions were given in \cite{Henneaux:2010xg, Campoleoni:2010zq, Sharapov:2024euk}.
It would be particularly interesting to study holomorphic splitting e.g. in the covariant formulation presented in \cite{Chen:2025xlo}.
For instance, the fact that the holomorphic splitting \eqref{eq: holomorphic splitting} can be written in terms of a rank-$k$ tensor might have a geometric interpretation within the higher-spin gravity bulk theory.
Relatedly, the equations \eqref{eq: shift eqns ads} and \eqref{eq: shift eqns ds} which characterise the shift-modes in $\text{AdS}_2$ and $\text{dS}_2$ respectively might shed some light on the origin of the (anti-)holomorphicity of the field strength. 
Regardless, a possible geometric explanation of the rich structures we have found would be tantalising. 

\section*{Acknowledgments}

It is our pleasure to thank Priyesh Chakraborty, Harry Goodhew, Yu-tin Huang, Austin Joyce, Shu-Heng Shao, and Jasper Roosmale Nepveu for illuminating discussions, and Samanta Saha, Kurt Hinterbichler, and Thomas Yan on feedback on an earlier draft.
We would further like to thank the anonymous referee for pointing out interesting questions and providing useful feedback.
C.Y.R.C. would further like to thank Karapet Mkrtchyan and Euihun Joung for discussions and collaborations on related topics.

The research of C.Y.R.C. was funded by Grant 115L104044 (Max Planck-IAS-NTU Center) supported by National Taiwan University under the framework of Higher Education SPROUT Project by the Ministry of Education in Taiwan. 
The research of L.W.L. is supported by the Taiwan NSTC Grant No. 113-2811-M-002 -167 -MY3 and the Yushan Young Fellowship. 
M.P. is supported in part by NSF grant PHY-2210349.

\bibliography{Bibliography}

\appendix

\section{Differential geometry on Riemann surfaces \label{app: riemann surfaces}}

In this appendix, we will briefly review some differential geometry on Riemann surfaces, in order to define certain differential operators and useful identities that they satisfy.
This is mostly following \cite{DHoker:1986eaw, Nakahara:2003nw, Blumenhagen:2009zz}.

Consider a Riemann surface $M$.
Locally, we can choose isothermal (conformally flat) coordinates, in which the metric takes the form 
\begin{equation}
    \dd s^{2} = \Omega(z,\zb)^{2} \dd z \,\dd \zb
    \label{eq: general 2d metric}
\end{equation}
for an arbitrary defining function $\Omega(z,\zb)^2$ which is real and positive
\begin{equation}
    \ol{\Omega(z,\zb)^2} = \Omega(\zb,z)^2 = \Omega(z,\zb)^2.
\end{equation}
In this appendix, we will define and collect useful identities for various differential operators on $M$.
When possible, we will keep the choice of $\Omega$ general.
However, we are ultimately interested in Euclidean (A)dS mapped to the complex UHP and Riemann sphere respectively, for which 
\begin{equation}
    \Omega(z,\zb)^{2} = 
    \begin{cases}
        \frac{-4L^{2}}{(z-\zb)^{2}}, & \text{AdS}\\
        \frac{4H^{-2}}{(1+z\zb)^{2}}, & \text{dS}.
    \end{cases}
    \label{eq: conformal factors (a)ds}
\end{equation}
It will be useful to make this choice explicit.  

For \eqref{eq: general 2d metric}, the metric determinant is
\begin{equation}
    \det g = -\frac{\Omega^{4}}{4},
\end{equation}
and the non-trivial Christoffel symbols are
\begin{equation}
    \Gamma^{z}_{zz} = 2\frac{\pp \Omega}{\Omega},\quad \Gamma^{\zb}_{\zb\zb} =2 \frac{\ol{\pp}\Omega}{\Omega}.
\end{equation}
In two dimensions, the Riemann tensor is fully determined by the Ricci scalar
\begin{equation}
    R = -8\Omega^{-4}\(-\pp\Omega \ol{\pp}\Omega + \Omega \pp\ol{\pp}\Omega\).
\end{equation}
As expected, when $\Omega$ is chosen to describe E(A)dS as in \eqref{eq: conformal factors (a)ds}, this will be constant:
\begin{equation}
    R = 
    \begin{cases}
        -2L^{-2}, & \text{AdS}\\
        2H^{2}, & \text{dS}.
    \end{cases}
\end{equation}

Let us now consider tensor fields on $M$.
The (co-)tangent bundle of $M$ is endowed with an automorphism induced by the complex structure on $M$, which defines a natural decomposition of the (co-)tangent bundle into its eigenspaces.
We may therefore write a generic tensor as $T^{q,\ol{q}}_{p,\ol{p}}$, which has $p$/$\ol{p}$ and $q$/$\ol{q}$ contravariant and covariant $z$-/$\zb$-indices respectively.
We can however further use the metric to convert $\zb$-indices to $z$-indices, resulting in a so-called holomorphic tensor with $p' = p + \ol{q}$ and $q' = q + \ol{p}$ covariant and contravariant $z$-indices respectively.
We will henceforth set $\ol{p} = \ol{q} = 0$ without loss of generality.
The rank of a holomorphic tensor with $p$ and $q$ contravariant and covariant indices is given by $h=p-q$\footnote{Here, we follow the convention set out in \cite{Blumenhagen:2009zz}.
Note this is opposite to the one used in \cite{Nakahara:2003nw}.
}.
Let us now denote the space of rank-$h$ holomorphic tensors on $M$ by $T^{(h)}(M)$.
The metric on $M$ naturally endows this with an invariant inner product.
For $V,W \in T^{(h)}$, this inner product is defined as
\begin{equation}
    \langle V,W \rangle =\int \dzz \sqrt{|g|} \(g^{z\zb}\)^{h} V\,\ol{W}.
\end{equation}
The action of the holomorphic covariant derivative a holomorphic tensor on $T^{(h)}$ is specified by
\begin{equation}
    \begin{aligned}
        \nabla_{z}^{(h)}:& T^{(h)} \rightarrow T^{(h+1)},\quad \nabla_{z}^{(h)} \cdot= \(\pp - h \frac{2\pp\Omega}{\Omega}\)\cdot,\\
        \nabla^{z}_{(h)}:& T^{(h)} \rightarrow T^{(h-1)},\quad \nabla^{z}_{(h)} \cdot= g^{z\zb} \ol{\pp}  \cdot\,.
    \end{aligned}
    \label{eq: complex cov d}
\end{equation}
It is useful to note that the covariant derivative $\nabla$ satisfies the following conjugation identity
\begin{equation}
    \nabla_{z}^{(h)} g_{z\zb}^{m} \cdot = g_{z\zb}^{m} \nabla_{z}^{(h-m)} \cdot\,.
    \label{eq: cov d conjugation}
\end{equation}
We define the adjoint to the covariant derivative to be an operator $(\nabla_{z}^{(h)})^{\dag}: T^{(h+1)} \rightarrow T^{(h)}$ satisfying
\begin{equation}
    \langle V, \nabla_{z}^{(h)} W \rangle = \langle (\nabla_{z}^{(h)})^{\dag}V,W\rangle,
\end{equation}
for $V \in T^{(h+1)}$ and $W \in T^{(h)}$.
Explicitly,
\begin{equation}
    \(\nabla_{z}^{(h)}\)^{\dag} = -\nabla^{z}_{(h+1)}.
\end{equation}

We will be particularly interested in cases where we consecutively apply covariant derivatives on a scalar test distribution. 
For any $\alpha \in \mathbb{N}_{> 0}$, we denote this by $\nabla^{\alpha}: T^{(0)} \rightarrow T^{(\alpha)}$, with  
\begin{equation}
    \begin{aligned}
        \nabla^{\alpha} \cdot &\equiv \nabla^{(\alpha-1)}_{z} \dots \nabla^{(0)}_{z} \cdot\\
        &= \(\pp - 2(\alpha-1)\frac{\pp\Omega}{\Omega} \) \dots \(\pp - 2\frac{\pp\Omega}{\Omega} \)\pp \;\cdot \\
        &=  \Omega^{2\alpha}  \(\Omega^{-2}\pp\cdot \)^{\alpha} \; \cdot 
    \end{aligned}
    \label{eq: nabla expressions}
\end{equation}
The adjoint to this operator is
\begin{equation}
    \begin{aligned}
        (\nabla^{\alpha})^{\dag}\cdot &= (-1)^{\alpha} \nabla^{z}_{(1)} \dots \nabla^{z}_{(\alpha+1)} \cdot \\
        &= \(-g_{z\zb}\)^{\alpha}\[g^{z\zb} \ol{\pp}\cdot\]^{\alpha} \cdot \\
        &= (-1)^{\alpha} \nabla_{\zb}^{(\alpha-1)}\dots \nabla_{\zb}^{(0)},
    \end{aligned}
\end{equation}
as expected (up to a sign).
Let us now define the operator $\mathcal{D}_{\alpha}: T^{(-\alpha)} \rightarrow T^{(0)}$, with
\begin{equation}
    \ol{\(\nabla^{\alpha}\)^{\dag}} g_{z\zb}^{\alpha} \cdot \equiv (-1)^{\alpha}g_{z\zb}^{\alpha} \mathcal{D}_{\alpha} \cdot .
\end{equation}
Using \eqref{eq: cov d conjugation}, we find that
\begin{equation}
    \begin{aligned}
        \mathcal{D}_{\alpha} \cdot &\equiv \nabla_{z}^{(-1)} \dots \nabla_{z}^{(-\alpha)} \cdot \\
        &= \(\pp + 2\frac{\pp\Omega}{\Omega}\) \dots \(\pp + 2\alpha\frac{\pp\Omega}{\Omega}\)\cdot \\
        &= \[\Omega^{-2}\pp \cdot\]^{\alpha}\(\Omega^{2\alpha}\cdot\).
    \end{aligned}
    \label{eq: mathcalD expressions}
\end{equation}
We recognise that this is just the covariant $z$-derivative acting on a tensor with $\alpha$ \textit{contravariant} $z$-indices.
Using \eqref{eq: cov d conjugation}, we also see that the operator $\mathcal{D}_{\alpha}$ is related to $\nabla^{\alpha}$ via integration by parts as
\begin{equation}
    \int \dzz \sqrt{|g|} \nabla^{\alpha}f(z,\zb) \,g(z,\zb) = (-1)^{\alpha} \int \dzz \sqrt{|g|} f(z,\zb) \,\mathcal{D}_{\alpha} g(z,\zb),
    \label{eq: nabla mathcalD relation}
\end{equation}
for some test functions $f$ and $g$.


\subsection{Useful identities}

We will now derive various identities which will prove useful later. 

\begin{identity}
\label{idty: commutator cov d ol pp}
The operator $\nabla^{\alpha}$ obeys the following commutation relation with $\ol{\pp}$
\begin{equation}
    \[\nabla^{\alpha},\ol{\pp}\] \cdot = \frac{\alpha(\alpha-1)}{2}  \(-\frac{R}{2}g_{z\zb} \) \nabla^{\alpha-1}_{z} \cdot 
\end{equation}
for Einstein manifolds.
\end{identity}
\begin{proof}
First note that the commutator of covariant derivatives acting on a rank-$h$ holomorphic tensor $T$
\begin{equation}
    \[\nabla_{z},\nabla_{\zb}\] T_{z \dots z} = h\(-\frac{1}{2}Rg_{z\zb}\) T_{z \dots z},
\end{equation}
and we therefore have
\begin{equation}
    \[\nabla_{z}^{(h)},\ol{\pp}\] \cdot = h \(-\frac{1}{2}Rg_{z\zb}\) \cdot .
\end{equation}
Now consider a scalar test distribution $\phi$.
For Einstein manifolds, $R$ is constant, and 
\begin{equation}
    \begin{aligned}
        \[\nabla^{\alpha},\ol{\pp}\]&= \sum_{n=0}^{\alpha-1} \nabla_{z}^{(\alpha-1)} \dots \nabla_{z}^{(\alpha-n)} \[\nabla_{z}^{\alpha-1-n},\ol{\pp}\] \nabla_{z}^{(\alpha-2-n)} \dots \nabla^{(0)}_{z} \phi \\
        &= -\frac{R}{2} \sum_{n=0}^{\alpha-1}(\alpha-1-n) \nabla_{z}^{(\alpha-1)} \dots \nabla_{z}^{(\alpha-n)} \(g_{z\zb}\nabla_{z}^{(\alpha-2-n)} \dots \nabla^{(0)}_{z} \phi\) \\
        &= -\frac{R}{2} \[\sum_{n=0}^{\alpha-1}(\alpha-1-n)\] \nabla_{z}^{(\alpha-2)} \dots  \nabla^{(0)}_{z} \phi \\
        &= -\frac{R}{2}g_{z\zb}\frac{\alpha(\alpha-1)}{2} \nabla^{\alpha-1} \phi,
    \end{aligned}
\end{equation}
where we have made repeated use of \eqref{eq: cov d conjugation}.
\end{proof}

\begin{identity}
\label{idty: commutator mathcalD ol pp}
The operator $\mathcal{D}_{\alpha}$ obeys the following commutation relation with $\ol{\pp}$
\begin{equation}
    \[\ol{\pp},\mathcal{D}_{\alpha} \] \cdot =  \frac{\alpha(\alpha+1)}{2} \(-\frac{R}{2}g_{z\zb}\)   \nabla_{z}^{(-2)}  \dots \nabla_{z}^{(-\alpha)} \cdot.
\end{equation}
Further, when acting on a holomorphic function $f(z)$, the latter takes the following simple form
\begin{equation}
    \(\ol{\pp}\pp + \frac{\alpha(\alpha+1)}{4} R g_{z\zb} \)\mathcal{D}_{\alpha} f(z) =  0.
\end{equation}
\end{identity}
\begin{proof}
Similar to identity \ref{idty: commutator cov d ol pp}, we have
\begin{equation}
    \begin{aligned}
        \[\ol{\pp},\mathcal{D}_{\alpha}\] \phi &= \sum \nabla_{z}^{(-1)} \dots \nabla_{z}^{(-n+1)} \[\ol{\pp},\nabla_{z}^{(-n)}\] \nabla_{z}^{(-n-1)} \dots \nabla_{z}^{(-\alpha)} \phi \\
        &= -\frac{R}{2}\sum_{n=1}^{\alpha} n \nabla_{z}^{(-1)} \dots \nabla_{z}^{(-n+1)} \( g_{z\zb}\nabla_{z}^{(-n-1)} \dots \nabla_{z}^{(-\alpha)} \phi\)  \\
        &= -\frac{R}{2}\(\sum_{n=1}^{\alpha} n \) g_{z\zb} \nabla_{z}^{(-2)}  \dots \nabla_{z}^{(-\alpha)} \phi  \\
        &= -\frac{R}{2} \frac{\alpha(\alpha+1)}{2} g_{z\zb}  \nabla_{z}^{(-2)}  \dots \nabla_{z}^{(-\alpha)} \phi.
    \end{aligned}
\end{equation}
This shows the first part of the identity.
Now, let us act on this with $\nabla_{z}^{(0)} = \pp $.
Then,

\begin{equation}
    \begin{aligned}
        \pp\[\ol{\pp},\mathcal{D}_{\alpha}\] \phi &= \frac{\alpha(\alpha+1)}{2} \(-\frac{R}{2}\) \nabla_{z}^{(0)} \[ g_{z\zb}   \nabla_{z}^{(-2)}  \dots \nabla_{z}^{(-\alpha)} \phi \] \\
        &= \frac{\alpha(\alpha+1)}{2} \(-\frac{R}{2}\) g_{z\zb} \nabla_{z}^{(-1)}  \dots \nabla_{z}^{(-\alpha)} \phi \\
        &= \frac{\alpha(\alpha+1)}{2} \(-\frac{R}{2}\) g_{z\zb} \mathcal{D}_{\alpha}\phi.
    \end{aligned}
\end{equation}
The second part of the identity follows straightforwardly when $\phi$ is a holomorphic function $f(z)$.
\end{proof}
There are additional identities which we only hold in $\text{(A)dS}_{2}$, which we will show separately. 

\subsection{Identities in $\text{EAdS}_{2}$ \label{sapp: eads2 identities}}

After mapping $\text{EAdS}_{2}$ to the complex UHP, we land on the metric \eqref{eq: general 2d metric} with the conformal factor $\Omega$ given in \eqref{eq: conformal factors (a)ds}.
We collect some useful expressions and identities for the covariant derivative and its dual here.

First, note that for $\text{EAdS}_{2}$, the conformal factor $\Omega$ satisfies 
\begin{equation}
    \pp^{n}\Omega^{m} = \frac{(m+n-1)!}{(m-1)!}\Omega^{m} \left(\frac{-1}{z-\zb}\)^{n}.
\end{equation}
Therefore, we find the following expressions
\begin{subequations}
    \begin{align}
        &\begin{aligned}
            \nabla^{\alpha+1}\cdot\, &=\sum_{n=0}^{\alpha}\frac{\alpha!(\alpha+1)!}{n!(n+1)!}\frac{1}{(\alpha-n)!}\left(\frac{1}{z-\zb}\right)^{\alpha-n}\partial^{n+1} \\
            &= \(\frac{1}{z-\zb}\)^{\alpha+1}\pp^{\alpha}\bigg[\(\frac{1}{z-\zb}\)^{-\alpha-1}\pp\;\cdot\bigg]
        \end{aligned}\\
        &\begin{aligned}
            \mathcal{D}_{\alpha} \cdot\,&=\sum_{n=0}^{\alpha}\frac{(2\alpha-n)!}{n!(\alpha-n)!}(-1)^{\alpha-n}\frac{1}{\(z-\zb\)^{\alpha-n}}\pp^{n} \cdot \\
            &= \(\frac{1}{z-\zb}\)^{-\alpha-1}\partial^\alpha\bigg[\(\frac{1}{z-\zb}\)^{\alpha+1} \;\cdot\bigg].
        \end{aligned}
    \end{align}
    \label{eq: nabla mathcalD ads}%
\end{subequations}
We can now use these expressions to derive identities which will prove useful later.
The following identity has a $\text{dS}_{2}$ analogue in identity \ref{idty: nabla mathcalD composition ds}, although we use different methods to prove them.
We would like to assure the reader that these two approaches are interchangeable---we include both just for the reader's enjoyment.
\begin{identity}
\label{idty: nabla mathcalD ads}
The operators $\nabla^{\alpha+1}$ and $\mathcal{D}_{\alpha}$ satisfy the following identity:
\begin{equation}
    \nabla^{\alpha+1}\mathcal{D}_{\alpha}\cdot = \partial^{2\alpha+1}\cdot .
\end{equation}
\end{identity}
\begin{proof}
To begin, note that because
\begin{equation}
    \nabla^{\alpha+1}=(z-\zb)^{-2\alpha}\[\partial(z-\zb)^2\]^\alpha\partial,\quad \mathcal{D}_\alpha=\[(z-\zb)^2\partial\]^\alpha(z-\zb)^{-2\alpha},
\end{equation}
we have 
\begin{equation}
    \nabla^{\alpha+1}\mathcal{D}_\alpha=(z-\zb)^{-2\alpha}\partial\[(z-\zb)^{2}\partial\]^{2\alpha}(z-\zb)^{-2\alpha}.
\end{equation}
To prove that this is equal to $\partial^{2\alpha+1}$, we will first show that
\begin{equation}
    \label{eq: nabla mathcalD ads identity 1}
    \[(z-\zb)^{2}\partial\]^{n}(z-\zb )^{-n}=n!(-1)^n\sum_{m=0}^{n}\frac{1}{m!}(\zb-z)^{m}\partial^m.
\end{equation}
Given this, and acting from the left with $\partial$, we have
\begin{equation}
    \begin{aligned}
        \partial\[(z-\zb)^{2}\partial\]^{n}(z-\zb)^{-n} &=n!(-1)^{n}\sum_{m=0}^{n}\frac{1}{m!}\left[-m(\zb-z)^{m-1}\partial^m+(\zb-z)^{m}\partial^{m+1}\right] \\
        &=-n!(-1)^{n}\sum_{m=0}^{n-1}\frac{1}{m!}(\zb-z)^{m}\partial^{m+1}+n!(-1)^{n}\sum_{m=0}^{n}\frac{1}{m!}(\zb-z)^{m}\partial^{m+1} \\
        &=(z-\zb)^{n}\partial^{n+1}.
    \end{aligned}
\end{equation}
Setting $n=2\alpha$, we get 
\begin{equation}
    \begin{aligned}
        \nabla_z^{\alpha+1}\mathcal{D}_\alpha &=(z-\zb)^{-2\alpha}\partial\[(z-\zb)^{2}\partial\]^{2\alpha}(z-\zb)^{-2\alpha} \\
        &=(z-\zb)^{-2\alpha}(z-\zb)^{2\alpha}\partial^{2\alpha+1} \\
        &=\partial^{2\alpha+1},
    \end{aligned}
\end{equation}
as required.

To show \eqref{eq: nabla mathcalD ads identity 1}, we first rewrite 
\begin{equation}
    \begin{aligned}
        \[(z-\zb)^{2}\partial\]^n(z-\zb)^{-n}&=[(z-\zb)^{2}\partial]^{n-1}(z-\zb)^2\partial(z-\zb)^{-n} \\
        &=[(z-\zb)^{2}\partial]^{n-1}(z-\zb)^{-(n-1)}\left(-n+(z-\zb)\,\partial\right)\\
        &\;\;\vdots\\
        &=\left(-1+(z-\zb)\,\partial\right)\left(-2+(z-\zb)\,\partial\right)\cdots \left(-n+(z-\zb)\,\partial\right).
    \end{aligned}
\end{equation}
Now consider the operator
\begin{equation}
    \mathcal{O}_{n,\ell}=\left(-(n-\ell+1)+(z-\zb)\,\partial\right)\left(-(n-\ell+2)+(z-\zb)\,\partial\right)\cdots \left(-n+(z-\zb)\,\partial\right),
\end{equation}
with $\ell=n$ corresponding to $\[(z-\zb)^{2}\partial\]^n(z-\zb)^{-n}$. 

We wish to write $\mathcal{O}_{n,\ell}$ as a sum of derivatives, which we take to be of the form
\begin{equation}
    \label{eq: nabla mathcalD ads identity 2}
    \mathcal{O}_{n,\ell}=\sum_{m=0}^{\ell}a_{m,\ell}(z-\zb)^{m}\partial^m
\end{equation}
for some undetermined coefficients $a_{m,\ell}$. 
Using the fact
\begin{equation}
    \left(-(n-\ell)+(z-\zb)\,\partial\right)\mathcal{O}_{n,\ell}=\mathcal{O}_{n,\ell+1}
\end{equation}
along with the definition (\ref{eq: nabla mathcalD ads identity 2}), we can derive a recursion relation for the coefficients $a_{m,\ell}$:
\begin{align}
    \label{eq: nabla mathcalD ads identity 3}
    a_{m,\ell+1}=-(n-\ell-m)a_{m,\ell}+a_{m-1,\ell}.
\end{align}
Given the conditions $a_{\ell,\ell}=1$ and $a_{0,\ell}=(-1)^\ell n!/(n-\ell)!$, (\ref{eq: nabla mathcalD ads identity 3}) has a unique solution
\begin{align}
    a_{m,\ell}=(-1)^{\ell-m}(\ell-m)!\binom{\ell}{m}\binom{n-m}{\ell-m}.
\end{align}
Taking $\ell=n$, we get (\ref{eq: nabla mathcalD ads identity 1}), thereby proving that $\nabla_z^{\alpha+1}\mathcal{D}_\alpha=\partial^{2\alpha+1}$.
\end{proof} 

\begin{identity}
\label{idty: ads cov d anti composition identity}
The operators $\nabla_z^{\alpha+1}$ and $\ol{\mathcal{D}}_{\alpha}$ satisfy the identity,
\begin{equation}
    \nabla_z^{\alpha+1}\ol{\mathcal{D}}_{\alpha}\tilde{f}(\zb) = 0
\end{equation}
for any anti-holomorphic function $\tilde{f}(\zb)$.
\end{identity}
\begin{proof}
It is easiest to demonstrate this using the representations of $\nabla^{\alpha+1}$ and $\ol{\mathcal{D}}_\alpha$ in \eqref{eq: nabla mathcalD ads}.
Because $\tilde{f}(\zb)$ is anti-holomorphic, we have
\begin{equation}
    \begin{aligned}
        \nabla^{\alpha+1}\ol{\mathcal{D}}_{\alpha}\tilde{f}(\zb) &=\sum_{n,m=0}^{\alpha}\frac{\alpha!(\alpha+1)!(2\alpha-m)!}{n!(n+1)!(\alpha-n)!m!(\alpha-m)!}\left(z-\zb\right)^{n-\alpha}\partial^{n+1}\left(z-\zb\right)^{m-\alpha}\ol{\pp}^{m}\tilde{f}(\zb)\nonumber\\
        &=\sum_{m=0}^{\alpha}\frac{\alpha!(\alpha+1)!(2\alpha-m)!}{m!(\alpha-m)!(\alpha-m-1)!}(z-\zb)^{m-2\alpha-1}\ol{\pp}^m\tilde{f}(\zb)\sum_{n=0}^{\alpha}\frac{(n+\alpha-m)!}{n!(n+1)!(\alpha-n)!}(-1)^{n+1}\\
        &=-\sum_{m=0}^{\alpha}\frac{(2\alpha-m)!}{(\alpha-m-1)!(m-\alpha)!}(z-\zb)^{m-2\alpha-1}\ol{\pp}^m\tilde{f}(\zb) \\
        &=0.
    \end{aligned}
\end{equation}
\end{proof}

\begin{identity}
\label{idty: ads shift mode identity}
When $|n|\leq k$, we have
\begin{equation}
    \mathcal{D}_k\frac{1}{z^{n-k}}=(-1)^k\ol{\mathcal{D}}_k\frac{1}{\zb^{n-k}}.
    \label{eq: AdS shift identity}
\end{equation}
\end{identity}
\begin{proof}
Using the explicit expressions \eqref{eq: nabla mathcalD ads} and the corresponding expression for $\ol{\mathcal{D}}_k$, we may write the left- and right- hand sides of \eqref{eq: AdS shift identity} in terms of Jacobi polynomials
\begin{equation}
    \label{eq: explicit Dk zn-k AdS}
    \mathcal{D}_k\frac{1}{z^{n-k}}= k!z^{-n}P_{k}^{(-n,n)}\(-\frac{z+\zb}{z-\zb}\),\quad \ol{\mathcal{D}}_k\frac{1}{\zb^{n-k}}= k!\zb^{-n}P_{k}^{(-n,n)}\(\frac{z+\zb}{z-\zb}\).
\end{equation}
The Jacobi polynomial $P_m^{(\alpha,\beta)}(x)$ has a special series representation when $n$, $n+\alpha$, $n+\beta$, and $n+\alpha +\beta$ are all non-negative integers \cite{Askey:1975}:
\begin{equation}
    P_m^{(\alpha,\beta)}(x) = (m+\alpha)!(m+\beta)!\sum_{s=0}^m\frac{1}{s!(m+\alpha-s)!(\beta+s)!(m-s)!}\(\frac{x-1}{2}\)^{m-s}\(\frac{x+1}{2}\)^s.
    \label{eq: Jacobi poly identity}
\end{equation}
Because $|n|\leq k$, the polynomials appearing in \eqref{eq: explicit Dk zn-k AdS} can be written in this way. 
Using this series representation, it is straightforward to show that,  
\begin{subequations}
    \begin{align}
        P_k^{(-n,n)}\(-\frac{z+\zb}{z-\zb}\)&=\(\frac{z}{\zb-z}\)^n\frac{(k+n)!(k-n)!}{(k!)^2}P_{k-n}^{(n,n)}\(-\frac{z+\zb}{z-\zb}\)\\
        P_k^{(-n,n)}\(\frac{z+\zb}{z-\zb}\)&=\(\frac{\zb}{z-\zb}\)^n\frac{(k+n)!(k-n)!}{(k!)^2}P_{k-n}^{(n,n)}\(\frac{z+\zb}{z-\zb}\).
    \end{align}
\end{subequations}
Together with the symmetry property $P_m^{(\alpha,\beta)}(-x)=(-1)^mP_m^{(\beta,\alpha)}(x)$, the result \eqref{eq: AdS shift identity} follows.
\end{proof}

\subsection{Identities in $\text{EdS}_{2}$ \label{sapp: eds2 identities}}

First, note that for $\text{EdS}_{2}$, the conformal factor $\Omega$ satisfies 
\begin{equation}
    \pp^{n}\Omega^{m} = \frac{(m+n-1)!}{(m-1)!}\Omega^{m} \left(\frac{-\zb}{1+z\zb}\)^{n}.
\end{equation}
Therefore, we have the following expressions for the covariant derivative and its dual
\begin{subequations}
    \begin{align}
        &\begin{aligned}
            \nabla^{\alpha+1}\cdot\, &=\sum_{n=0}^{\alpha}\frac{\alpha!(\alpha+1)!}{n!(n+1)!}\frac{1}{(\alpha-n)!}\left(\frac{\zb}{1+z\zb}\right)^{\alpha-n}\partial^{n+1} \\
            &= \(\frac{1}{1+z\zb}\)^{\alpha+1}\pp^{\alpha}\bigg[\(\frac{1}{1+z\zb}\)^{-\alpha-1}\pp\;\cdot\bigg]
        \end{aligned} \\
        & \begin{aligned}
            \mathcal{D}_{\alpha} \cdot\,&=\sum_{n=0}^{\alpha}\frac{(2\alpha-n)!}{n!(\alpha-n)!}\(\frac{-\zb}{1+z\zb}\)^{\alpha-n}\pp^{n} \cdot \\
            &= \(\frac{1}{1+z\zb}\)^{-\alpha-1}\partial^\alpha\bigg[\(\frac{1}{1+z\zb}\)^{\alpha+1} \;\cdot\bigg]
        \end{aligned}
    \end{align}
    \label{eq: nabla mathcalD ds}%
\end{subequations}
We can now use these expressions to derive the $\text{EdS}_{2}$ version of identity \ref{idty: nabla mathcalD ads}.
As promised, we will do this differently from before, although both derivations are completely interchangeable.
\begin{identity}
\label{idty: nabla mathcalD composition ds}
The operators $\nabla^{\alpha+1}$ and $\mathcal{D}_{\alpha}$ satisfy the following identity:
\begin{equation}
    \nabla_z^{\alpha+1}\mathcal{D}_{\alpha}\cdot = \partial^{2\alpha+1}\cdot .
\end{equation}
\end{identity}
\begin{proof}
To see this, we will make repeated use of the methods described in \cite{Egorychev:1984}.
First note that using \eqref{eq: nabla mathcalD ds}, we have
\begin{equation}
    \begin{aligned}
        \nabla_{z}^{\alpha+1}\mathcal{D}_{\alpha}\phi &= \sum_{i=0}^{\alpha} \frac{\alpha!(\alpha+1)!}{i!(i+1)!(\alpha-i)!}\(\frac{\zb}{1+z \zb}\)^{\alpha-i} \pp^{i+1} \[\sum_{j=0}^{\alpha} \frac{(\alpha+j)!}{j!(\alpha-j)!} \(\frac{-\zb}{1+z\zb}\)^{j}\pp^{\alpha-j}\phi \] \\
        &= \sum_{i,j=0}^{\alpha} (-1)^{j+1-\ell} \sum_{\ell=0}^{i+1} \lambda_{i,j}^{\ell+\alpha-j} \(\frac{\zb}{1+z\zb}\)^{j+\alpha+1-\ell}  \pp^{\alpha-j+\ell}\phi \\
        &= \sum_{i,j=0}^{\alpha} \sum_{\ell=\alpha-j}^{\alpha-j+i+1} (-1)^{\alpha+1-\ell} \lambda^{\ell}_{i,j} \(\frac{\zb}{1+z\zb}\)^{2\alpha+1-\ell}  \pp^{\ell}\phi \\
        & = \sum_{\ell=0}^{2\alpha+1}\sum_{i=\text{max}(0,\ell-\alpha-1)}^{\alpha}\sum_{j=\text{max}(\alpha-\ell,0)}^{\text{min}(\alpha+1+i-\ell,\alpha)} (-1)^{\alpha+1-\ell}\lambda_{i,j}^{\ell} \(\frac{\zb}{1+z\zb}\)^{2\alpha+1-\ell} \pp^{\ell}\phi,
    \end{aligned}
\end{equation}
with 
\begin{equation}
    \begin{aligned}
        \lambda_{i,j}^{\ell} = (-1)^{i} (\alpha+1)!\binom{\alpha}{i}\binom{\alpha+i-\ell}{j-1} \frac{(\alpha+j)!}{j!(\alpha-j)!(\ell+j-\alpha)!}.
    \end{aligned}
\end{equation}
In particular, 
\begin{equation}
     \nabla_{z}^{\alpha+1}\mathcal{D}_{\alpha}\phi = \pp^{2\alpha+1}\phi + R,
\end{equation}
and our claim is that
\begin{equation}
    R = \sum_{\ell=0}^{2\alpha} (-1)^{\alpha+1-\ell}R_{\ell}\(\frac{\zb}{1+z\zb}\)^{2\alpha+1-\ell} \pp^{\ell}\phi,\quad R_{\ell} = \sum_{i=0}^{\alpha}\sum_{j=0}^{\alpha+1-\ell+i} \lambda_{i,j}^{\ell} 
\end{equation}
vanishes.
In particular, this has to hold for each individual $\ell$, so we need $R_{\ell} =0$.

Let us first consider $0 \leq \ell \leq \alpha$.
In this case, 
\begin{equation}
    R_{\ell} = \sum_{j=0}^{\alpha} \frac{(\alpha+1)!(\alpha+j)!}{j!(\alpha-j)!(\ell+j-\alpha)!} \sum_{i=0}^{\alpha} (-1)^{i} \binom{\alpha}{i} \binom{\alpha+i-\ell}{j-1}.
\end{equation}
Now note that
\begin{equation}
    \begin{aligned}
        \sum_{i=0}^{\alpha} (-1)^{i} \binom{\alpha}{i} \binom{\alpha+i-\ell}{j-1} &= \oint_{|z|=\epsilon} \frac{\dd z}{2\pi i}\sum_{i=0}^{\alpha} (-1)^{i} \binom{\alpha}{i} \frac{(1+z)^{\alpha+i-\ell}}{z^{j}} \\
        &= \oint_{|z|=\epsilon} \frac{\dd z}{2\pi i}\frac{(1+z)^{\alpha-\ell}}{z^{j}} \[1-(1+z)\]^{\alpha} \\
        &= \oint_{|z|=\epsilon} \frac{\dd z}{2\pi i} (1+z)^{\alpha-\ell} z^{\alpha-j}  \\
        &= 0.
    \end{aligned}
\end{equation}
The individual terms in the sum over $j$ therefore vanish in the expression above, showing that $R_{\ell\leq \alpha} =0 $.

Next, let us consider $\alpha+1 \leq \ell \leq 2\alpha$.
Now, let us write
\begin{equation}
    R_{\ell} = \ell! (\alpha+1) \sum_{i=0}^{\alpha}(-1)^{i}\binom{\alpha}{i} \sum_{j=-1}^{\alpha+i-\ell} \binom{\alpha+i-\ell}{j} \binom{\ell}{\alpha-j-1}\binom{\alpha+j+1}{j+1}.
\end{equation}
It is useful to separate the contributions to $R_{\ell}$ for $j=-1$ and the sum over $j \geq 0$. 
Let us denote these $S_{1}$ and $S_{2}$ respectively.
The only term which survives in $S_{1}$ is 
\begin{equation}
    \begin{aligned}
        \frac{1}{(\alpha+1)\ell!}S_{1} &= \sum_{i=0}^{\alpha}(-1)^{i}\binom{\alpha}{i}  \binom{\alpha+i-\ell}{-1} \binom{\ell}{\alpha}\binom{\alpha}{0} \\
        &= (-1)^{\ell-(\alpha+1)} \binom{\ell}{\alpha} \binom{\alpha}{\ell-\alpha-1} .
    \end{aligned}
\end{equation}
We are left with showing that this precisely cancels the contribution from $S_{2}$.
To see this, let us first rewrite it as the following contour integral
\begin{equation}
    \begin{aligned}
        \frac{1}{(\alpha+1)\ell!}S_{2} &= \sum_{i=0}^{\alpha} (-1)^{i} \binom{\alpha}{i} \sum_{j=0}^{\alpha+i-\ell} \binom{\alpha+i-\ell}{j} \binom{\ell}{\alpha-j-1}\binom{\alpha+j+1}{j+1} \\
        &= \oint_{|z|=\epsilon} \frac{\dd z}{2\pi i}\oint_{|w|=\epsilon} \frac{\dd w}{2\pi i}\oint_{|u|=\epsilon} \frac{\dd u}{2\pi i} \frac{1}{1-u}w^{\ell-2-2\alpha}u^{\ell-2\alpha-1}(1+z)^{\ell}(1+w)^{\alpha+1} \\
        &\qquad \quad (w+z+wz)^{\alpha-\ell}(uw-w-z-zw)^{\alpha} z^{-\alpha}.
    \end{aligned}
\end{equation}
Then, first evaluating the $u$- then $z$-residues, 
\begin{equation}
    \begin{aligned}
        \frac{1}{(\alpha+1)\ell!}S_{2} &=  \oint_{|z|=\epsilon} \frac{\dd z}{2\pi i}\oint_{|w|=\epsilon} \frac{\dd w}{2\pi i} \sum_{n=0}^{2\alpha-\ell} \binom{\alpha}{2\alpha-\ell-n} (-1)^{\ell+n-\alpha} \frac{(1+z)^{\ell}(1+w)^{\alpha+1}(w+z+zw)^{m}}{z^{\alpha}w^{2+n}} \\
        &= \oint_{|w|=\epsilon} \frac{\dd w}{2\pi i} (1+w)^{\alpha+1} \sum_{m=0}^{\alpha-1} \frac{(1+w)^{m}}{w^{m+2}} \binom{\ell}{\alpha-1-m} \sum_{n=0}^{2\alpha-\ell}(-1)^{\ell+n-\alpha} \binom{\alpha}{2k-\ell-n}\binom{n}{m} \\
        &= \sum_{m=0}^{\alpha-1} (-1)^{\ell-\alpha-m}\binom{\ell}{\alpha-1-m} \binom{\alpha-m-1}{\ell-\alpha-1}\binom{m+\alpha+1}{m+1} \\
        &= (-1)^{\alpha-\ell} \binom{\ell}{\alpha}\binom{\alpha}{\ell-\alpha-1}.
    \end{aligned}
\end{equation}
This shows $R_{\ell \geq \alpha+1 } = 0$. 
Together, this shows that $R=0$.
\end{proof}

\begin{identity}
\label{idty: cov d anti composition identity ds}
The operators $\nabla_z^{\alpha+1}$ and $\ol{\mathcal{D}}_{\alpha}$ satisfy the identity,
\begin{equation}
    \nabla_z^{\alpha+1}\ol{\mathcal{D}}_{\alpha}\tilde{f}(\zb) = 0
\end{equation}
for any anti-holomorphic function $\tilde{f}(\zb)$.
\end{identity}
\begin{proof}
It is easiest to demonstrate this using the representations of $\nabla^{\alpha+1}$ and $\ol{\mathcal{D}}_\alpha$ in \eqref{eq: nabla mathcalD ds}.
Because $\tilde{f}(\zb)$ is anti-holomorphic, we have
\begin{equation}
    \begin{aligned}
        \nabla^{\alpha+1}\ol{\mathcal{D}}_{\alpha}\tilde{f}(\zb) &=\sum_{n,m=0}^{\alpha}\frac{\alpha!(\alpha+1)!(2\alpha-m)!}{n!(n+1)!(\alpha-n)!m!(\alpha-m)!}\(\frac{\zb}{1+z\zb}\)^{\alpha-n}\partial^{n+1}\(\frac{-z}{1+z\zb}\)^{\alpha-m}\ol{\pp}^{m}\tilde{f}(\zb) \\
        &=  \sum_{n,m=0}^{\alpha}\sum_{i=0}^{n+1}\frac{(m+i-1)!}{i!(n+1-i)!(m+i-n-1)!}\frac{\alpha!(\alpha+1)!(2\alpha-m)!}{n!(\alpha-n)!(m-1)!(\alpha-m)!} (-1)^{\alpha+i+m} \\
        &\qquad \frac{z^{m+i-n}\zb^{\alpha+i-n}}{(1+z \zb)^{\alpha+i+m-n}}\ol{\pp}^{m}\tilde{f}(\zb) \\ 
        &= \sum_{m=0}^{\alpha}\sum_{i=0}^{\alpha+1}\sum_{n=\text{max}\{0,i-1\}}^{\alpha}(-1)^{\alpha+m+n+1-i}\frac{(m+n-i)!}{n!(\alpha-n)!(n+1-i)!}\frac{ \alpha!(\alpha+1)! (2\alpha-m)!}{i!(m-1)!(m-i)!(\alpha-m)!} \\
        &\qquad \frac{z^{m+2-i} \zb^{\alpha+1-i}}{(1+z \zb)^{\alpha+1+m-i}} \ol{\pp}^{m}\tilde{f}(\zb) \\
        &= \sum_{m=0}^{\alpha}\sum_{i=0}^{\alpha+1}(-1)^{\alpha+m+1}\frac{(\alpha+1)! (2\alpha-m)!}{(\alpha+1-i)!( i-1 - m)!i!(m-i)!} \frac{z^{m+1-i}\zb^{\alpha+1-i}}{(1+z \zb)^{\alpha+1+m-i}} \ol{\pp}^{m}\tilde{f}(\zb) \\
        &= 0,
    \end{aligned}
\end{equation}
as required.
\end{proof}

\begin{identity}
\label{idty: ds shift mode identity}
When $|n|\leq k$, 
\begin{equation}
    \mathcal{D}_k\frac{1}{z^{n-k}}=(-1)^n\ol{\mathcal{D}}_k\frac{1}{\zb^{-n-k}}.
\label{eq: dS shift identity}
\end{equation}
\end{identity}
\begin{proof}
Using the explicit expressions \eqref{eq: nabla mathcalD ds} and the corresponding expression for $\ol{\mathcal{D}}_k$, we may write the left- and right-hand sides of \eqref{eq: dS shift identity} in terms of Jacobi polynomials
\begin{equation}
    \label{eq: explicit Dk zn-k dS}
    \mathcal{D}_k\frac{1}{z^{n-k}}= k!z^{-n}P_{k}^{(-n,n)}\(\frac{1-z\zb}{1+z\zb}\),\quad  \ol{\mathcal{D}}_k\frac{1}{\zb^{-n-k}}=k!\zb^{n}P_{k}^{(n,-n)}\(\frac{1-z\zb}{1+z\zb}\).
\end{equation}
As in the $\text{EAdS}_2$ case, because $|n|\leq k$ we may use \eqref{eq: Jacobi poly identity} and the symmetry of Jacobi polynomials to write,
\begin{subequations}
    \begin{align}
        P_{k}^{(-n,n)}\(\frac{1-z\zb}{1+z\zb}\)&=\(-\frac{z \zb}{1+z \zb}\)^n\frac{(k-n)!(k+n)!}{(k!)^2}P_{k-n}^{(n,n)}\(\frac{1-z\zb}{1+z\zb}\),\\
        P_{k}^{(n,-n)}\(\frac{1-z\zb}{1+z\zb}\)&=\(\frac{1}{1+z \zb}\)^n\frac{(k-n)!(k+n)!}{(k!)^2}P_{k-n}^{(n,n)}\(\frac{1-z\zb}{1+z\zb}\).
    \end{align}
    \label{eq:simplified Jacobi ds}
\end{subequations}
Plugging these into \eqref{eq: explicit Dk zn-k dS}, we get \eqref{eq: dS shift identity}.
\end{proof}

\section{(Anti-)holomorphicity of $\text{EAdS}_{2}$ field strength \label{app: holomorphicity}}

In this appendix, we show that the conserved current $F$ of the discrete series scalar in $\text{EAdS}_{2}$ satisfies the continuity equation as in \eqref{eq: F (anti-)holomorphicity}.
The analogous statement for $\text{EdS}_{2}$ follows similarly.

By definition of the field strength $F$ in \eqref{eq: field strength components}, 
\begin{equation}
    \ol{\pp} F = \[\ol{\pp},\nabla^{k+1}\] \phi + \nabla^{k+1}\ol{\pp}\phi.
\end{equation}
The first term is given by identity \ref{idty: commutator cov d ol pp}.
For the second term, using the definition \eqref{eq: nabla mathcalD ads} and the equation of motion (for generic masses for now), we find that
\begin{equation}
    \begin{aligned}
        \nabla^{k+1}\ol{\pp}\phi &= \sum_{n=0}^{k} \frac{k!(k+1)!}{n!(n+1)!(k-n)!}(z-\zb)^{n-k} \ol{\pp}\pp^{n+1}\phi \\
        &= \sum_{n=0}^{k} \frac{k!(k+1)!}{n!(n+1)!(k-n)!}(z-\zb)^{n-k} \pp^{n}\(-\frac{L^{2}m^{2}}{(z-\zb)^{2}}\phi \) \\
        &= -L^{2}m^{2} \sum_{i=0}^{k} (-1)^{i}\frac{k!(k+1)!}{i!} (z-\zb)^{i-k-2}\pp^{i}\phi \sum_{n=i}^{k} (-1)^{n}\frac{n-i+1}{(n+1)!(n-k)!} \\
        &= -L^{2}m^{2} \sum_{n=0}^{k} \frac{k!(k-1)!}{n!(n-1)!(k-n)!}(z-\zb)^{n-k-2}\pp^{n}\phi.
    \end{aligned}
\end{equation}
Comparing with the definition \eqref{eq: nabla mathcalD ads}, we then have
\begin{equation}
    \ol{\pp}F = \frac{k(k+1)-L^{2}m^{2}}{(z-\zb)^{2}} \nabla^{k}\phi
\end{equation}
on-shell, so holomorphicity follows for the discrete series scalar fields.


\section{Action of $\text{dS}_{2}$ isometries \label{app: ds2 isometry action}}

Let us collect the expressions for the action of the generators of the $\text{dS}_{2}$ isometry group $\text{SO}(3)$ given in \eqref{eq: ell dS op} on the operators in the mode expansion of the field \eqref{eq: ds varphi mode expansion}.

First, let us define
\begin{equation}
    \mu_{n} = \sqrt{\frac{(n-k)(n+k+1)}{n(n+1)}},\quad \nu_{n} = \sqrt{\frac{(n-k-1)(n+k)}{n(n-1)}}, \quad \kappa = \frac{2k+1}{\sqrt{(k+1)(2k+1)!}},
    \label{eq: mu nu definition}
\end{equation}
such that \eqref{eq: ell dS op} becomes
\begin{subequations}
    \begin{align}
        \ell_{1}=&\sum_{n=k+1}^\infty \nu_{n}\alpha_{1-n} \alpha_{n} +\sum_{n=k+1}^\infty \mu_{n}\tilde{\alpha}_{-1-n} \tilde{\alpha}_{n} +\sum_{n=-(k-1)}^{k}(k+1-n)x_{1-n}p_{n}+\kappa \left(i^{k}\tilde{\alpha}_{-k-1}+i^{-k}\alpha_{k+1}\right)p_{-k} \\
        \ell_0=&\sum_{n=k+1}^\infty \alpha_{-n} \alpha_n-\sum_{n=k+1}^\infty \tilde{\alpha}_{-n} \tilde{\alpha}_{n} +i\sum_{n=-k}^knx_{-n}p_{n}\\
        \ell_{-1}=&\sum_{n=k+1}^\infty \mu_{n}\alpha_{-1-n} \alpha_{n} +\sum_{n=k+1}^\infty \nu_{n}\tilde{\alpha}_{1-n} \tilde{\alpha}_{n}+\sum_{n=-k}^{k-1}(k+1+n)x_{-1-n}p_{n}+\kappa\left(i^{-k}\tilde{\alpha}_{k+1}+i^k\alpha_{-k-1}\right)p_k.
    \end{align}
\end{subequations}
Then, using the canonical commutation relations \eqref{eq: ds ccr}, it is easy to verify that for $m \geq k+1$ 
\begin{subequations}
    \begin{alignat}{3}
        [\ell_{1},\alpha_{m}] &= -m\nu_{m+1} \alpha_{m+1},\\
        [\ell_{1},\alpha_{-m}] &= m\nu_{m} \alpha_{-m+1} +(k+1) \kappa i^{-k} \delta_{m,k+1} p_{-k} \\
        [\ell_{1},\tilde{\alpha}_{-m}] &= m \mu_{m}\tilde{\alpha}_{-m-1},\\
        [\ell_{1},\tilde{\alpha}_{m}]  &= -m\mu_{m-1}\tilde{\alpha}_{m-1} -(k+1)\kappa i^{k}\delta_{m,k+1}p_{-k} 
    \end{alignat}
\end{subequations}
and
\begin{subequations}
    \begin{align}
        [\ell_{-1},\alpha_{-m}] &= m \mu_{m}\alpha_{-m-1} \\
        [\ell_{-1},\alpha_{m}] &= -m\mu_{m-1}\delta_{-1+m,n} \alpha_{m-1} -(k+1)\kappa i^k \delta_{m,k+1} p_k \\
        [\ell_{-1},\tilde{\alpha}_{m}] &=  -m \nu_{m+1}\tilde{\alpha}_{m+1} \\
        [\ell_{-1},\tilde{\alpha}_{-m}] &=  m \nu_{m}\tilde{\alpha}_{1-m} +\kappa i^{-k}(k+1)\delta_{m,k+1} p_k.
    \end{align}
\end{subequations}
Further, for all signs of $m$, 
\begin{subequations}
    \begin{align}
        [\ell_{\pm 1},x_{m}] &= -i (k+1\pm m)x_{m\pm 1}-i\kappa \(i^{\pm k}\tilde{\alpha}_{\mp(k+1)}+i^{\mp k}\alpha_{\pm(k+1)}\right) \delta_{m,\pm k} \\
        [\ell_{\pm 1},p_{m}] &= i(k\mp m)p_{m \pm 1} 
    \end{align}
\end{subequations}
and
\begin{subequations}
    \begin{align}
        [\ell_{0},\alpha_{m}] &= -m \alpha_{m}\\
        [\ell_{0},\tilde{\alpha}_{m}] &=m\tilde{\alpha}_{m}\\
        [\ell_{0},x_{m}] &= -m x_{m}\\
        [\ell_{0},p_{m}] &= - m p_{m}.
    \end{align}
\end{subequations}
%


\section{Explicit invariance of action \label{app: explicit invariance}}

In this appendix, we show that the action \eqref{eq: isothermal action} is explicitly invariant under the transformations \eqref{eq: phi chiral trans}.

First note that the variation of the massive scalar action in isothermal coordinates is
\begin{equation}
    \begin{aligned}
        \delta_\ell S &= -\int \dzz 2\(\pp\ol{\pp} \phi - \frac{\Omega^{2}m_{k}^{2}}{4} \phi\) \delta_\ell \phi \\
        &= - (-1)^{k}\int \dzz \(\pp\ol{\pp} \phi - \frac{\Omega^{2}m_{k}^{2}}{4} \phi\) \mathcal{D}_{k} \mathcal{E}(\ell).
    \end{aligned}
\end{equation}
By repeatedly integrating by parts and ignoring any resulting boundary terms, the first term can be manipulated to yield
\begin{equation}
    \begin{aligned}
       -(-1)^{k} \int \dzz  \pp\ol{\pp} \phi\, \mathcal{D}_{k} \mathcal{E}(\ell) &= -\int \dzz \nabla^{k+1}\ol{\pp} \phi \, \mathcal{E}(\ell)\\
       &= -\int \dzz \(  \frac{\Omega^{2}m_k^2}{4}\nabla^{k} \phi  \mathcal{E}(\ell)+ \ol{\pp}\nabla^{k+1}\phi   \mathcal{E}(\ell)\) \\
       &= -\int \dzz \( (-1)^{k}\frac{\Omega^{2}m_k^2}{4} \phi  \, \mathcal{D}_{k}\mathcal{E}(\ell) + \ol{\pp}\nabla^{k+1}\phi \, \mathcal{E}(\ell) \),
    \end{aligned}
\end{equation}
where we used \eqref{eq: nabla mathcalD relation} in the first and third equalities and identity \ref{idty: commutator cov d ol pp} in the second equality.
For the discrete series scalar we have \eqref{eq: shift pt}, so it follows that
\begin{equation}
    \begin{aligned}
        \delta_\ell S &= -\int \dzz \, \ol{\pp}F \, \mathcal{E}(\ell) \\
        &=  -\int \dzz \, \ol{\pp}F\sum_{i=0}^\infty(-1)^i\pp^i\frac{\pp}{\pp F_i}\ell(\{F_n\},z) \\
        &=-\int \dzz \, \sum_{i=0}^\infty\ol{\pp}F_i\frac{\pp}{\pp F_i}\ell(\{F_n\},z)\\
        &=-\int\dzz\,\ol{\pp}\ell\\
        &=0,
    \end{aligned}
\end{equation}
where in the third equality used repeated integration by parts, and in the fourth equality we used the chain rule. The action \eqref{eq: isothermal action} is therefore invariant under \eqref{eq: phi chiral trans}. 

\end{document}

%% file: figures/ads.tikz
\begin{tikzpicture}

\begin{scope}[shift={(-5,0)}]

\draw [very thick, dashed, color=red!70!magenta] (0,-2) -- (0,-3) node[below, color=black] {$\rho = -\pi/2$};
\draw [very thick, color=red!70!magenta] (0,-2) -- (0,2);
\draw [very thick, dashed, color=red!70!magenta] (0,2) -- (0,3);

\draw [very thick, dashed, color=blue!75] (3,-2) -- (3,-3) node[below, color=black] {$\rho = \pi/2$};
\draw [very thick, color=blue!75] (3,-2) -- (3,2);
\draw [very thick, dashed, color=blue!75] (3,2) -- (3,3);

\draw[thick, -latex] (-0.5,-1.5) -- (-0.5,1.5) node[above] {$\tau$};

\draw[thick] (0,-1) -- (3,-1);
\draw[thick] (0,0) -- (3,0);
\draw[thick] (0,1) -- (3,1);

\end{scope}

\draw [thick, -latex] (-0.5,0) -- (2.5,0) node[above] {};

\begin{scope}[shift={(7,0)}]

\fill [black!15] (-3,-3) rectangle (3,0);

\draw [thick, dashed] (-3,-3) rectangle (3,3);
\draw [anchor=base] (2.5, 2.6) -- (2.5, 3) node[right, yshift=-0.2cm] {$z$};
\draw (2.5, 2.6) -- (3, 2.6);

\draw [very thick, color=red!70!magenta] (-3,0) -- (0,0);
\draw [very thick, color=blue!75] (0,0) -- (3,0);
\filldraw [black] (0,0) circle (1.5pt);

\draw [thick] (1,0) arc (0:180:1);
\draw [thick] (2,0) arc (0:180:2);

\end{scope}

\end{tikzpicture}

%% file: figures/modemapping.tikz
\begin{tikzpicture}[
space/.style={circle, fill=black!5, minimum size=1cm},
]

\node[space] (pos) {$\mathcal{V}_{+}$};
\node[space] (nullp) [below right=of pos, xshift=1cm, yshift=1cm] {$\mathcal{N}$};
\node[space] (nullx) [above right=of pos, xshift=1cm, yshift=-1cm] {$\mathcal{Z}$};
\node[space] (neg) [above right=of nullp, xshift=1cm, yshift=-1cm] {$\mathcal{V}_{-}$};

\draw[thick, -latex] (pos) -- (nullp); 
\draw[thick, -latex] (pos.150) arc (30:330:0.5cm);
\draw[thick, -latex] (neg) -- (nullp); 
\draw[thick, -latex] (neg.30) arc (150:-150:0.5cm);
\draw[thick, -latex] (nullx) -- (pos); 
\draw[thick, -latex] (nullx) -- (neg); 
\draw[thick, -latex] (nullx.150) arc (210:-30:0.5cm);
\draw[thick, -latex] (nullp.-30) arc (30:-210:0.5cm);

\end{tikzpicture}

%% file: figures/ds.tikz
\begin{tikzpicture}

\begin{scope}[shift={(-4,0)}]

\draw [very thick] (0,-2) arc (0:-180:2 and 0.5) node[left] {$\eta = -\pi/2$};
\draw [very thick, dashed] (0,-2) arc (0:180:2 and 0.5);
\draw [very thick] (-4,2) arc (-180:180:2 and 0.5) node[left] {$\eta = +\pi/2$}; 
\draw [thick] (-4,-2) -- (-4,2);
\draw [thick] (0,-2) -- (0,2);

\draw [thick] (0,-0.6) arc (0:-180:2 and 0.5);
\draw [thick, dashed] (0,-0.6) arc (0:180:2 and 0.5);
\draw [thick] (0,0.6) arc (0:-180:2 and 0.5);
\draw [thick, dashed] (0,0.6) arc (0:180:2 and 0.5);

\draw [thick, -latex] (-4,-2.5) arc (-165:-15:2 and 0.5) node[right] {$\vartheta$};

\end{scope}

\draw [thick, -latex] (-2.5,0) -- (0.5,0);

\begin{scope}[shift={(4.5,0)}]

\draw [thick, dashed] (-2.5,-2.5) rectangle (2.5,2.5);
\draw [anchor=base] (2, 2.1) -- (2, 2.5) node[right, yshift=-0.2cm] {$z$};
\draw (2, 2.1) -- (2.5, 2.1);
\filldraw [black] (0,0) circle (1.5pt);

\draw [thick] (0,0) circle (1cm);
\draw [thick] (0,0) circle (2cm);

\end{scope}

\end{tikzpicture}

%% file: figures/virasoro.tikz
\begin{equation*}
    F(z) \sim \dots\tikzmark{leftdots} + \alpha_{-k-2}\tikzmark{-k-2}z + \alpha_{-k-1}\tikzmark{-k-1} + \textcolor{gray!50}{\alpha_{-k}\tikzmark{-k}z^{-1} + \dots\tikzmark{middledots} + \alpha_{k}\tikzmark{2k+1}z^{-2k-1}}  + \alpha_{k+1}\tikzmark{k+1}z^{-2k-2} + \alpha_{k+2}\tikzmark{k+2}z^{-2k-3} + \dots\tikzmark{rightdots} 
\end{equation*}

\begin{tikzpicture}[remember picture, overlay]


\draw[thick, blue!85, -latex, bend right=66] ([xshift=-0.5em,yshift=-0.4em]pic cs:leftdots) to ([xshift=-1.5em,yshift=-0.4em]pic cs:-k-2);
\draw[thick, blue!85, -latex, bend right=53] ([xshift=-0.5em,yshift=-0.4em]pic cs:-k-2) to ([xshift=-1.5em,yshift=-0.4em]pic cs:-k-1);
\draw[thick, blue!85, -latex, bend right=27] ([xshift=-0.2em,yshift=-0.4em]pic cs:k+1) to ([xshift=-0.6em,yshift=-0.4em]pic cs:k+2);
\draw[thick, blue!85, -latex, bend right=35] ([xshift=0.3em,yshift=-0.4em]pic cs:k+2) to ([xshift=-0.5em,yshift=-0.4em]pic cs:rightdots);


\draw[dashed, thick, red!70!magenta, -latex, bend left=66] ([xshift=-0.5em,yshift=0.7em]pic cs:leftdots) to ([xshift=-1.5em,yshift=0.7em]pic cs:-k-1);
\draw[dashed, thick, red!70!magenta, -latex, bend left=55] ([xshift=-1em,yshift=0.7em]pic cs:-k-2) to ([xshift=-0.2em,yshift=0.7em]pic cs:-k);
\draw[dashed, thick, red!70!magenta, -latex, bend left=35] ([xshift=-0.2em,yshift=0.7em]pic cs:k+1) to ([xshift=-0.5em,yshift=0.7em]pic cs:rightdots);

\end{tikzpicture}